%% file: aa.tex
\documentclass{aa}  
\usepackage{graphicx}
\usepackage{aalongtable}
\usepackage{txfonts}
\usepackage{natbib}
\bibpunct{(}{)}{;}{a}{}{,}

\newcommand{\cps}{ct~s$^{-1}$}

\newcommand{\ergps}{erg~s$^{-1}$}

\newcommand{\cxc}{\textit{Chandra}}
\newcommand{\xmm}{\textit{XMM-Newton}}

\begin{document}
   \title{The XMM-Newton Optical Monitor Survey of the Taurus Molecular Cloud}

   \titlerunning{{\it XMM-Newton} OM survey of the TMC}
   \authorrunning{Audard et al.}

   \author{Marc Audard\inst{1}\fnmsep\thanks{\emph{New address (since September 2006): }Integral Science Data
   Centre, Ch. d'Ecogia 16, CH-1290 Versoix, Switzerland \& Geneva Observatory, University of Geneva, Ch. des Maillettes 51, 1290 Sauverny, 
   Switzerland} 
          \and
          Kevin R. Briggs\inst{2}
	  \and
	  Nicolas Grosso\inst{3}
	  \and
	  Manuel G\"udel\inst{2}
	  \and
	  Luigi Scelsi\inst{4}
	  \and
	  J\'er\^ome Bouvier\inst{3}
	  \and
	  Alessandra Telleschi\inst{2}
          }

   \offprints{M. Audard}

   \institute{Columbia Astrophysics Laboratory, Columbia University, 550 West 120th Street, Mail code 5247, New York, NY
   10027, USA\\
   	\email{audard@astro.columbia.edu}
         \and
             Paul Scherrer Institut, W\"urenlingen and Villigen, CH-5232 Villigen PSI, Switzerland\\
             \email{briggs@astro.phys.ethz, guedel@astro.phys.ethz.ch, atellesc@astro.phys.ethz.ch}
	 \and
		Laboratoire d'Astrophysique de Grenoble, Universit\'e Joseph Fourier, F-38041
		Grenoble Cedex 9, France\\
		\email{Nicolas.Grosso@obs.ujf-grenoble.fr, Jerome.Bouvier@obs.ujf-grenoble.fr}
	\and
		Dipartimento di Scienze Fisiche ed Astronomiche, Universit\`a di Palermo, Piazza del
		Parlamento 1, I-90134 Palermo, Italy\\
		\email{scelsi@astropa.unipa.it}
             }

   \date{Received August 30, 2006; accepted November 8, 2006}

 
  \abstract
   {The Optical Monitor (OM) on-board \xmm\ obtained optical/ultraviolet data for the \xmm\ Extended Survey of the Taurus Molecular Cloud (XEST), 
   simultaneously with the X-ray detectors.}
   {With the XEST OM data, we aim to study the optical and ultraviolet properties of TMC members, and to do correlative
   studies between the X-ray and OM light curves.    In particular, we aim to determine whether accretion plays a significant role in the 
   optical/ultraviolet and X-ray emissions. The Neupert effect in stellar flares is also investigated.}
   {Coordinates, average count rates and magnitudes were extracted from OM images, together with light curves with low time resolution (a few kiloseconds). 
   For a few sources, OM FAST mode data were also available, and we extracted OM light curves with high time resolution. The OM data were correlated with Two Micron
   All Sky Survey (2MASS) data and with the XEST catalogue in the X-rays.}
   {The XEST OM catalogue contains 2,148 entries of which 1,893 have 2MASS counterparts. However, only
   98 entries have X-ray counterparts, of which 51 of them are
   known TMC members and 12 additional are TMC candidates. The OM data indicate that accreting stars are statistically brighter in the $U$ band 
   than non-accreting stars after correction for extinction, and have $U$-band excesses, most likely due to accretion.
   The OM emission of accreting stars is variable, probably due to accretion spots, but it does not correlate
   with the X-ray light curve, suggesting that accretion does not contribute significantly to the X-ray emission of most accreting stars.
   In some cases, flares were detected in both X-ray and OM light curves and followed a Neupert effect pattern, in which the optical/ultraviolet emission precedes
   the X-ray emission of a flare, whereas the X-ray flux is proportional to the integral of the optical flux.}
   {}


   \maketitle
   
%

\section{Introduction}

The \xmm\  Extended Survey of the Taurus Molecular Cloud (XEST)  \citep{guedel06a} is primarily
focused on the X-ray emission of young stellar and subtellar objects (YSOs). However, since \xmm\  \citep{jansen01} is capable of
observing simultaneously in the X-rays with the European Photon Imaging Cameras \citep{strueder01,turner01} and 
the Reflection Grating Spectrometers \citep{dherder01}, and in the optical and ultraviolet regimes 
with the Optical Monitor \citep{mason01}, we have obtained deep images and sensitive light curves, primarily in 
the $U$ band, but also in the near-ultraviolet regime in some cases.

Simultaneous X-ray and optical/UV coverage of young stars provides the ideal means to study the physical processes
occurring in the stellar upper atmospheres \citep[e.g.,][]{stelzer03} or the interactions between an accretion disk and a star. 
For example, our understanding of the physics of flares in magnetically active stars can benefit from simultaneous monitoring. 
In the chromospheric evaporation model \citep[e.g.,][]{antonucci84}, magnetic reconnection in the corona injects accelerated
particles propagating along the magnetic field lines, heating the chromosphere and the
transition region through collisions (evident in emission in the optical/ultraviolet). Heated material (visible in soft X-rays) 
then moves up along the magnetic field lines into the corona where it cools. A ``Neupert'' effect \citep{neupert68}
should, therefore, be observed in the optical/UV and X-ray light curves, in which the time profile of the 
X-ray light curve is proportional to the time integral of the optical light curve (conversely, the optical flux 
has the same time profile as the derivative of the X-ray flux). This effect has already been observed in main-sequence
active stars \citep{hawley95,guedel02a,hawley03,guedel04,mitrakraev05}. Stellar Neupert effects were also reported from X-rays and 
radio gyrosynchrotron emission of accelerated particles gyrating along the magnetic fields 
\citep{guedel96,guedel02b,smith05}, although it should be mentioned that not all multi-wavelength observations 
necessarily detect typical Neupert effects \citep[e.g.,][]{stelzer03,osten05}, probably because of different conditions
in the flaring source.

Optical studies of young stars have also provided evidence of rotational modulation of active regions, 
suggesting that a solar-type magnetic activity operates in T Tauri stars
\citep[e.g.,][etc.]{rydgren83,rydgren84,bouvier88,bouvier89,bouvier90,bouvier93,vrba93,bouvier95,bouvier97}. 
Detailed analyses indicate that weak-line T Tauri stars (WTTS) on average rotate faster than the still accreting
classical T Tau stars (CTTS), possibly due to disk locking or loss of angular momentum due to 
stellar winds in CTTS \citep{bouvier93}. In addition, rotational modulation in some CTTS can be
dominated by hot spots due to accreting material falling from the accretion disk onto the stellar 
surface \citep{vrba86,bouvier89,vrba89,bouvier93,herbst94,bouvier95}.

Although most low-mass accreting stars display X-ray spectra
typical of hot coronal plasma, at least a few of them display either lower temperature spectra or densities atypical of coronal, or both,
indicating that accretion could play a significant role in the production of X-rays
\citep{kastner02,stelzer04,schmitt05,ness05,robrade06,telleschi06a}. In some cases, the soft X-ray component could
arise from shocks in jets originating from the young accreting stars \citep{guedel05,kastner05,guedel06b}. 
In low-mass young stars with optical outbursts, the sudden rise in mass accretion rate also appears
to impact on the X-ray emission. During the accretion outburst of V1647 Ori, the mean X-ray flux
closely tracked  the
near-infrared luminosity, and the X-ray spectrum hardened \citep{kastner04,grosso05,kastner06}. On the other hand, \citet{audard05} observed
little X-ray flux variability in the early epochs of the outburst, in contrast to the optical/near-infrared flux enhancements, but they found
a change in the X-ray spectrum of the young star V1118 Ori from a dominant hot plasma
pre-outburst to a cool plasma during the outburst. The coronal hot plasma essentially disappeared, probably
because the inner accretion disk disrupted the coronal loops. Finally, accretion may also be a
dominant mechanism in the more massive, accreting Herbig Ae stars \citep{swartz05},
although magnetically confined winds may be a good alternative as well \citep{telleschi06b}.

Optical and X-ray correlations of young stars can, therefore, provide important information on flare physics and
rotational modulation due to spots and active regions in magnetically active stars \citep[e.g.,][]{flaccomio05} 
or due to hot accretion spots. Multi-wavelength studies are ideal to determine to which extent accretion plays a significant
role in the production of X-rays in young accreting stars. Recently, \citet{stassun06} studied the correlation
between the optical ($BVRI$) and X-ray light curves of young stars in the Orion Nebula Cluster. The optical observations
covered the 13-day \cxc\  observation of the cluster, and the typical exposure times ranged from 5~s for short exposures
to 420-720~s for long exposures. However, the observing cadence was about 1 per hour for each filter. \citet{stassun06} found little
evidence of correlations between the optical and X-ray variability, although there were some exceptions.

In this paper, we report on the Optical Monitor (OM) data obtained as part of the XEST. Section~\ref{sect:OMdata} describes the OM data,
whereas we introduce the OM catalogue in Section~\ref{sect:OMcat}. A vast majority of our
optical (and UV in a few cases) detections is of sources which are probably foreground or background sources.
We provide the full OM catalogue of detected sources as online material, although this paper focuses specifically on known or
probable TMC members (including new membership candidates identified by \citealt{scelsi06}). A separate paper focuses on the
OM survey of brown dwarfs \citep{grosso06a}. Section~\ref{sect:OMprop} makes use of the OM data and
data compiled by \citet{guedel06a} in order to derive basic properties of the OM TMC sample. 
Finally, since the OM and X-ray detectors observed simultaneously, we also
study correlations between the optical/ultraviolet and X-ray light curves for TMC members in Section~\ref{sect:OMXcorr}.


\section{Optical Monitor data}
\label{sect:OMdata}

%
\begin{table*}
\caption{OM-specific Observation Log}
\label{table:log}      
\centering          
\begin{tabular}{c l c c r c c c l}     
\hline\hline       
XEST	&  \multicolumn{1}{c}{ObsID} & RA$^a$ & $\delta$$^a$ & \multicolumn{1}{c}{PA$^a$ } & OM    & OM  & Fast & Image\\ 
{}      &        &  h m s      &  $\degr$ $\arcmin$ $\arcsec$ & \multicolumn{1}{c}{($\degr$)} & Filter & Mode & Data & Exposures \\
\hline                    
01	&	0301500101	& 04 21 59.4  & +19 32 06  & $80.01$	&      UVW1    & Image Fast	       &       Y       &       $18 \times 3540$s\\
02	&	0203540201	& 04 27 19.6  & +26 09 25  & $82.05$	&      U       & Full-Frame Low Res$^b$        &       N       &       $4 \times 5000$s, $3170$s\\
03	&	0203540301	& 04 32 18.9  & +24 22 28  & $81.92$	&      U       & Image Fast	       &       Y       &       $20 \times 1500$s\\
04	&	0203540401	& 04 33 34.4  & +24 21 08  & $261.88$	&      U       & Image Fast	       &       Y       &       $12 \times 1640$s\\
05	&	0203540501	& 04 39 34.9  & +25 41 46  & $262.68$	&      U       & Full-Frame Low Res$^b$        &       N       &       $4000$s, $4479$s, $2 \times 5000$s\\
06	&	0203540601	& 04 04 42.9  & +26 18 56  & $79.03$	&      U       & Full-Frame Low Res$^b$        &       N       &       $4 \times 5000$s\\
07	&	0203540701	& 04 41 12.5  & +25 46 37  & $262.61$	&      U       & Full-Frame Low Res$^b$        &       N       &       $3 \times 5000$s, $1679$s\\
08	&	0203540801	& 04 35 52.9  & +22 54 23  & $81.90$	&      U       & Image Fast	       &       Y       &       $10 \times 1260$s, $10 \times 1240$s\\
09	&	0203540901	& 04 35 55.1  & +22 39 24  & $261.85$	&      U       & Full-Frame Low Res$^b$        &       N       &       $5 \times 5000$s\\
10	&	0203542201	& 04 42 20.9  & +25 20 35  & $262.27$	&      U       & Image Fast	       &       Y       &       $15 \times 1780$s\\
11	&	0203541101	& 04 21 51.1  & +26 57 33  & $81.69$	&      U       & Full-Frame Low Res$^b$        &       N       &       $5 \times 5000$s\\
12	&	0203542101	& 04 35 17.4  & +24 15 00  & $261.59$	&      U       & Image Fast	       &       Y       &       $15 \times 1500$s\\
13	&	0203541301	& 04 29 52.0  & +24 36 47  & $81.55$	&      U       & Full-Frame Low Res$^b$        &       N       &       $3 \times 5000$s, $1679$s\\
14	&	0203541401	& 04 30 30.6  & +26 02 14  & $262.72$	&      U       & Full-Frame Low Res$^b$        &       N       &       $2\times 5000$s, $4179$s\\
15	&	0203541501	& 04 29 42.4  & +26 32 51  & $262.81$	&      U       & Full-Frame Low Res$^b$        &       Y       &       $12 \times 1260$s, $5 \times 1380$s\\
16	&	0203541601	& 04 19 43.0  & +27 13 34  & ---		&      ---     & ---		       &       ---     &       ---\\
17	&	0203541701	& 04 33 21.2  & +22 52 41  & $261.86$	&      U       & Full-Frame Low Res$^b$        &       N       &       $3 \times 4170$s\\
18	&	0203541801	& 04 33 54.7  & +26 13 28  & $83.07$	&      U       & Image Fast	       &       Y       &       $5 \times 1360$s, $5 \times 1260$s, $10 \times 1200$s\\
19	&	0203541901	& 04 32 43.0  & +25 52 32  & $82.83$	&      U       & Image Fast	       &       Y       &       $15 \times 1680$s, $5 \times 1660$s\\
20	&	0203542001  	& 04 14 12.9  & +28 12 12  & $78.06$	&      U       & Image Fast	       &       Y       &       $10 \times 2480$s\\
21	&	0101440701	& 04 21 59.0  & +28 18 08  & --- 	&      ---     & ---		       &       ---     &       ---\\
22	&	0109060301	& 04 31 39.0  & +18 10 00  & $83.58$	&      UVW2    & Image$^c$   	       &       N       &       $5 \times 1000$s\\
23	&	0086360301	& 04 18 31.2  & +28 27 16  & $259.74$	&      UVW2    & Image  	       &       N       &       $5 \times 1000$s\\
24	&	0086360401	& 04 18 31.2  & +28 27 16  & $259.67$	&      UVW2    & Image  	       &       N       &       $5 \times 1000$s\\
25	&	0152680201$^d$	& 04 34 55.5  & +24 28 54  & $244.71$	&      UVW2    & Image  	       &       N       &       $10 \times 980$s\\
26	&	0101440801	& 04 55 59.0  & +30 34 02  & ---		&      ---     & ---		       &       ---     &       ---\\
27	&	0201550201	& 03 54 07.9  & +31 53 01  & ---		&      ---     & ---		       &       ---     &       ---\\
28	&	0200370101	& 04 19 15.8  & +29 06 27  & $82.13$	&      UVW1    & Science User Defined$^{c,e}$  &       Y       &       $25 \times 4000$s, $2 \times 2200$s\\
\hline                  
\multicolumn{9}{l}{$^a$ Nominal boresight coordinates (J2000.0) and average spacecraft position angle}\\
\multicolumn{9}{l}{$^b$ $1024 \times 1024$~pixel images with $1\arcsec$ pixel size ($17\arcmin \times 17\arcmin$)}\\
\multicolumn{9}{l}{$^c$ Grism exposures taken after the imaging exposures}\\
\multicolumn{9}{l}{$^d$ This observation is part of a campaign on AA Tau that will be presented extensively elsewhere}\\
\multicolumn{9}{l}{$^e$ $624 \times 624$~pixel images with $0\farcs 5$ pixel size ($5\arcmin \times 5\arcmin$) together with a $10\farcs 5 \times 10\farcs 5$ window in
Fast mode}\\
\end{tabular}
\end{table*}
%

The Optical Monitor (OM; \citealt{mason01}) is a 30-cm optical/UV telescope that can provide
coverage in the optical and UV regimes (bandwidth $180-600~$nm) simultaneously
with the X-ray cameras. The OM detector is a micro-channel plate (MCP) intensified
CCD. The final array has a format of $2048 \times 2048$ pixels, each pixel having a size of about $0\farcs 48$,
leading to a square field-of-view (FOV) of about $17\arcmin \times 17\arcmin$, which covers the central part of the X-ray cameras (15\arcmin-radius FOV). 
The OM carries a wheel of filters ($V$, $U$, $B$, UVW1, UVM2, UVW2, and a broad white light filter) 
and 2 grisms for the visible and UV ranges. In this paper, we report on OM data taken
with the $U$ ($\approx 300-400$~nm) and UVW2 ($\approx 175-250$~nm) filters. The OM
point-spread-function varies from $1\farcs 4$ to $2\arcsec$, depending on the filter.
The OM can operate in the ``Imaging'' or ``Fast'' modes; of particular interest here, 
in the default Imaging mode (``Image'' in Table~\ref{table:log}), a set of 5 {\em consecutive} exposures is taken, each covering a
different portion of the FOV.
In each of the 5 exposures, a large window (W$_{1,...,5}$; with $1\arcsec$ pixels)
is complemented by a smaller central imaging mode window of size $2\arcmin \times 2\arcmin$ 
with $0\farcs 5$ pixel resolution (W$_0$).
Thus, the sequence is exposure 1: W$_0$+W$_1$, exposure 2: W$_0$+W$_2$, etc. until exposure 5:
W$_0$+W$_5$, after which the sequence can start again. 
An example of the configuration can be found in the \xmm\  Users' Handbook \citep{ehle05}.
In the default Fast mode  (``Image Fast'' in Table~\ref{table:log}), a similar set of images is obtained together with an additional
central window (CW; $22\times 23$ pixels, i.e., $10\farcs 5 \times 10\farcs 5$) that is operated
in fast mode with a time resolution of $0.5$~s. Consequently, the sequence of images is
exposure 1: W$_0$+W$_1$+CW, exposure 2: W$_0$+W$_2$+CW, etc. until exposure 5: W$_0$+W$_5$+CW. 
This means that the on-axis target
should be monitored continuously with high time resolution, and 5 times in the small central $2\arcmin \times 2\arcmin$
$W_0$ imaging window (with each an exposure equal to the integration time of the exposure), and
once in the large, central $1\arcsec$-pixel window. If a secondary target is located in
a sky area covered by the small central imaging window $W_0$, it is observed in a similar
fashion as above, except that no fast mode data is available. If it is located in an area
outside the $2\arcmin \times 2\arcmin$ window, it will be observed only in one of the 5 consecutive large imaging
window, i.e., once per OM exposure. Clearly, the best time coverage for a secondary target
is when it is located in a sky area covered by the small central window. Note that other 
imaging modes can be used as well. For example, in ``Full-Frame'' imaging mode, images of the whole
OM FOV can be obtained either in full resolution ($0\farcs 5$ pixel size; ``High Resolution'') or in low 
($1\arcsec$ pixel size; ``Low Resolution''). Additional user-specified mode can be used as well (see XEST-28 in Table~\ref{table:log}).
We remind that, because the OM FOV is smaller ($17\arcmin \times 17\arcmin$) than the EPIC field-of-view 
($15\arcmin$ radius), several interesting X-ray sources did not fall on the OM detector, and, 
therefore, no photometry or light curve could be obtained.

The OM data were reduced with the \xmm\  Science Analysis System (SAS) 6.1. We have processed the ``Imaging'' and
``Fast'' mode data using the metatasks \emph{omichain} and \emph{omfchain}, respectively, with default parameters, except for
the \emph{omdetectminsignificance} parameter of \emph{omichain}, which determines the minimum significance of a 
source to be included in the source-list file and which we changed from 1.0 to 3.0. In Table~\ref{table:log}, we
provide an observation log of the XEST fields \citep{guedel06a}, with an emphasis on the OM-specific parameters.
In particular, we provide the imaging mode and emphasize when a Fast mode window was available, and we provide the
exposures of the individual imaging windows. For example, for XEST-01, for the first exposure, the imaging windows $W_0$ and $W_1$
observed simultaneously for 3540~s, then (after some overhead delay), $W_0$ and $W_2$ observed for another 3540~s, etc.
Since the XEST-01 OM data were obtained in FAST mode as well, the CW window was operating in parallel. For XEST-02, $W_0+W_1$, 
$W_0+W_2$, $W_0+W_3$, $W_0+W_4$ each observed for 5000~S, while the last exposure with windows $W_0$ and $W_5$ observed for 3170~s only.

   \begin{figure}
   \centering
   \resizebox{\hsize}{!}
   {\includegraphics[width=\textwidth]{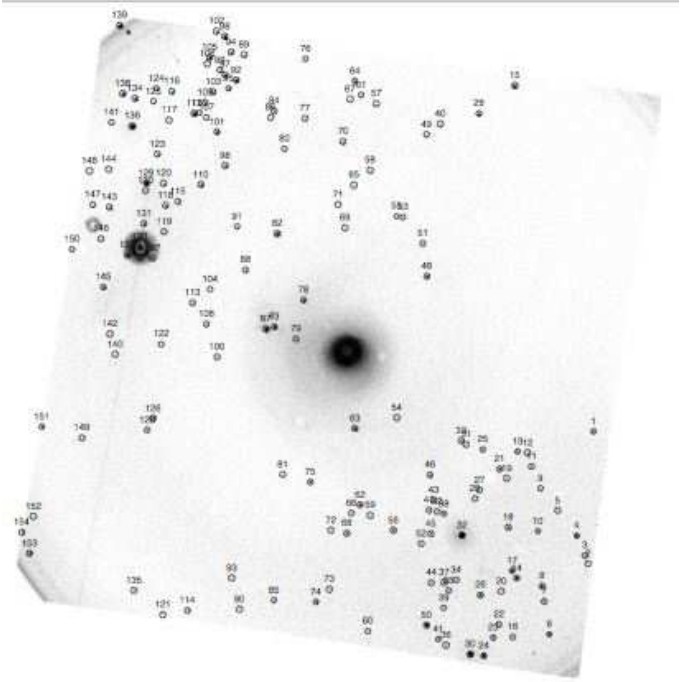}}
      \caption{Example of an OM image. This rich field (XEST-11, toward FS Tau) shows artefacts, such 
      as the ``ring'' of emission at the center (due to the reflection of diffuse sky light from outside the FOV), fixed pattern noise around 
      the bright source XEST-11-OM-133 (HD 283579, which is 
      not classified as a TMC member), and a
      ghost ring below source XEST-11-OM-147 which is also caused by XEST-11-OM-133. Other artefacts can be seen
      (trail along the bright source, defects on the detector).
          }
         \label{fig:OMim}
   \end{figure}
%

For ``Imaging'' mode data, we obtained images and source lists for individual exposures, a mosaicked image, and a
merged source list. Each individual source list included, in particular, the exposure source number, 
sky and galactic coordinates, raw and corrected count rates, magnitudes, quality flags, and the source 
identification in the final source list. Consequently, it was possible, knowing the source identification,
to obtain exposure-specific count rates and magnitudes, i.e., low time-resolution light
curves.\footnote{For point sources identified in both the small $0\farcs 5 \times 0\farcs 5$ 
and the larger $1\arcsec \times 1\arcsec$ central windows, we used count rates and magnitudes from the
small window because of the smaller pixel bin size and of the more frequent time sampling.} 
However, SAS 6.1 identified a significant number of spurious sources, mostly in the ghost image
near a bright star produced by internal reflection of light within the detector window, in the enhanced
``ring'' of emission near the center of the detector (due to the reflection of diffuse sky light from outside the FOV),
and in fixed
pattern noise around bright sources \citep{ehle05}. Consequently, we manually inspected the mosaicked
image, merged source list identifications, and removed obvious misidentifications. We also modified
the merged source list to ensure that the same source had the same source identification in each
exposure; indeed, the \emph{omsrclistcomb} task did not combine the source lists accurately in a few
cases, i.e., mostly bright sources for which strong fixed pattern noise introduced small positional
inaccuracies that were not properly identified by SAS 6.1. We note that SAS 6.5 eliminates most of the
above issues; however, this version of the SAS incorrectly combines the source lists, and, therefore, 
we preferred not to use it. Figure~\ref{fig:OMim} shows an example of an OM image with
some artefacts.

For ``Fast'' mode data, in addition to individual images, we obtain event lists and 
time series files with high time resolution (the default value is $10$~s with \emph{omfchain}) for each exposure. We merged the
different time series files into a single time series file for further analysis.

\section{Optical Monitor catalogue}
\label{sect:OMcat}

\begin{table}
\caption{OM astrometric correction vectors in arcseconds}
\label{table:astrometry}      
\centering          
\begin{tabular}{c r r l}     
\hline\hline       
XEST	&  \multicolumn{1}{c}{$\Delta X$} & \multicolumn{1}{c}{$\Delta Y$} & \multicolumn{1}{l}{r.m.s.}\\ 
\hline
01 & $ 3.086$ & $-2.462$ & $0.90$\\
02 & $ 4.338$ & $-1.016$ & $0.62$\\
03 & $ 3.320$ & $-0.970$ & $0.52$\\
04 & $-1.471$ & $ 3.632$ & $0.83$\\
05 & $-3.561$ & $ 4.579$ & $0.52$\\
06 & $ 3.948$ & $-3.227$ & $0.57$\\
07 & $-3.240$ & $ 4.412$ & $0.99$\\
08 & $ 2.955$ & $-0.800$ & $0.81$\\
09 & $-2.728$ & $ 5.588$ & $0.62$\\
10 & $-2.219$ & $ 2.623$ & $0.61$\\
11 & $ 3.847$ & $-0.137$ & $0.61$\\
12 & $-2.675$ & $ 3.093$ & $0.61$\\
13 & $ 2.143$ & $-1.162$ & $0.86$\\
14 & $-4.529$ & $ 3.239$ & $0.68$\\
15 & $-3.891$ & $ 5.163$ & $0.71$\\
17 & $-3.824$ & $ 5.114$ & $0.58$\\
18 & $ 2.893$ & $-1.391$ & $0.92$\\
19 & $ 3.617$ & $-0.044$ & $0.62$\\
20 & $ 6.201$ & $-5.232$ & $0.48$\\
22 & $ 1.586$ & $-1.008$ & $0.06$\\
23 & $ 0.368$ & $ 1.105$ & $0.00^a$\\
24 & $-1.363$ & $ 0.720$ & $0.00^a$\\
25 & $-3.065$ & $ 4.090$ & $1.63$\\
28 & $ 5.519$ & $-0.994$ & $0.77$\\
\hline                    
\multicolumn{4}{l}{$^a$ {\footnotesize Only V410 Tau detected in the OM}}\\
\end{tabular}
\end{table}

\input{OMcat_paper.tex}

A catalogue of OM sources was compiled, cross-checked with the Two Micron
      All Sky Survey (2MASS) and the XEST X-ray sources. We provide average
photometry and source coordinates as compiled in the combined source lists in the imaging mode data. We have calculated 
boresight corrections to OM coordinates by cross-correlating with the 2MASS catalogue (search radius of 
$3\arcsec$) and by means of iterative steps until the median position offsets became zero. Table~\ref{table:astrometry}
provides the position offsets ($\Delta X, \Delta Y$) in arcseconds 
($\mathrm{RA}_\mathrm{corr} = \mathrm{RA}_\mathrm{OM} + \Delta X / \cos (\delta_{\mathrm{2MASS}})$ and 
$\delta_\mathrm{corr} = \delta_\mathrm{OM} + \Delta Y$). We also cross-correlated the OM and the X-ray 
catalogues \citep{guedel06a} using 2MASS-corrected coordinates (search radius of 
$3\arcsec$). Minor manual interventions were
necessary
when 2 close sources were separated in the OM but not in X-rays (UZ Tau E+W(AB), HP Tau/G2 and HP Tau/G3 AB).
Finally, we remind that we used the same detection parameters for all observations. Consequently, the source 
detection process is not optimized to each exposure; however, a visual inspection showed that the chosen parameters
were generally sufficient, although we do not claim completeness in the XEST OM catalogue.

\begin{table*}
\normalsize
\caption{Detection statistics}
\label{table:detections}      
\centering          
\begin{tabular}{l r r r r r r r r}
\hline\hline       
{}					& \# & \multicolumn{7}{c}{Type$^a$} \\
{}					& {} & 0 & 1 & 2 & 3 & 4 & 5 & 9 \\
\hline
Detected OM sources 			&  2148 \\
X-ray sources in OM FOV 		&  916  \\
OM detections not detected in X-rays	&  2	& 0 & 0 & 1 & 0 & 1 & 0 & 0\\
OM detections also detected in X-rays	&  98   \\
TMC members detected in X-rays and OM	&  51	& 0 & 1 & 33&17 & 0 & 0 & 0\\
TMC candidates detected in X-rays and OM	&  12   \\
TMC members detected in X-rays but not in OM	&  34	& 0 & 7 & 8 & 8 & 4 & 0 & 7\\
TMC candidates detected in X-rays but not in OM	&  21   \\
TMC members detected in X-rays but outside OM FOV$^b$		&  55	& 0 & 1 &16 &32 & 4 & 2 & 0\\
\hline                    
\multicolumn{9}{l}{$^a$ See \citet{guedel06a} and text for the definition of the object type}\\
\multicolumn{9}{l}{$^b$ Statistics for XEST observations with OM data}\\
\end{tabular}
\end{table*}

Table~\ref{table:OMcat} provides the first 30 entries of the full OM catalogue, in increasing right ascension. In Column~1, we provide the
catalogue source number. Columns 2, 3, and 4 provide the 2MASS-corrected OM right ascension and declination
and the positional error $\Delta$ (including the RMS error on the boresight shift, which was added in quadrature to the statistical 1-sigma
positional error calculated for each source). 
Column 5 provides the XEST OM identification in the xx-OM-zzz notation in Column 2, where xx refers to the XEST exposure and zzz refers to the OM 
identification for this XEST exposure.  
Columns 6 and 7 give the 2MASS cross-identification, if found, and the offset $\rho_1$
between the OM corrected coordinates and the 2MASS coordinates. Columns 8 and 9 give the XEST X-ray identification (in xx-yyy notation,
see \citealt{guedel06a}), if found, and the offset $\rho_2$ between the OM and X-ray corrected
coordinates. 
Column 10 gives the average magnitude and its uncertainty, and Column 11 the detection significance. 
For clarity, we have added in the exponent a $\ddagger$ character if the filter was UVW2 
and a $\dagger$ character if the filter was UVW1. No special character was added for the commonly used $U$ filter. We provide the full OM catalogue
as online material (Table~\ref{table:OMcatfull}).

The XEST OM catalogue contains 2148 entries, among which 1893 have 2MASS counterparts and 98 have XEST X-ray counterparts.
Out of these 98 sources, 51 are TMC members according to the master list from \citet{guedel06a} and 12
are new TMC candidates identified by \citet{scelsi06}. Thus 35 OM sources had X-ray counterparts
but were not classified as TMC members or candidates. Note that V410 Tau ABC (XEST-23-032 \& XEST-24-028) was 
counted twice in the OM catalogue since it was observed in two different XEST observations. In addition, UZ Tau E+W(AB) (XEST-19-049) were not separated in
X-rays but were so in the OM (UZ Tau W = XEST-19-OM-092, UZ Tau E = XEST-19-OM-094). Similarly, the X-ray source XEST-08-051 was attributed to
both HP Tau/G3 AB and HP Tau/G2 (centroid fitting suggests that the X-rays come mostly from HP Tau/G2); however, they
were separated in the OM data (XEST-08-OM-038 for HP Tau/G3 AB and XEST-08-OM-040 for HP Tau/G2). 

There were 2 TMC members detected in the OM but not detected in X-rays with \xmm\  (FV Tau/c AB, XEST-02-OM-020
and 2MASS J04141188+2811535, XEST-20-OM-002). However, FV Tau/c AB is very close to FV Tau AB (XEST-02-013 =  XEST-02-OM-017) and
its faint X-ray emission may have been overshadowed by the stronger X-ray flux of FV Tau AB. Note that FV Tau/c AB was detected
by \cxc\  \citep{guedel06a}. A similar case occurred for the
brown dwarf 2MASS J04141188+2811535 \citep{luhman04}, which is close to the X-ray bright V773 Tau (XEST-20-042 = XEST-20-OM-003). The OM data
of the brown dwarf is addressed in a separate paper \citep{grosso06a}.

While the above detection numbers start from OM detections, we have also taken a different
approach, starting from the catalog of X-ray sources detected with XEST: we have determined that 916 X-ray
sources had coordinates such that they fell on the OM detector. Out of those, only 98 are detected in the OM 
(see above), while 34 TMC members of TMC detected in X-rays remained undetected in the OM.
In addition, 21 new TMC candidates detected in X-rays (\citealt{scelsi06}) were not detected in the OM.
Table~\ref{table:detections} summarizes the detection statistics
pertaining to the OM/X-ray correlations. The object type (as defined by \citealt{guedel06a}, i.e.,
`0' and `1' for a protostar of Class 0 or Class I, `2' for an accreting (classical) T Tau star
usually showing a Class-II IR spectrum, `3' for a weak-lined or Class-III object, `4' for a brown
dwarf, `5' for an Herbig Ae star, and `9' for uncertain classifications or other object types)
is given when relevant.

No Type 0 object was in our sample since the sole object surveyed 
was observed by \cxc\  only and remained undetected in X-rays (L1527 IRS; \citealt{guedel06a}).
For TMC members detected in the X-rays and falling on the OM FOVs, the OM detection rate was 12.5~\%
(1 out of 8) for Type 1 objects,
whereas this rate reached  80.5~\% in Type 2 objects. Type 3 objects had a high 68~\% detection rate. Interestingly,
only 1 brown dwarf was detected in the OM, whereas 11 such objects fell on the OM FOV and 4 of them were detected in 
X-rays \citep{grosso06a,grosso06b}. We remind that the numbers reported in Table~\ref{table:detections} makes
no attempt not to count duplicates. For example, CFHT-Tau 5, detected in X-rays in two XEST observations (XEST-03-031 = XEST-04-003), was
counted once in ``TMC X-ray OM non-detections'' since it fell on the OM detector but remained undetected in XEST-03, and
once in ``TMC X-ray outside OM FOV'' since it fell off the OM detector in XEST-04.

\input{OM_TMConly_paper.tex}
\input{OM_TMCcand_paper.tex}

Tables~\ref{table:OMTMConlycat} and \ref{table:OMTMCcandcat} repeat some of the columns in 
Table~\ref{table:OMcatfull} for TMC members and TMC candidates, respectively, but also
provide the name of the TMC source (from \citealt{guedel06a}) or ``(TMC cand)'' (from \citealt{scelsi06}). The average count rate $\mu$ (corrected for coincidence losses
and aperture radius) and its error $\Delta\mu$ are also given, after calculation from the magnitude and its 
error\footnote{$\log \mu = -0.4 (m - Z)$, $\Delta\mu / \mu = (1 - x) / (1 + x)$, where $\log x = -0.8 \Delta m$, $m$ and $\Delta m$ 
are the source's magnitude and its error, and $Z$ is the zero-point magnitude.}. We used the zero-point magnitudes
given in the \xmm\  UHB ($Z_U = 18.2593$, $Z_\mathrm{UVW1} = 17.1882$, $Z_\mathrm{UVW2} = 14.8026$; \citealt{ehle05}). Note 
that this calibration does not take into account corrections specific to the source's spectral type.

\section{Properties of the OM Taurus sample}
\label{sect:OMprop}

We focus in this section on some of the OM properties of the TMC sample as defined in \citet{guedel06a} that were detected
in both X-rays and in the OM. Specifically, we focus on whether correlations can be found, e.g., between the X-ray luminosity and
the OM magnitude, and on the $U$-band properties of weak-lined T Tau stars and classical T Tau stars. 
The low number of detections of Type 1 sources is such that we will not discuss them.

\begin{figure}
   \centering
   \resizebox{\hsize}{!}
   {\includegraphics[width=\textwidth]{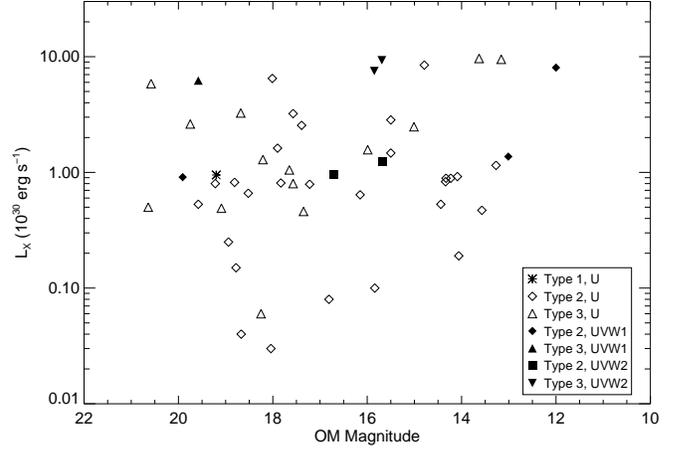}}
      \caption{X-ray luminosity ($L_\mathrm{X}$) in $10^{30}$~\ergps\  as a function of the observed OM magnitude.
      	Various types of stars are identified (see text). Ultraviolet magnitudes are identified with filled symbols.
	      }
         \label{fig:ommag_lx}
   \end{figure}

\begin{figure}
   \centering
   \resizebox{\hsize}{!}
   {\includegraphics[width=\textwidth]{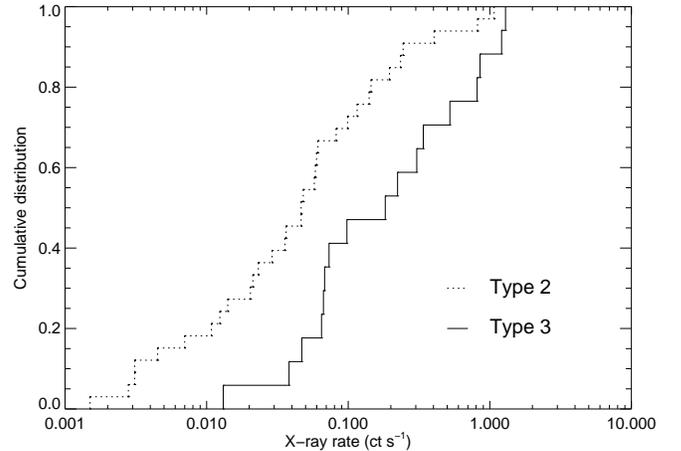}}
      \caption{Cumulative distributions of X-ray count rates of Type 2 and 3 stars. 
	      }
         \label{fig:cumdis_ratex}
   \end{figure}

Figure~\ref{fig:ommag_lx} shows the scatter plot of X-ray luminosities (as derived from the DEM method; see \citealt{guedel06a})
as a function of the observed OM magnitude ($U$-band or ultraviolet UVW1 and UVW2 bands). We used different symbols for the various
types of stars. There is no strong evidence that any correlation exists in any type or any OM filter. 
However, at least one trend can be noticed: Type 2 stars (usually CTTS) are
overall {\em fainter}  in X-rays than Type 3 stars (usually WTTS).
In fact, at a given OM magnitude, there is a wide range of X-ray
luminosities below $10^{31}$~\ergps\  in Type 2 stars. Fig.~\ref{fig:cumdis_ratex} shows the cumulative distribution of X-ray
count rates in Type 2 and 3 stars that were detected in both OM and X-rays. A Kolmogorov-Smirnov (K-S) test gives 
a significance level of $P = 0.0055$ that both distributions are compatible ($D=0.49$; the $D$ statistic is the maximum vertical deviation between the 
two curves), indicating that the two cumulative distributions differ at a high significance level.
This result is discussed in detail in \citet{guedel06a} and \citet{telleschi06a}. 

\begin{figure}
   \centering
   \resizebox{\hsize}{!}
   {\includegraphics[width=\textwidth]{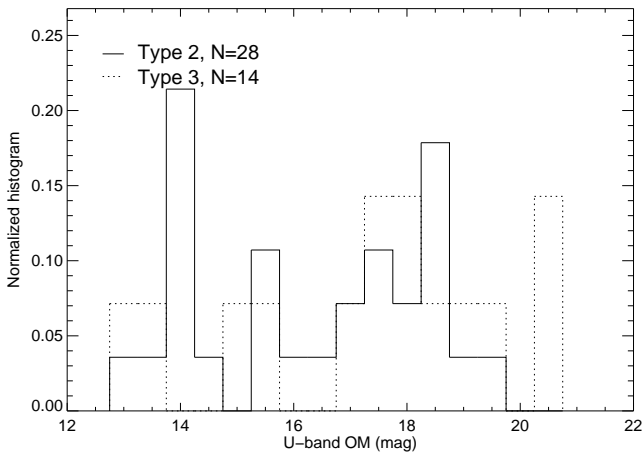}}
      \caption{Histogram of OM $U$ magnitudes for Type 2 and 3 stars.
	     }
	 \label{fig:magOM_histo}
   \end{figure}

Despite the clear difference in X-ray luminosity between Type 2 and 3 stars, there is little evidence that the observed 
OM magnitude of Type 2 stars are much different from those of Type 3 stars (Fig.~\ref{fig:magOM_histo}). Indeed, a K-S test 
gives a significance level of $P=0.54$ ($D=0.25$). 
If the observed OM magnitudes scale with $L_*$, then the similar distributions for Type 2 and 3 stars would not be surprising, 
and again, we would find the CTTS to be fainter only in the X-rays (Fig.~\ref{fig:cumdis_ratex}), pointing at a true 
deficiency of emission in  that wavelength range \citep{telleschi06a}. However, it is important to keep in mind that the observed magnitudes could be biased by certain factors.
Therefore, we aim to test the following hypotheses that could produce an observational bias: I) 
Bolometric luminosities in Type 2 and Type 3 stars are different; II) The binary fraction is higher
in one object type than another; III) Extinction in the $U$ band is stronger for one type than for the other; 
IV) One object type has a significant excess in the $U$-band.

%

%

   \begin{figure}
   \centering
   \resizebox{\hsize}{!}
   {\includegraphics[width=\textwidth]{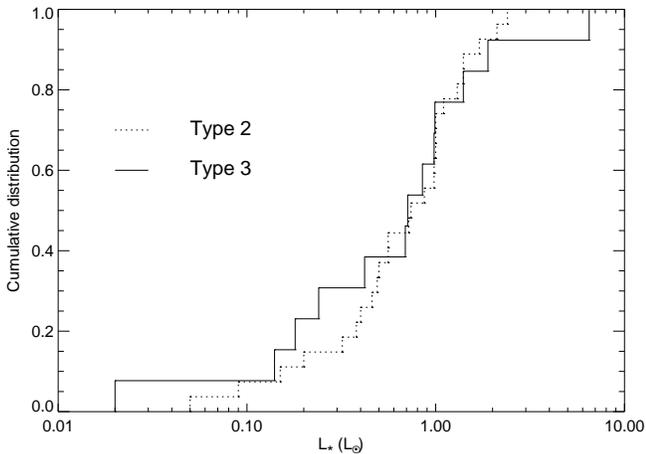}}
      \caption{Cumulative distributions of stellar luminosity of Type 2 and 3 objects detected in the $U$ filter. 
	      }
         \label{fig:cumdis_Lstar}
   \end{figure}
%

\subsection{Hypothesis I}

Figure~\ref{fig:cumdis_Lstar} shows the cumulative distribution of the star's luminosity
\citep[see][]{guedel06a} for Type 2 and 3 stars detected in the $U$ band. Note that we used the luminosity of the primary 
component if available for multiples (Table~9 in \citealt{guedel06a}). There is no evidence that the stellar luminosity
of Type 2 stars is different from that of Type 3 stars, as indicated by the high significance of a K-S test ($D=0.16, P=0.97$).
Therefore, hypothesis I can be safely rejected as a potential biasing factor. 

\subsection{Hypothesis II}
In our OM $U$-band sample, there are 14 and 9 TMC stars of Type 2 and 3, respectively,
with more than 1 component. Therefore, we created cumulative distributions of the OM magnitude for the 
remaining single components (14 and 5 stars for Type 2 and 3, respectively): 
There is no evidence of a difference in the OM magnitudes for single Type 2 and 3 stars
(K-S test: $D=0.37, P=0.58$). Therefore, hypothesis II is also rejected.
 
\subsection{Hypothesis III}
We used extinction magnitudes in the $J$ band ($A_J$) if available, and determined the $A_U$ magnitudes.
Our procedure was the following: first, we calculated  $A_J/A_V$ and $A_U/A_V$ ratios based on coefficients given by \citet{cardelli89}: $A_J/A_V = 0.4008 - 
0.3679 / R_V$ and $A_U/A_V = 0.9530 + 1.9090 / R_V$, where we used $R_V = 5.5$. Indeed such a value gives a better fit of the 
observed $A_J/A_V$ values.\footnote{When more than one extinction magnitude was available for multiples, we used the value from 
the primary.} Then, we calculate the estimated extinction magnitude in the $U$ band, $A_U$:

\begin{displaymath}
A_U = \frac{A_U}{A_V}  \frac{A_V}{A_J} {A_J}.
\end{displaymath}

We obtained the distributions of $A_U$ for Type 2 and 3 stars observed with the OM in the $U$ band. Unexpectedly, the cumulative 
distributions do not differ significantly. As a safety check, we calculated $A_U$ from $A_V$ ($A_U = \frac{A_U}{A_V}  {A_V}$) and 
found the same result. Since $A_V$ is more subject to optical veiling than $A_J$, we therefore preferred to use $A_U$ from $A_J$ in 
this paper.

\begin{figure}
   \centering
   \resizebox{\hsize}{!}
   {\includegraphics[width=\textwidth]{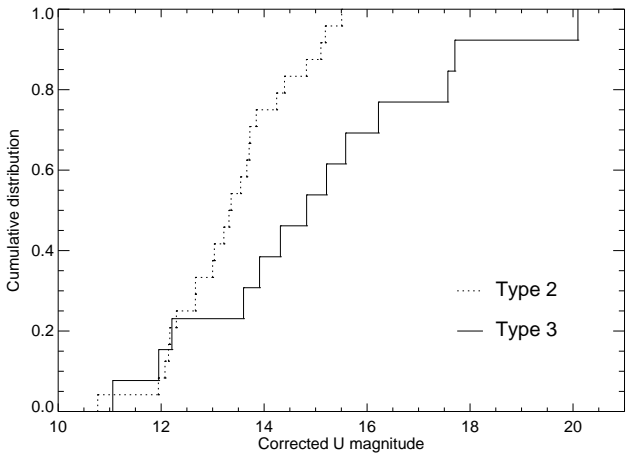}}
      \caption{Cumulative distributions of $U$-band OM magnitudes after correction for extinction. Type 2 stars have brighter
      	OM magnitudes than Type 3 stars. Stars with known $T_\mathrm{eff}$ and $A_J$ are shown.
	      }
         \label{fig:cumdis_omcorrmag}
   \end{figure}

Although the distributions of extinction magnitudes are overall similar in our sample, we have noticed that they are correlated with
the observed $U$-band magnitude in Type 2 stars, but not in Type 3 stars. It appears that, in our sample, Type 2 stars with fainter
$U$ magnitudes suffer from more extinction than brighter Type 2 stars. Therefore, we have used the above $A_U$ to correct the 
observed OM $U$-band magnitudes and we constructed the cumulative distribution (Fig.~\ref{fig:cumdis_omcorrmag}). The distributions are significantly different, 
with Type 2 stars being much brighter in the $U$-band than Type 3 stars (K-S statistic $D = 0.44$ with a significance level $P$ of 
$0.05$).

Therefore, hypothesis III appears to be correct in the sense that Type 2 stars are in fact brighter in the $U$-band but suffer 
more extinction than Type 3 stars, as could be expected based on observed properties of young stars. 
We emphasize, however, that this trend is opposite to the trend in X-rays where CTTS are {\it fainter}.
Nevertheless, as shown earlier, both types of stars have similar stellar luminosities, indicating that there must be a 
$U$-band excess in Type 2 stars. 
Therefore, we need to address hypothesis IV before we can draw any firm conclusion.

%

   \begin{figure*}
   \centering
   \resizebox{.475\textwidth}{!}{\includegraphics{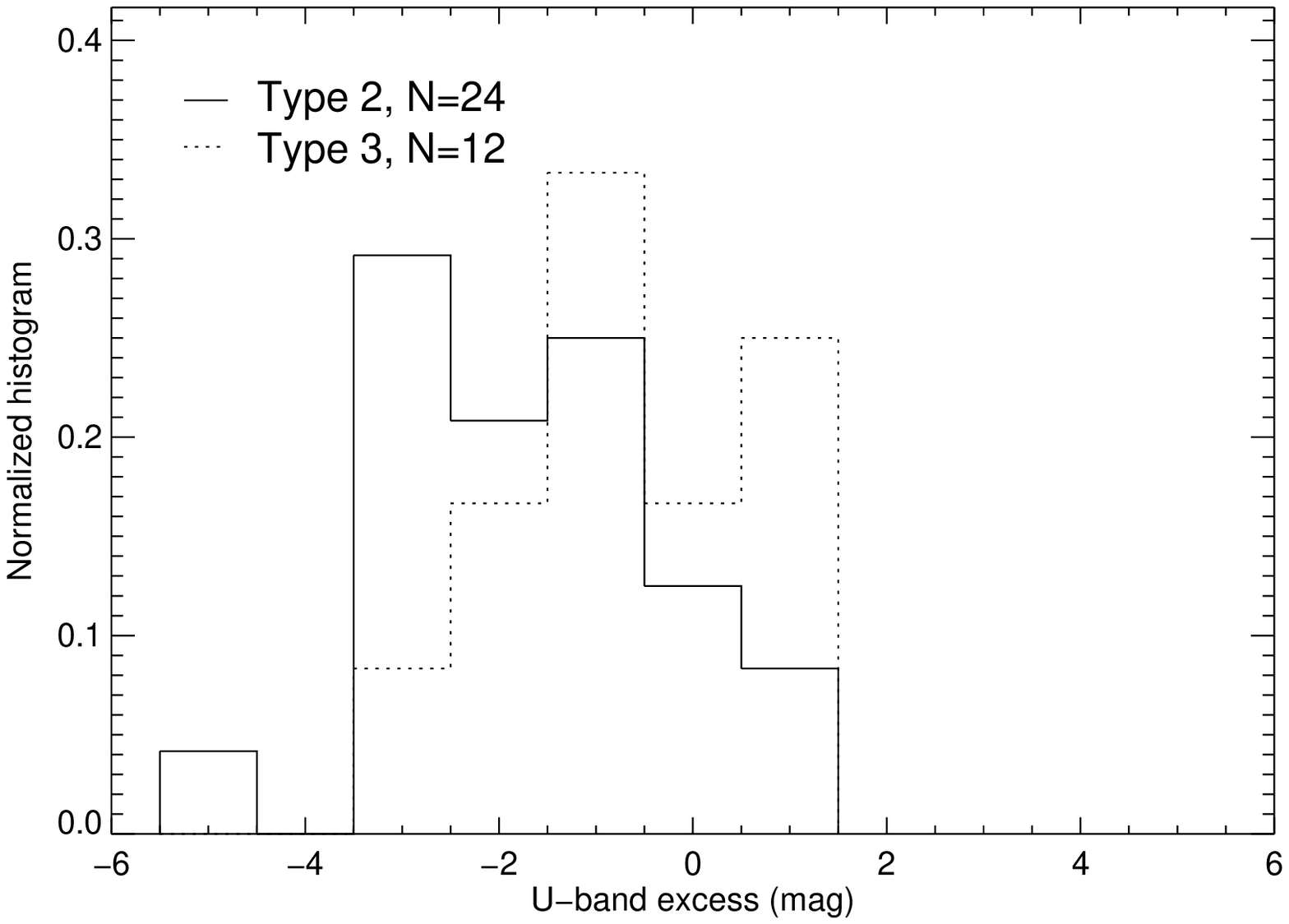}}
   \resizebox{.475\textwidth}{!}{\includegraphics{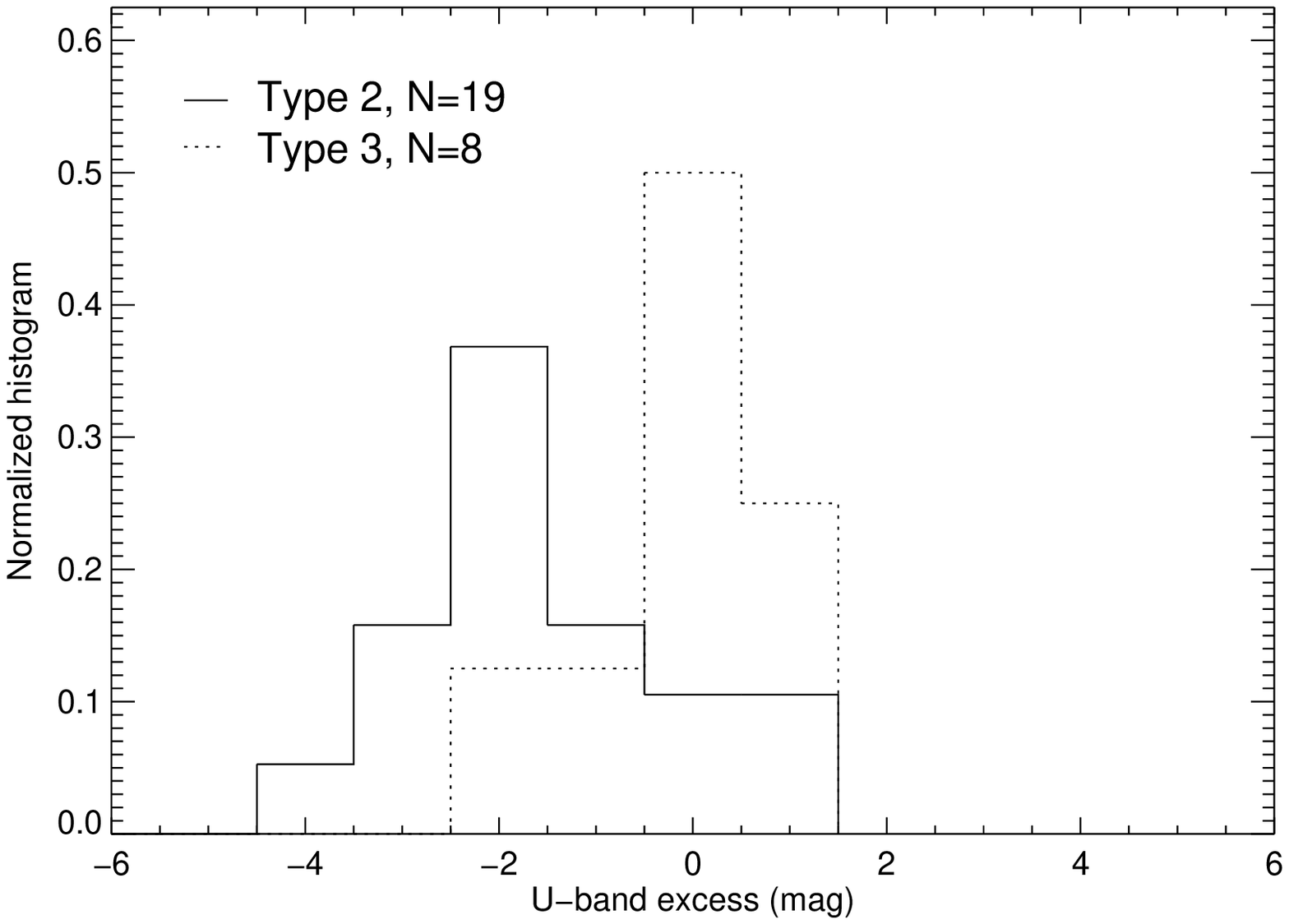}}
   \caption{Normalized distributions of $U$-band excesses for Type 2 and 3 stars. Excesses determined when $A_J$ was used to calculate
   $A_U$ are shown in the left figure, whereas excesses determined with $A_V$ used are shown in the right figure.
      }
         \label{fig:excess}
   \end{figure*}
%

\subsection{Hypothesis IV}
To determine the $U$-band excesses, we used the distance modulus ($m-M = 5.73$ for a distance to the TMC of 140~pc) to calculate 
the corrected absolute $U$-band magnitude, $M_U$, from the observed extinction-corrected $U$ magnitude.  We then determined the stellar photospheric 
absolute $U$ magnitude tabulated by \citet{siess00} (that use a $Z=0.02$ metallicity and the  conversion table of \citealt{kenyon95}). 
We finally compared the stellar photospheric $M_U$ with the
observed extinction-corrected absolute $U$ magnitude. Figure~\ref{fig:excess} shows the normalized histograms of the $U$-band excesses. 
We provide distributions determined from $A_U$ that was calculated either from $A_J$ (left figure) or $A_V$ (right figure).
The distributions of excesses are different, as shown from Kolmogorov-Smirnov tests of the cumulative distributions of 
excesses that give very low probabilities ($P=0.16$ and $P=0.04$ for excesses derived from $A_J$ and $A_V$, respectively).
Type 2 stars showing generally $U$-band excess by $0-3$~mag, whereas Type 3 stars either show no excess or small $U$-band 
deficiency. There are exceptions for stars of both types, probably due to 
inaccurate background subtraction (mostly due to the central ghost emission at the center of the OM image), inaccurate extinction correction,
binary effects, lower than expected count rate in bright stars due to instrumental effect (fixed pattern noise). 

In conclusion, we have shown that binarity and stellar luminosity in our sample cannot bias the observed OM magnitudes. However,
although the extinction magnitudes are statistically similar in Type 2 and 3 stars, they are correlated in Type 2 stars
with the observed magnitudes. After correction for extinction, Type 2 stars are found to be brighter than Type 3 stars, which 
can be explained by a $U$-band excess due to accretion.

   \begin{figure*}
   \centering
   \resizebox{\hsize}{!}
   {\includegraphics[width=\textwidth]{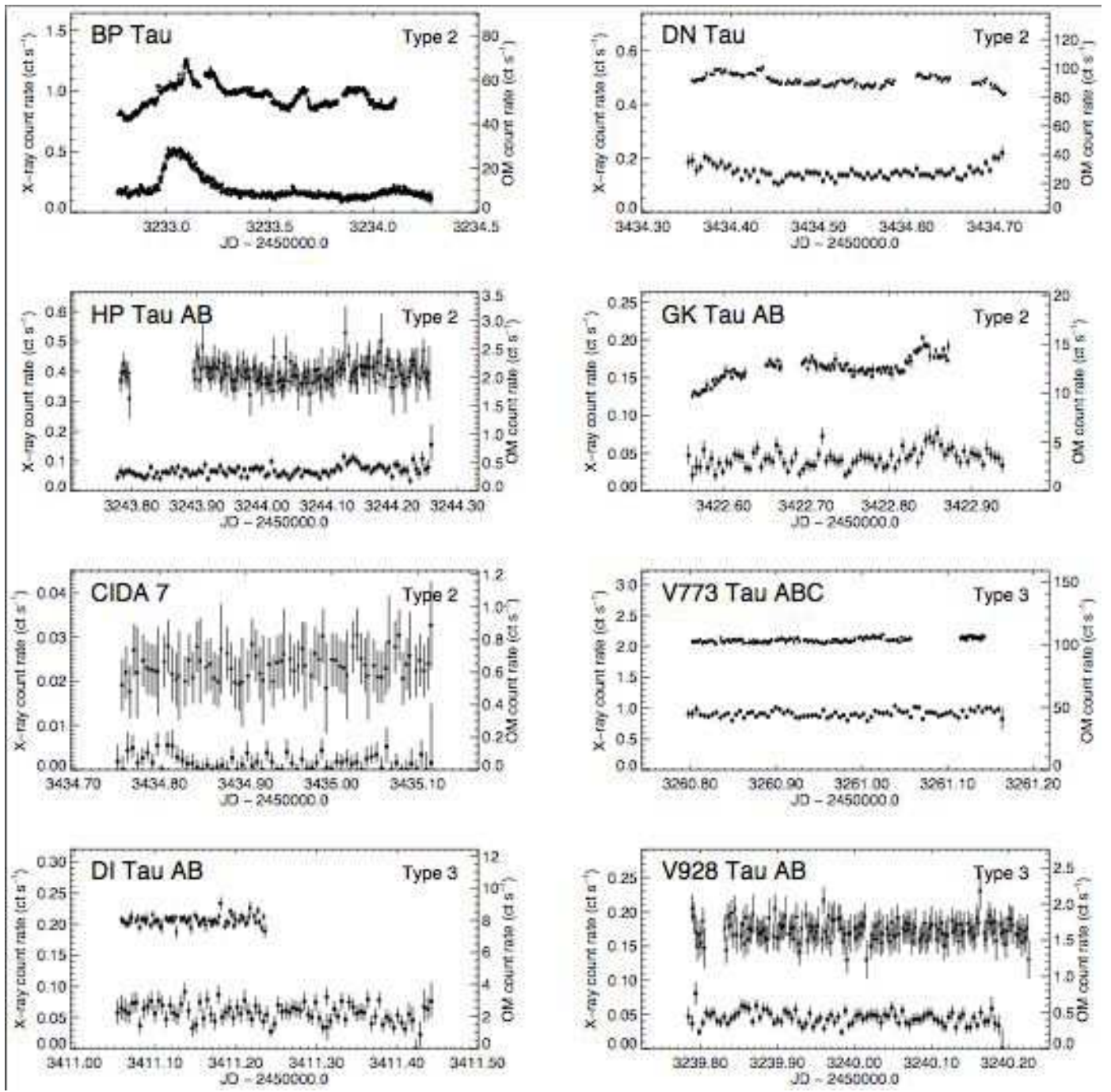}}
      \caption{The TMC sample of stars with OM FAST mode data. The upper curve in each panel is the OM light curve (right
      $y$-axis), whereas the lower curve is the X-ray light curve (left $y$-axis). The OM light curves were rebinned to 
      bin sizes of 200~s, except for CIDA 7, which required a bin size of 400~s. The type of the source is given together
      with the source name.
      The OM FAST data of IT Tau and UZ Tau EW are not shown since the sources were
      not properly placed on the small OM FAST window, and thus the light curves were unreliable. In addition, the T Tau
      OM FAST light curve is the subject of a separate paper and will be presented elsewhere (G\"udel et al. 2007, in
      preparation). Note that small manual 
      corrections were necessary for DN Tau, DI Tau AB, and BP Tau (see footnote~\ref{foot:OMcorr}).}
         \label{fig:omfast}
   \end{figure*}
%

\section{Optical and X-ray light curves}
\label{sect:OMXcorr}

Figure~\ref{fig:all_lc} presents the collection of simultaneous X-ray and optical/UV light curves of XEST sources detected 
in both X-rays and in the OM. The X-ray light curves are shown on the top panel with the XEST identification, while the OM
light curve is shown in the bottom panel with the XEST-OM identification. The OM light curves are taken from the IMAGING
mode data only. The $x$-axis OM error bars correspond to the length of an exposure identification. If a TMC member is associated with
the source, its name is given next to the XEST or XEST-OM identification. Special cases occur, e.g., when two nearby sources
were not separated in X-rays but they were in the OM. For example, the X-ray light curve of XEST-19-049 is associated with
both UZ Tau E and W, whereas we manually extracted (due to problems in the SAS software to properly extract both sources 
due to the close separation) the OM light curves for UZ Tau W (AB) and UZ Tau E (XEST-19-OM-092 and XEST-19-OM-094,
respectively)\footnote{The OM light curves were obtained by manually determining the position of the two stars in each 
exposure image, and then using an IDL routine to extract the light curve. We used radii of half the stars' separation
(3\farcs 48, i.e., a radius of 3.65 pixels for the central window), which minimizes the contamination of one source onto the second one. 
We corrected for aperture, applied the theoretical and empirical corrections, corrected for deadtime, and applied a
time-dependent sensitivity correction. The above procedure is similar to what the SAS task \emph{omsource} does; however, the
latter cannot yet extract for radii lower than 6 pixels. We defined an annulus of radii 20 and 30 pixels ($\approx
10\arcsec$  and $15\arcsec$) for the local background contribution. See \citet{grosso06a} for a detailed description of the
procedure.}. A similar case is reported for HP Tau/G3 AB and HP Tau/G2 (XEST-08-051). The X-ray light curve, although
dominated by HP Tau/G2, has some contribution (about $8$~\%; see \citealt{telleschi06a}) from HP Tau/G3 AB. In contrast, the 
OM cleanly separated their emission (HP Tau/G3 AB = XEST-08-OM-038; HP Tau/G2 = XEST-08-OM-40). TMC candidates \citep{scelsi06} 
are identified as well. 
X-ray and OM sources presently not identified as either TMC member or TMC candidate only have their XEST and XEST-OM identifications. 

As mentioned earlier, our database includes several cases where only a few OM points were available for a specific target
(e.g., CoKu Tau 3 AB; CW Tau; CIDA 1, etc.). However, there are several cases for which we have very good coverage even with
the OM IMAGING data (e.g., FM Tau; V773 Tau ABC; CIDA 7; etc.). DG Tau A (XEST-02-022 = XEST-02-OM-028) is a good example
where even low time resolution was able to catch the $U$-band emission of a flare, which peaked before the peak in the
X-rays, a signature of the Neupert effect. \citet{guedel06b} give a full analysis of the DG Tau A case 
and make use of the OM light curve.

In addition, some OM data were obtained in FAST mode, allowing us to obtain high time resolution light curves.
Figure~\ref{fig:omfast} shows the OM FAST light curves together with the X-ray light curves. The classification
type is also given on the upper right corner of each panel. The light curve
of T Tau will be presented elsewhere (G\"udel et al. 2007, in preparation). The OM FAST data of IT Tau
and UZ Tau EW were not useful since the sources fell on or slightly outside the OM FAST window, and thus the
light curves are not reliable. We also applied some manual corrections to the OM FAST light curves of DN Tau, DI Tau, and BP
Tau.\footnote{In the case of DN Tau, due to a feature in SAS 6.x, about half the counts were processed in the exposure S009 
(DATE-OBS='2005-03-05T01:37:35' and DATE-END='2005-03-05T02:02:35') only.
The average count rate in this exposure is 37.7~ct~s$^{-1}$, after correction for deadtime, aperture, and coincidence losses.
The detected count rate for S009 is 32.2~\cps\  (i.e., 1.17 times smaller), based on the theoretical correction in the \xmm\  Users Handbook (the empirical correction
is minimal). In contrast, the average count rate in the preceding exposure (S408) is 87.7~ct~s$^{-1}$. The latter corresponds
to a detected count rate of $63.4$~\cps\ (i.e., 1.38 times smaller). Thus, we multiplied the count rates in S009 by 
$2 / 1.17 \times 1.38 = 2.36$ to correct for the SAS feature. For DI Tau, the pointing of the satellite changed unexpectedly
after DATE-OBS='2005-02-09T19:30:13' (exposures S412, S009, S413-S416), and DI Tau fell just on the edge of the OM FAST window. We thus discarded count rates after
this epoch. For BP Tau, a problem with the time stamps of exposure S051 occurred. The DATE-OBS and DATE-END keywords
of the following exposure, S052, were incorrectly used, and thus the time stamps of events were incorrectly calculated
by the SAS software. We manually corrected these time stamps and keywords by using the correct keywords,
DATE-OBS='2004-08-16T06:45:56' DATE-END='2004-08-16T08:27:51', and incidentally shifting the time stamps by $-6256$~seconds. 
The same timing problem for the S051 IMAGING data of BP Tau needed a similar corrections of the date keywords.\label{foot:OMcorr}}

Focusing on the OM FAST light curves, it is clear that Type 3 stars show almost no variability in the OM, in
contrast to Type 2 stars. The classical T Tau star BP Tau shows significant UV variability (the BP Tau data
were originally presented by \citealt{schmitt05}). One large X-ray flare is visible at the start of the
observation, with two slowly varying flux enhancements that could be due to flares, but also rotational
modulation of bright active regions. The UVW1 light curve of BP Tau does not show direct evidence of
correlations with the X-ray light curve. In fact, the UV flux peaks \emph{after} the X-ray light curve, in
contradiction with the Neupert effect paradigm. However, such a peculiar behavior
is not uncommon in some flares observed in the Sun or in main-sequence or evolved magnetically 
active stars with no accretion \citep[e.g.,][]{ayres01,stelzer03,osten05}. 
Nevertheless, the strong variability in the OM light curve is generally uncorrelated with the X-ray light
curve, which strongly suggests that the OM light curve is completely dominated by the UV emission 
of accretion spots in BP Tau. The period of BP Tau being $7.60$~d, much longer than
the $1.5$~d duration of the observation or typical OM ``events'' ($0.1-0.2$~d), it
is more probable that such events are due to accretion events such as
those observed in the brown dwarf 2MASS J04141188+2811535 \citep{grosso06a}.

   \begin{figure}
   \centering
   \resizebox{\hsize}{!}
   {\includegraphics[width=\textwidth]{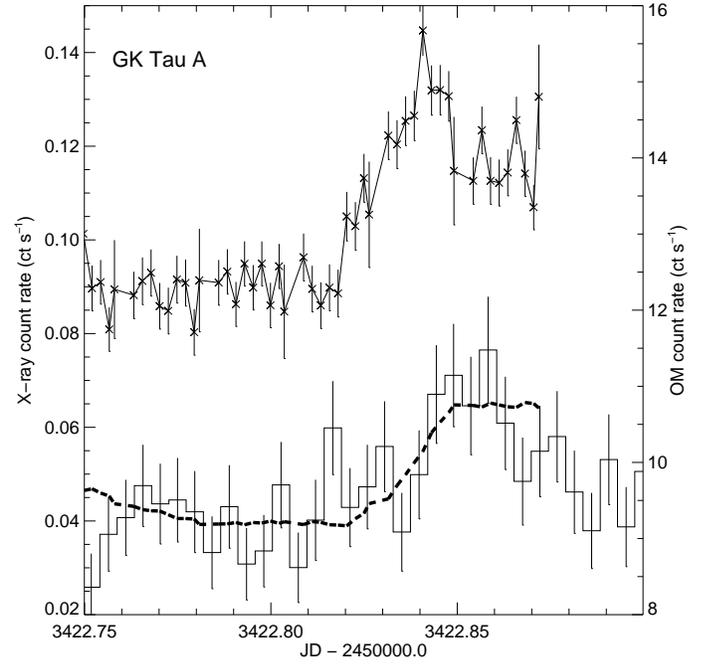}}
      \caption{Details of the GK Tau OM and X-ray light curves. The OM light curve (top curve) was rebinned to a bin size of
      200~s, while the X-ray light curve (bottom histogram) used a 400~s bin size. The thick dashed line corresponds
      to the OM light curve convolved through a kernel $K(t,t^\prime)$ (see text) but it was shifted down for clarity. 
      The GK Tau A event shows evidence of a Neupert-like effect.}
         \label{fig:gktaudetail}
   \end{figure}
%

The GK Tau light curves deserve special attention. Indeed, the OM light curve
shows evidence of two behaviors: a general slow variation uncorrelated with the
X-ray light curve that is probably due to the rotational modulation of accretion
spots ($P = 4.60$~d), and a short-duration event around JD 2453422.84 that appears related to a
small flare-like event in the X-rays. Figure~\ref{fig:gktaudetail} shows an
extract of the light curves. Note that the OM emission is from GK Tau A. The companion
at $2\farcs 5$ was not detected in the OM. In the X-rays, the binary cannot be resolved; nevertheless,
the A component should contribute the most to the X-ray emission. In addition, the X-ray emission
is blended with the X-ray emission of the nearby GI Tau. The X-ray light curves of GK Tau AB and
GI Tau shown in this paper take into account this blend since they were extracted from smaller
radii (160~pix, i.e., $8\arcsec$). We convolved the OM light curve through the kernel function, $K(t,t^\prime)$, where
$K(t,t^\prime) = e^{-(t-t^\prime)/\tau}$ if $t^\prime<t$ and $K(t,t^\prime) = 0$ otherwise,
and where $\tau$ was set to 2~ks; for clarity and comparison with the X-ray light curve, we introduced
a vertical offset to the convolved light curve. The convolution of the OM light curve is similar
to an integration; however, the use of the exponential function in the kernel allows us
to mimic the decaying effect of radiative and conductive cooling in X-rays.
This exercise is a good indication
that the event observed in X-rays and in the OM in GK Tau is a flare that displayed a Neupert-like
effect, with the optical emission peaking before the X-rays. A similar case is reported in DG Tau A
with OM IMAGING data \citep{guedel06b}. GK Tau's OM light curve can, therefore,
be seen as a combination of stellar photospheric emission, accretion spot emission producing
slow variability, and flare-like events. As seen from GK Tau, but most visibly BP Tau (and other Type 2 stars), 
the long-term variability due to accretion does not appear to correlate with the X-ray variability in general. 
This indicates that the contribution of accretion to the X-ray emission is limited, probably to the soft component
only, as suggested from high-resolution X-ray spectra of classical T Tau stars \citep{telleschi06a}.
In contrast, Type 3 stars with little or no accretion do not show evidence of slow variability in the OM.

%

\section{Summary and Conclusions}

We have presented results from the Optical Monitor data of the  \xmm\  Extended Survey of the Taurus Molecular 
Cloud. Optical ($U$) or ultraviolet (UVW1 or UVW2) magnitudes were obtained; in addition low time resolution light
curves from the IMAGING data were compiled and compared to the X-ray light curves. For a handful of sources, FAST mode
data with high time resolution were available as well.

The OM data are unique since they provide strictly simultaneous coverage in the optical (or ultraviolet) together with
X-rays. This \xmm\  capability allowed us to study light curves; we were able to detect Neupert-like effects in the
accreting stars GK Tau A and DG Tau A, but also slow variability in accreting stars that is most likely due to accretion
spots. A statistical analysis shows that, although observed $U$ magnitudes are similar in Type 2 and 3 stars, 
accreting stars are, in fact, brighter in the $U$ band than Type 3 stars after
correction for extinction. This excess emission most likely originates from accretion. 
This behavior, although not a new discovery, is now shown to be disconnected from the X-ray variability. Indeed,
for example, the light curve of BP Tau shows strong ultraviolet variability that is not reproduced in the X-ray light
curve. In contrast, non-accreting stars (or stars with low accretion rates) do not show evidence of slow variability. 
Their emission is therefore probably dominated by photospheric emission. Similar results were found in the optical ($BVRI$) and X-ray
data of Orion Nebula Cluster pre-main sequence stars \citep{stassun06}.

The XEST OM catalogue lists 2,148 detected sources and about 88~\% (1893) OM sources had 2MASS counterparts; in contrast, 916 sources were detected in the X-rays and had
coordinates such that they could be detected in the OM (not necessarily). However, only 98 X-ray sources matched
OM sources, after astrometric corrections applied to both the X-rays and the OM coordinates based on 2MASS coordinates.
And out of these 98 sources, only 51 are considered as TMC members (12 are bona-fide TMC candidates;
\citealt{scelsi06}). Therefore, the vast majority (about 97~\%) detected in the OM images are, in fact, not TMC members.
While it is hard to determine the origin of a target based on X-rays or optical alone, the OM data, together with 2MASS
and {\it Spitzer} data (Padgett et al. 2006, in preparation), should provide useful constraints to discriminate 
between stellar and non-stellar objects, and thus discover new members of the Taurus Molecular Cloud.

\begin{acknowledgements}
      We thank an anonymous referee for detailed comments and suggestions that helped to improve this manuscript.
      M.~A. acknowledges support by National Aeronautics and Space Administration (NASA) grant  NNG05GF92G and
      from a Swiss National Science Foundation Professorship (PP002--110504). 
      We acknowledge financial support by the International Space Science Institute
      (ISSI) in Bern to the {\it XMM-Newton} XEST team. The PSI group was supported by the Swiss National
      Science Foundation (grants 20-66875.01 and 20-109255/1). Part of this research is based on observations
      obtained with {\it XMM-Newton}, an ESA science mission with instruments and contributions directly funded
      by ESA member states and the USA (NASA). This publication makes use of data products from the Two Micron
      All Sky Survey (2MASS), which is a joint project of the University of Massachusetts and the Infrared
      Processing and Analysis Center/California Institute of Technology, funded by NASA and the National Science
      Foundation. Finally, our research made use of the SIMBAD database, operated at CDS, Strasbourg, France.

\end{acknowledgements}


\clearpage


\clearpage

\clearpage

\Online
\appendix

\section{Full OM catalogue}

\input{OMcatfull_paper.tex}

\clearpage
\section{X-ray and OM light curves}

\begin{figure*}
\centering
\resizebox{.9\textwidth}{!}{\includegraphics{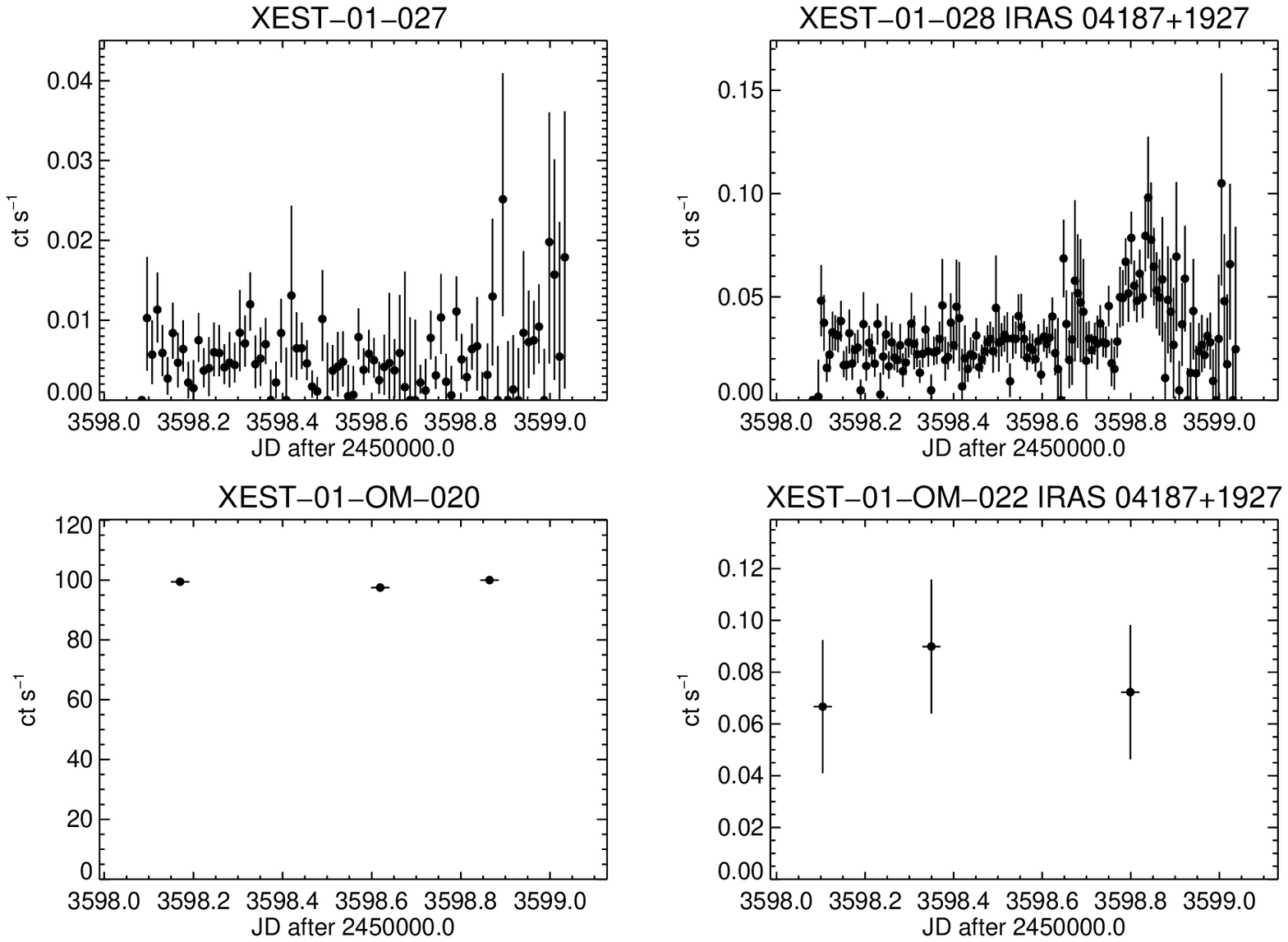}}
\resizebox{.9\textwidth}{!}{\includegraphics{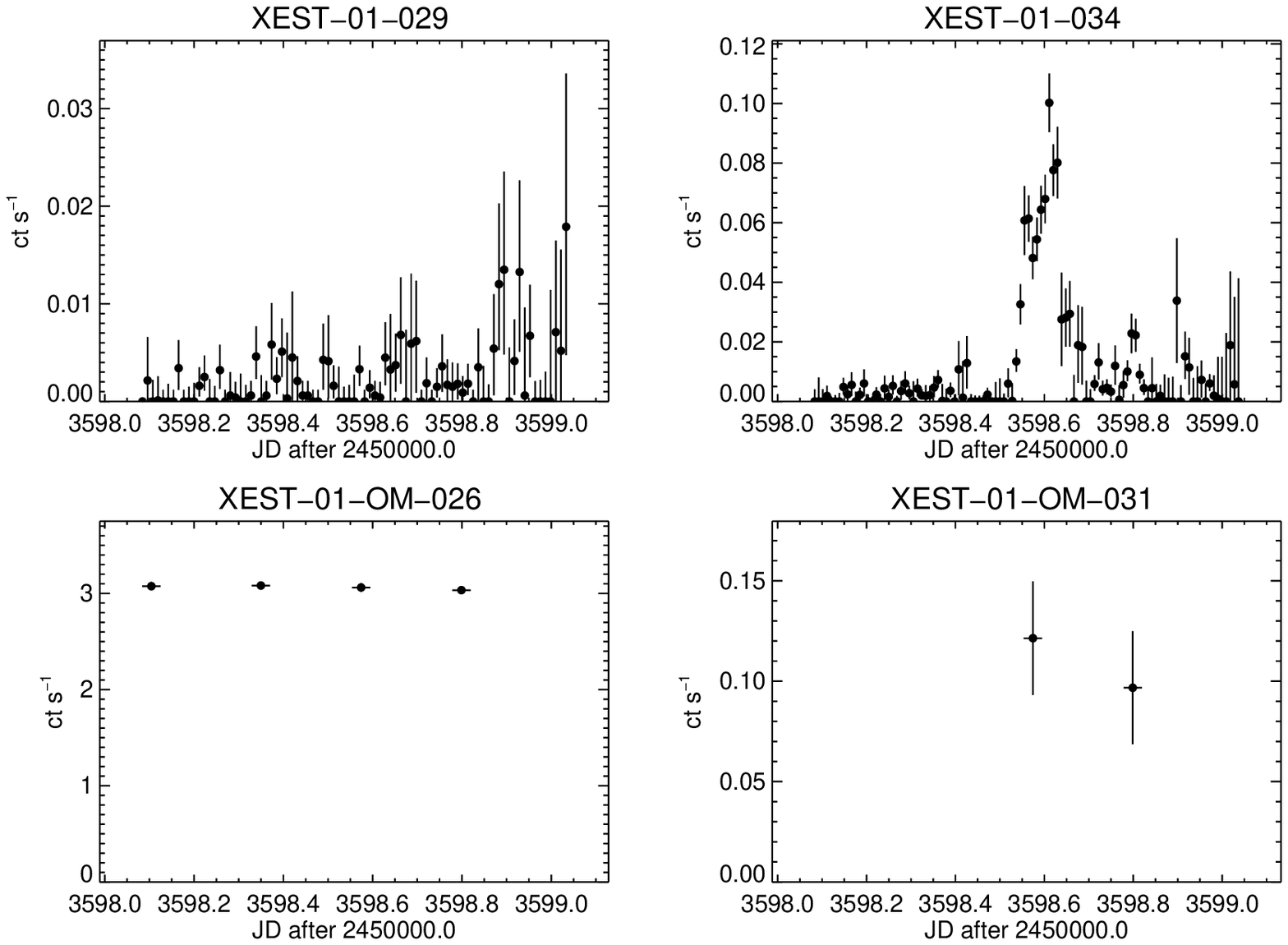}}
\caption{Light curves.}
	   \label{fig:all_lc}%
\end{figure*}

\clearpage\addtocounter{figure}{-1}

\begin{figure*}
\centering
\resizebox{.9\textwidth}{!}{\includegraphics{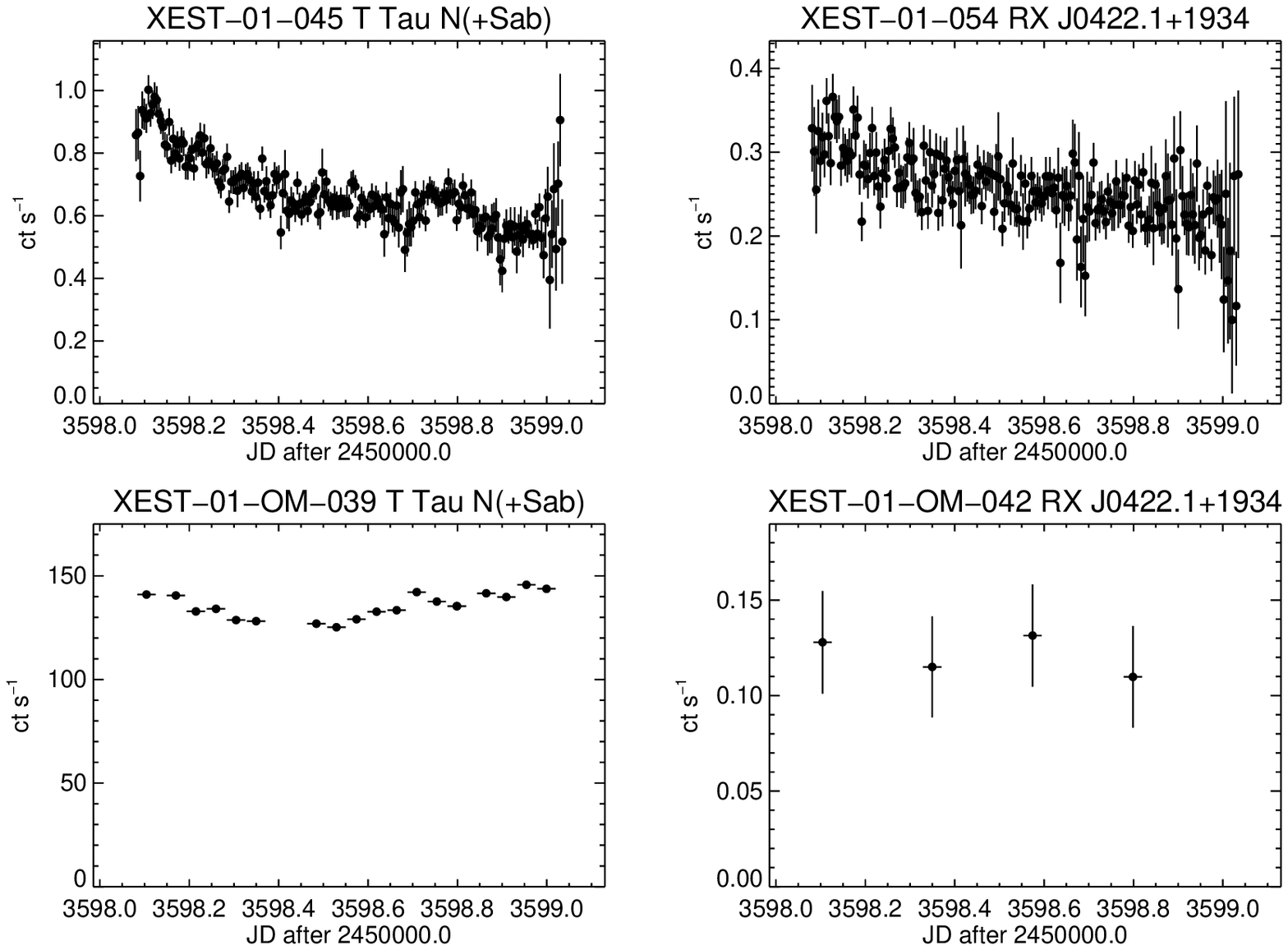}}
\resizebox{.9\textwidth}{!}{\includegraphics{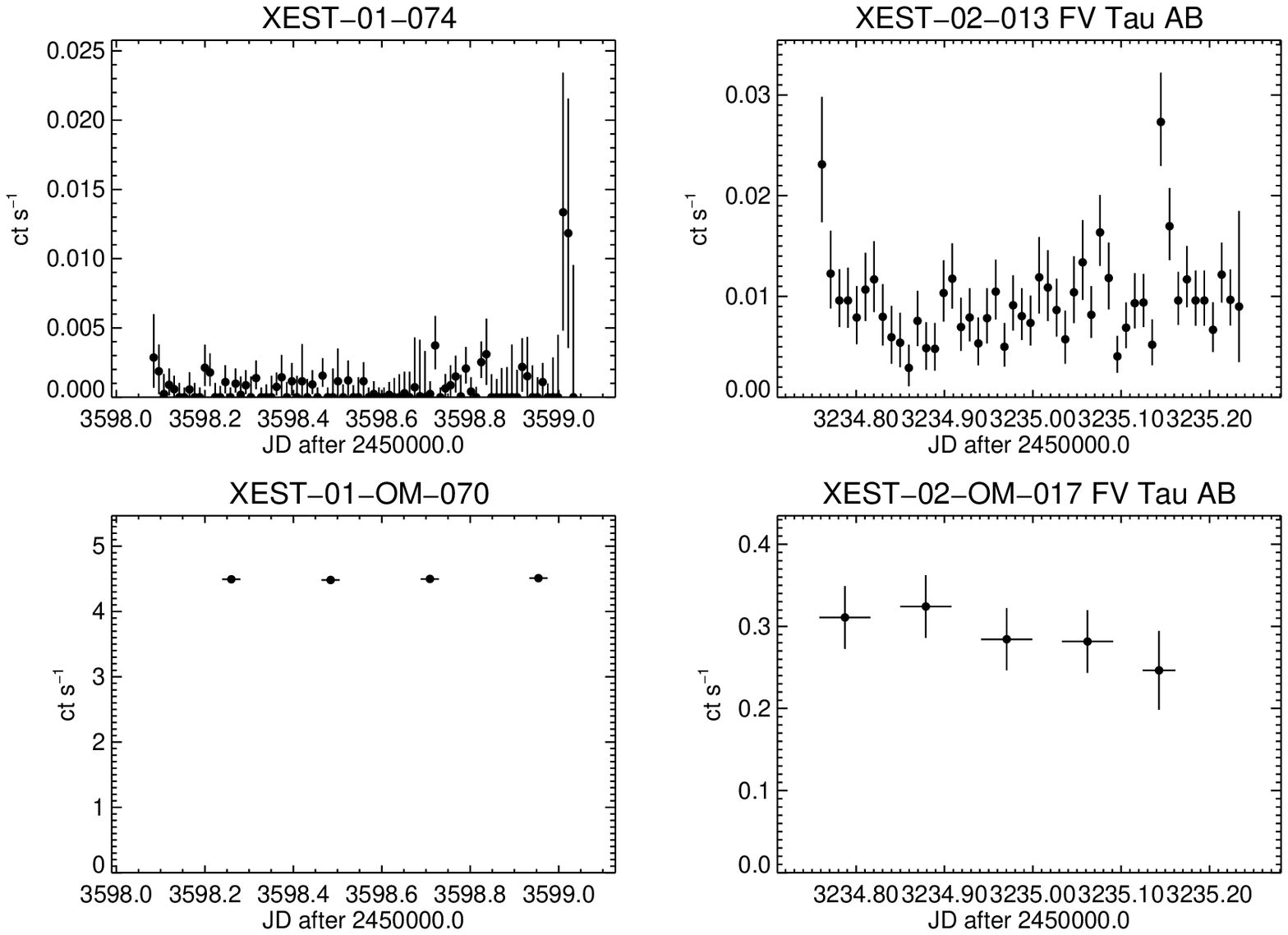}}
\caption{Light curves (continued).}
\end{figure*}

\clearpage\addtocounter{figure}{-1}

\begin{figure*}
\centering
\resizebox{.9\textwidth}{!}{\includegraphics{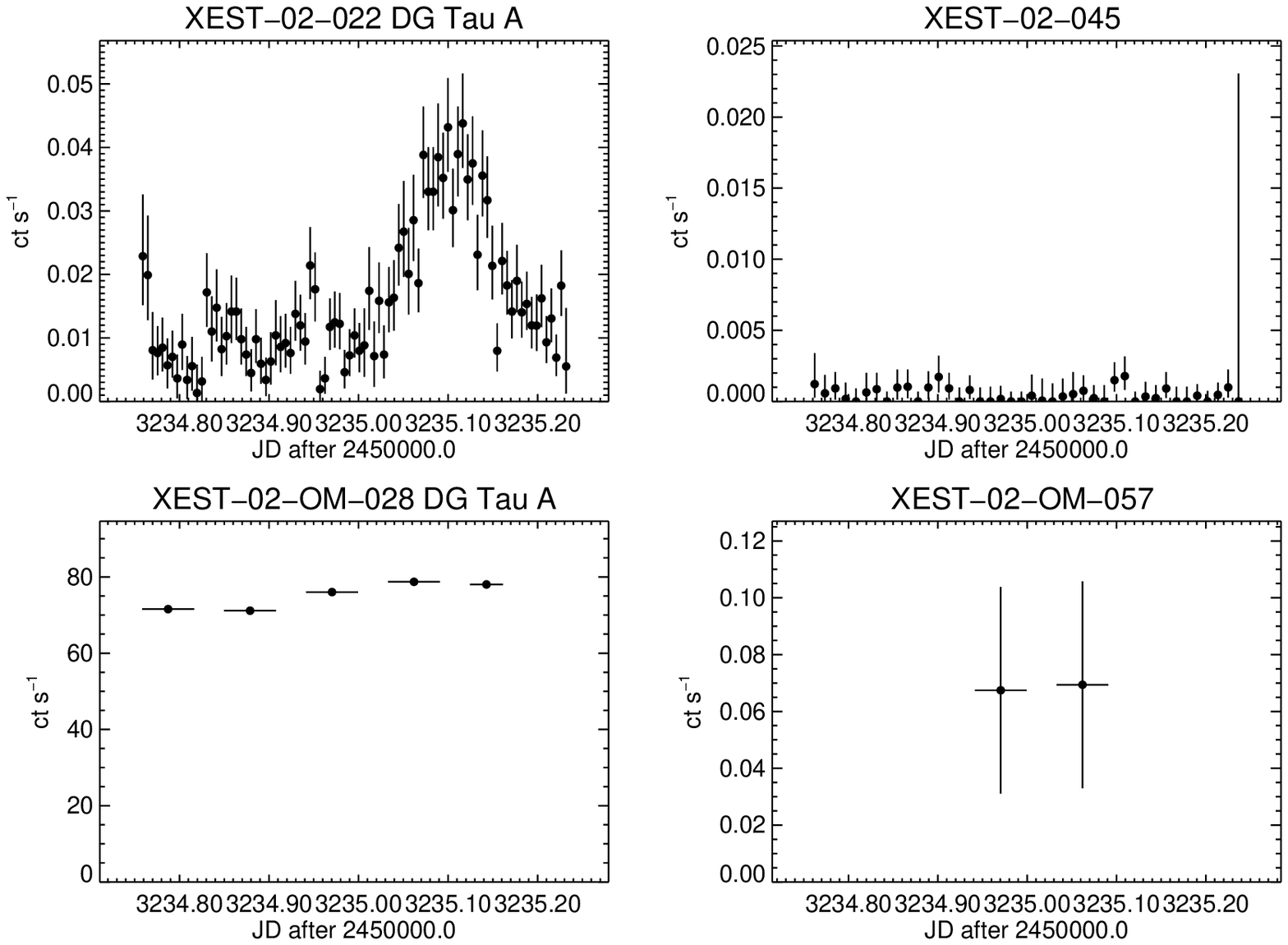}}
\resizebox{.9\textwidth}{!}{\includegraphics{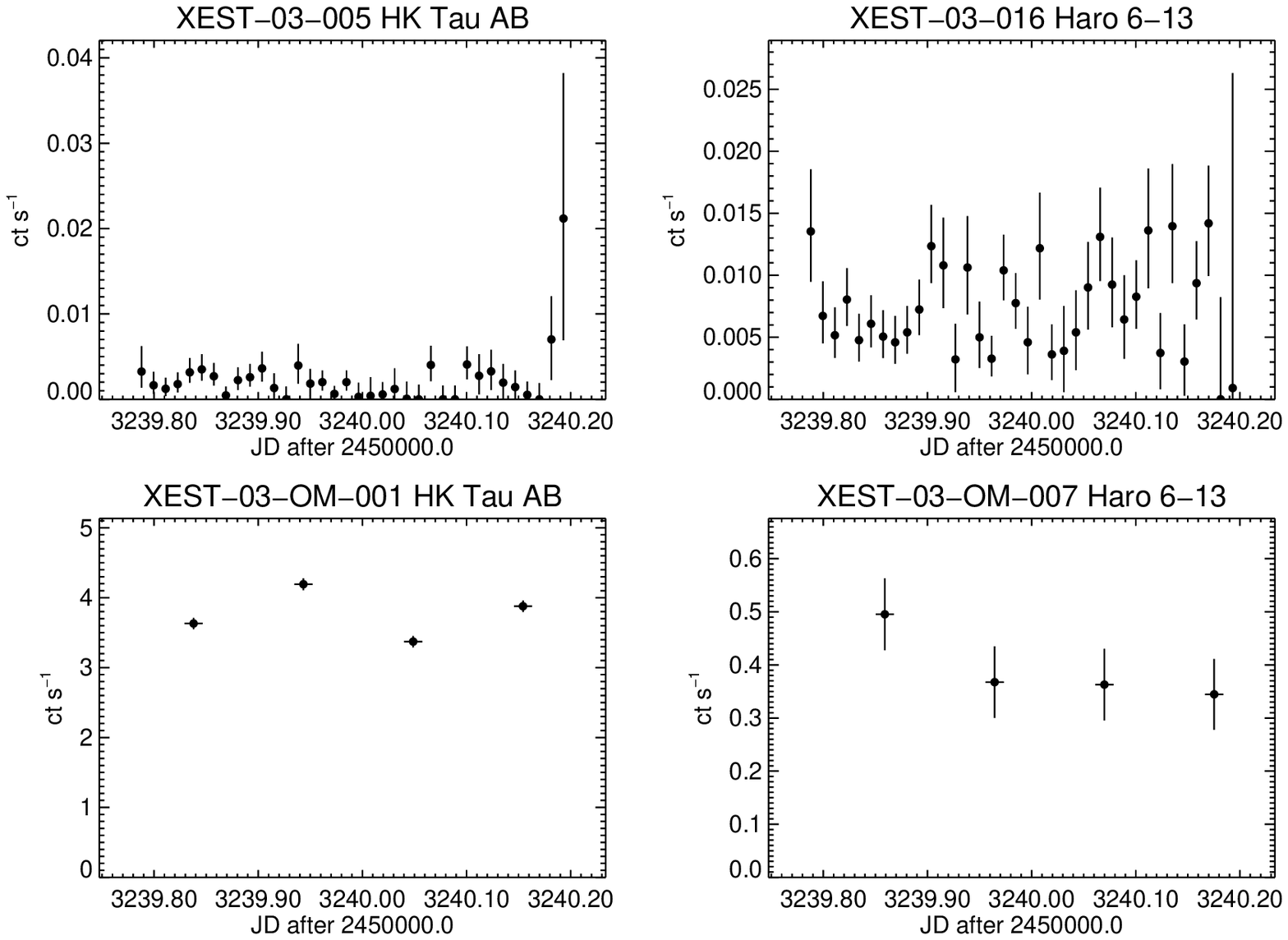}}
\caption{Light curves (continued).}
\end{figure*}

\clearpage\addtocounter{figure}{-1}

\begin{figure*}
\centering
\resizebox{.9\textwidth}{!}{\includegraphics{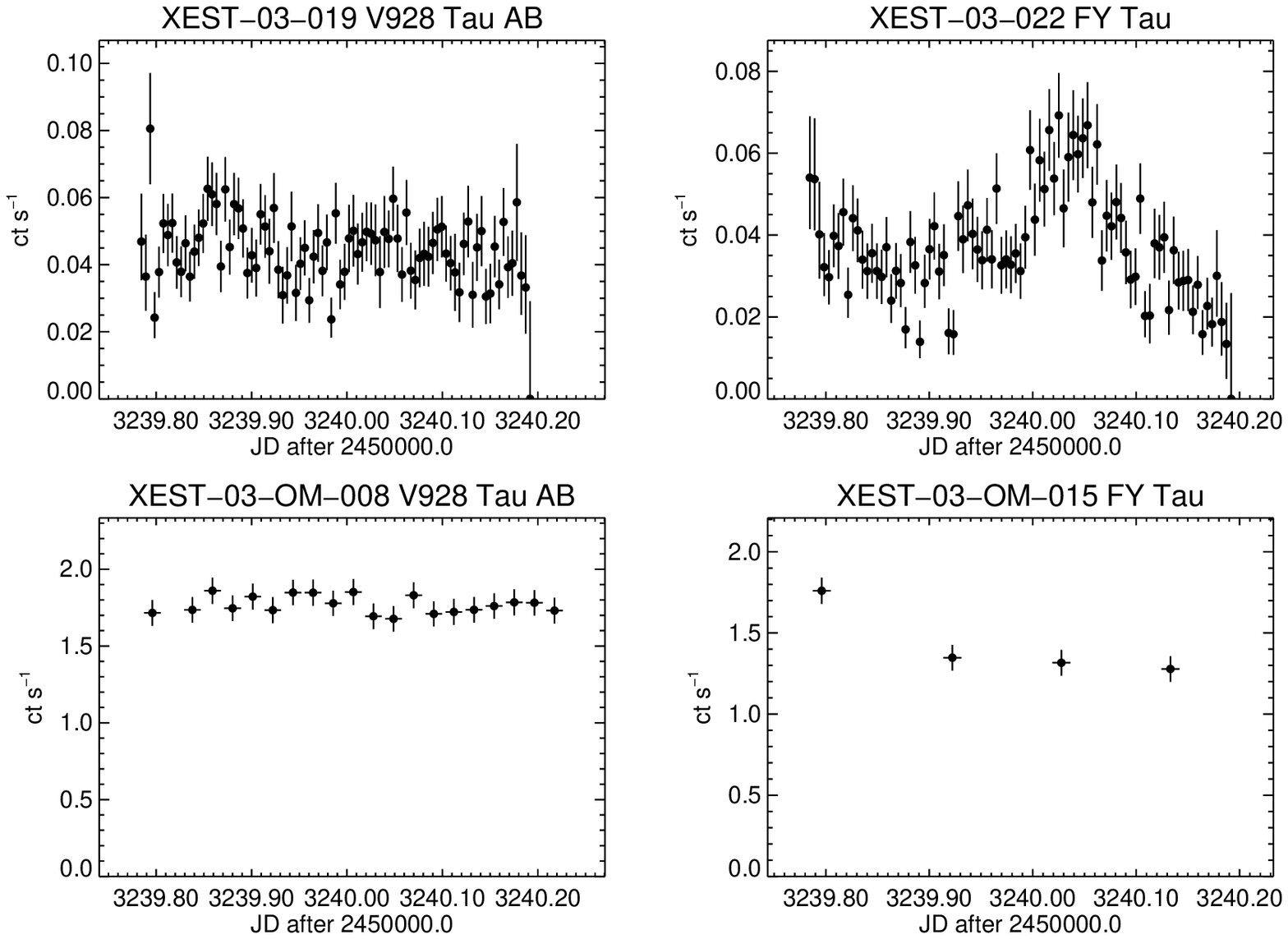}}
\resizebox{.9\textwidth}{!}{\includegraphics{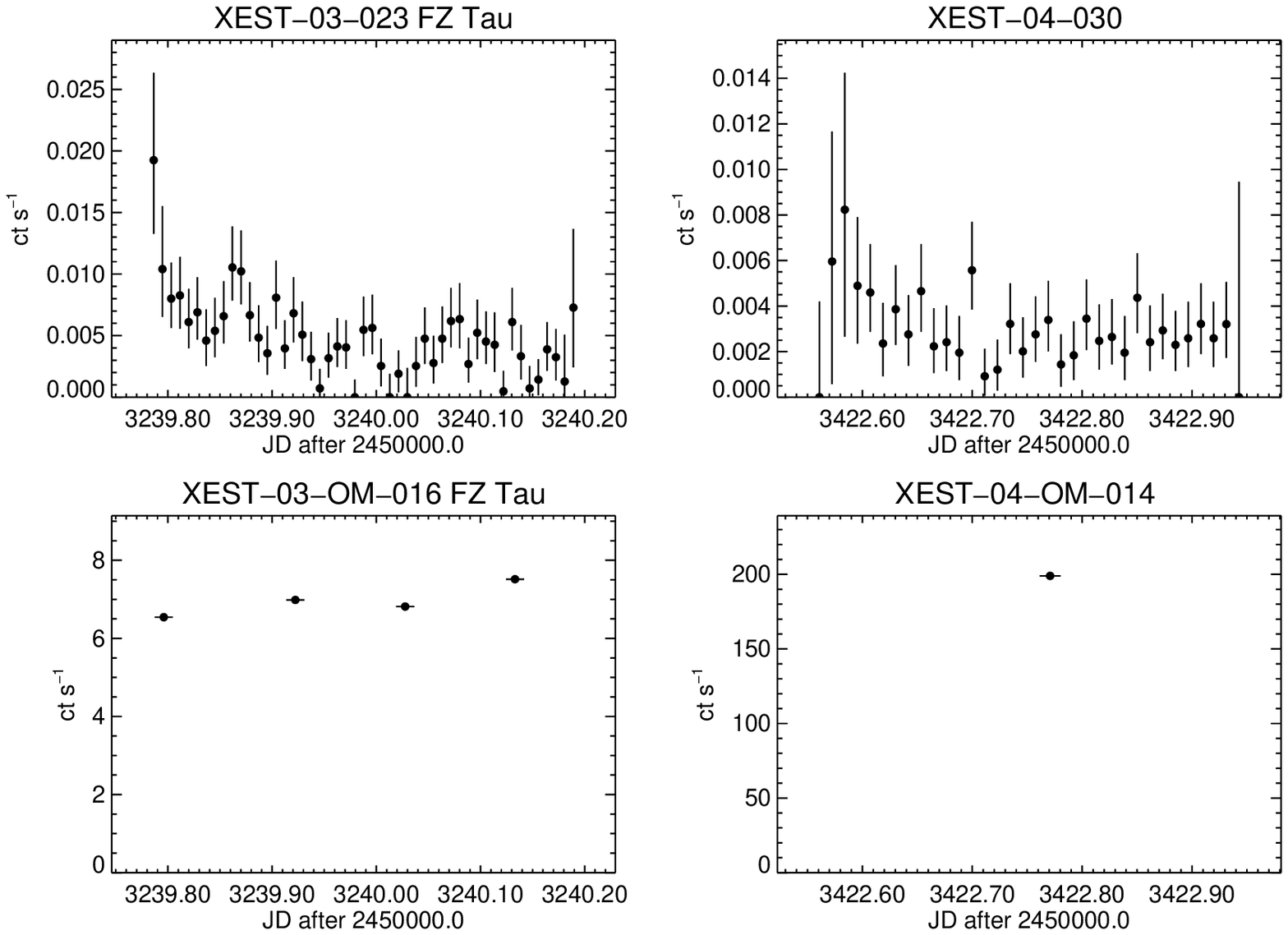}}
\caption{Light curves (continued).}
\end{figure*}

\clearpage\addtocounter{figure}{-1}

\begin{figure*}
\centering
\resizebox{.9\textwidth}{!}{\includegraphics{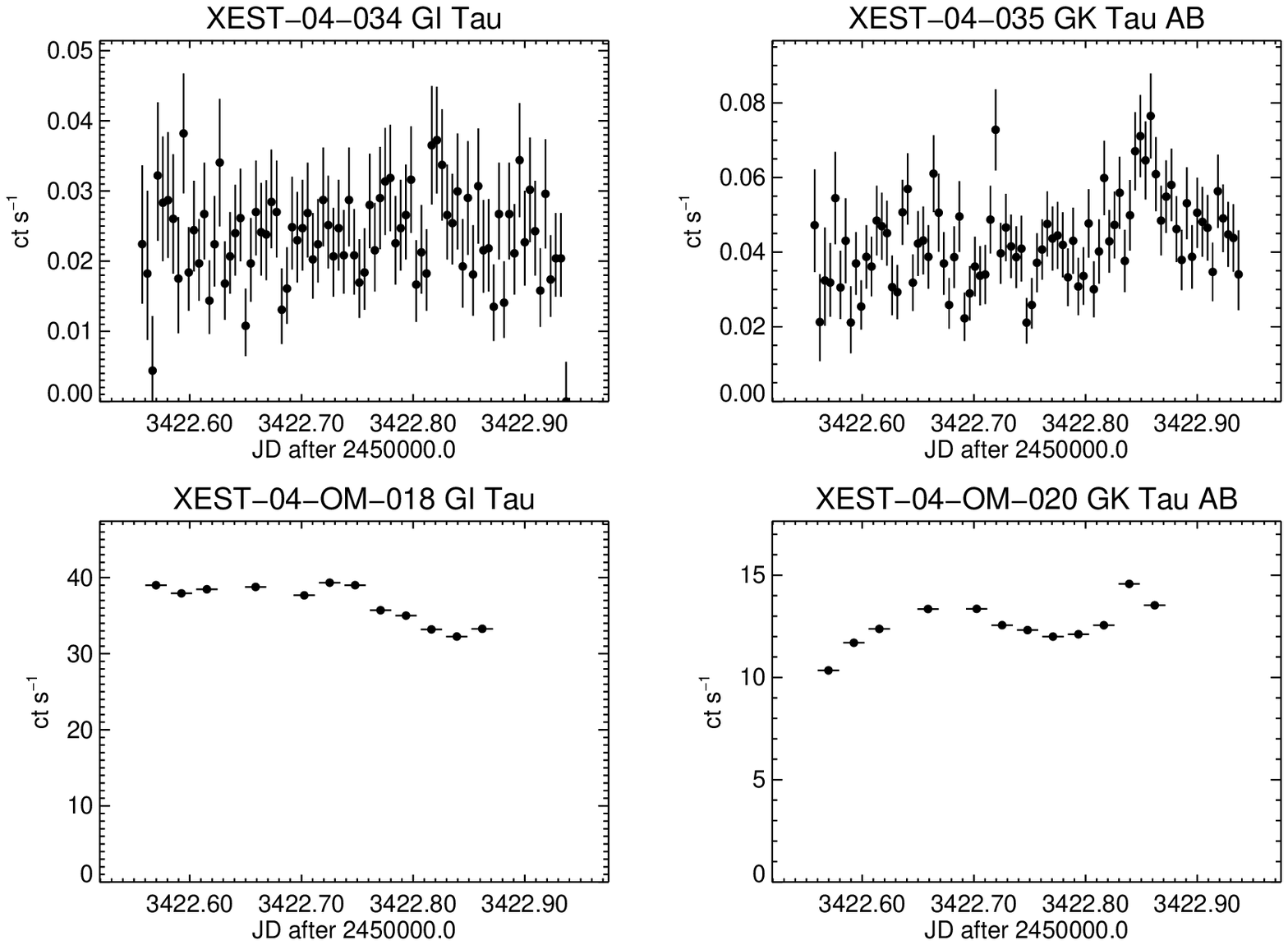}}
\resizebox{.9\textwidth}{!}{\includegraphics{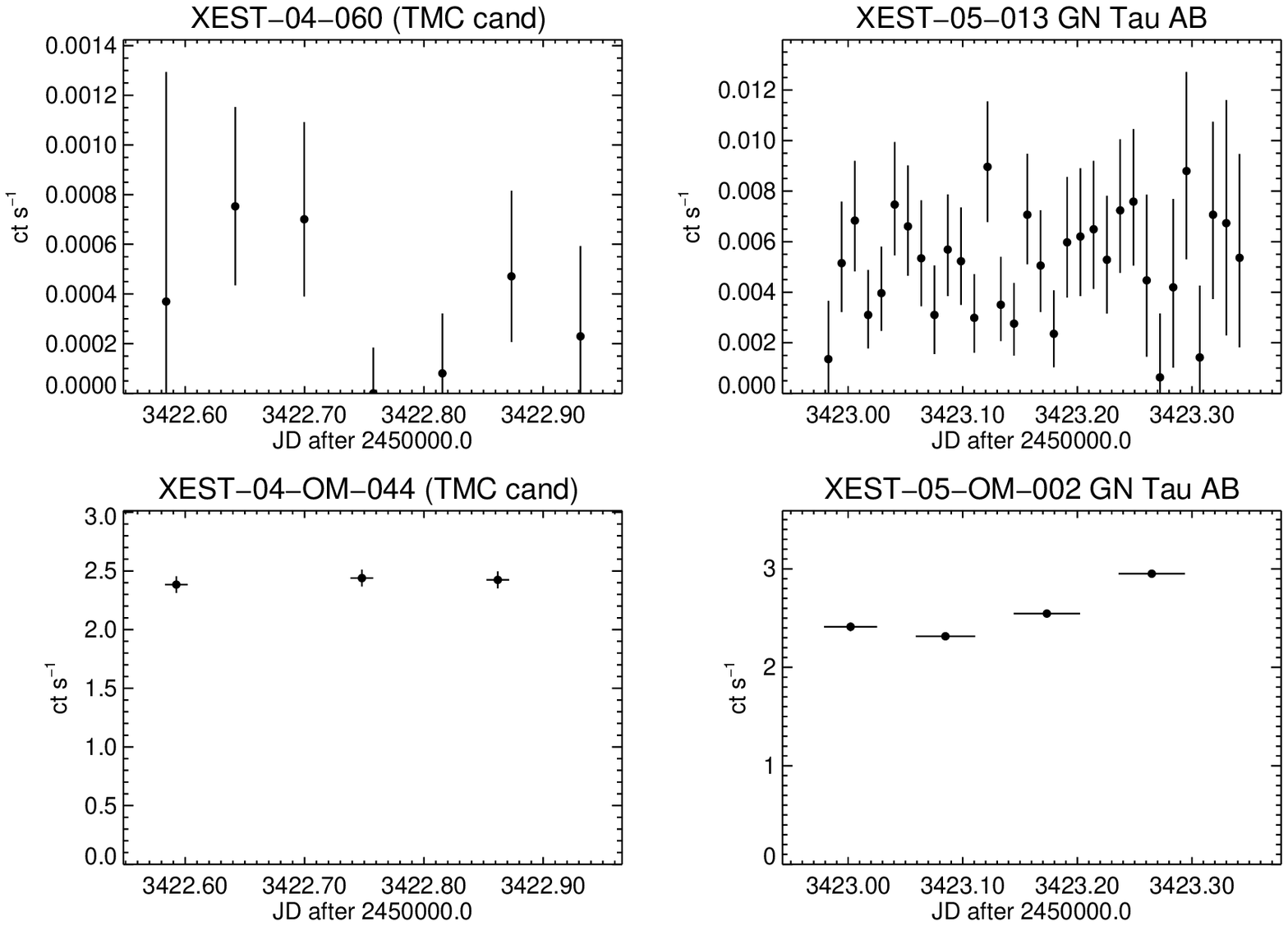}}
\caption{Light curves (continued).}
\end{figure*}

\clearpage\addtocounter{figure}{-1}

\begin{figure*}
\centering
\resizebox{.9\textwidth}{!}{\includegraphics{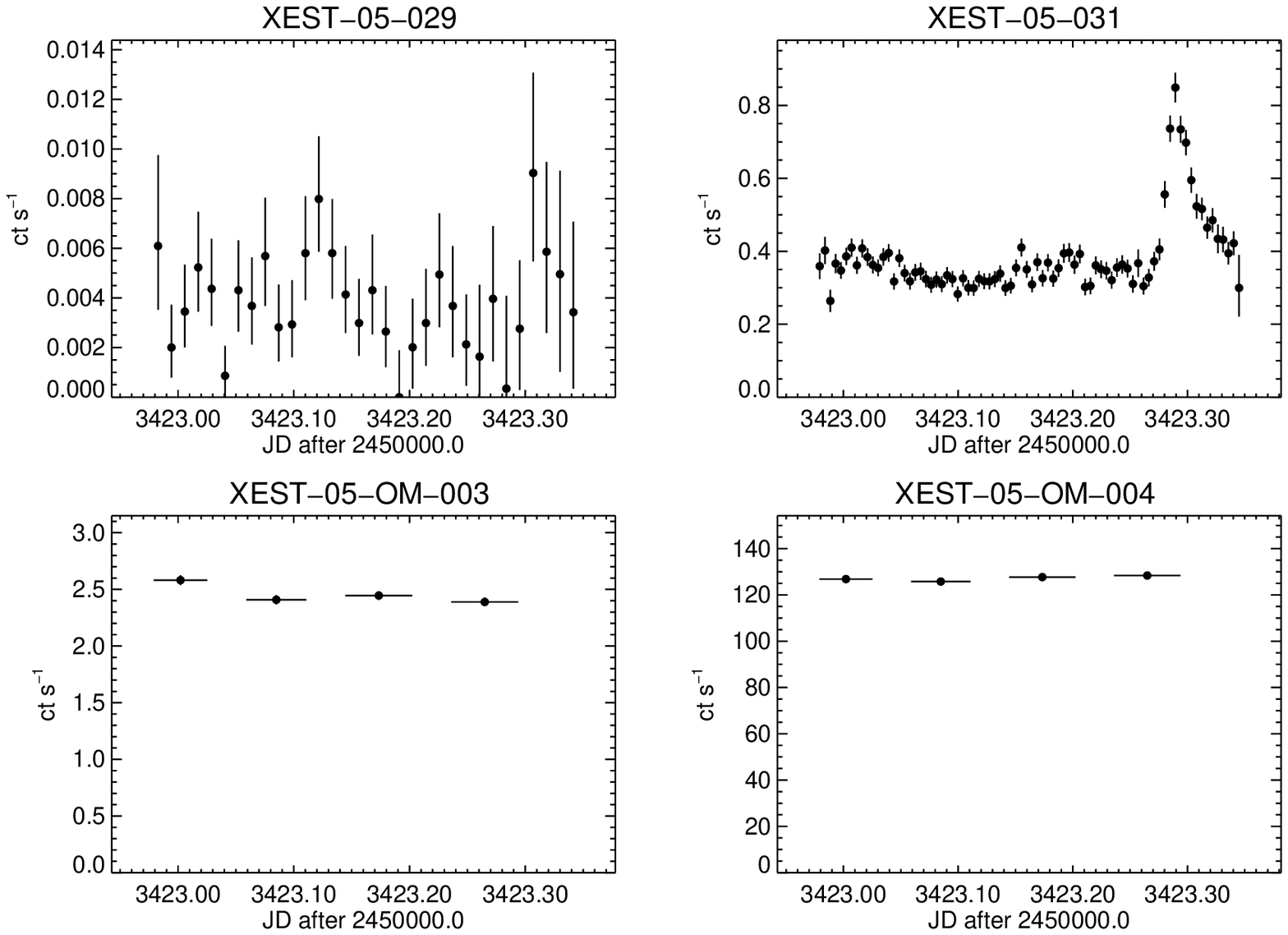}}
\resizebox{.9\textwidth}{!}{\includegraphics{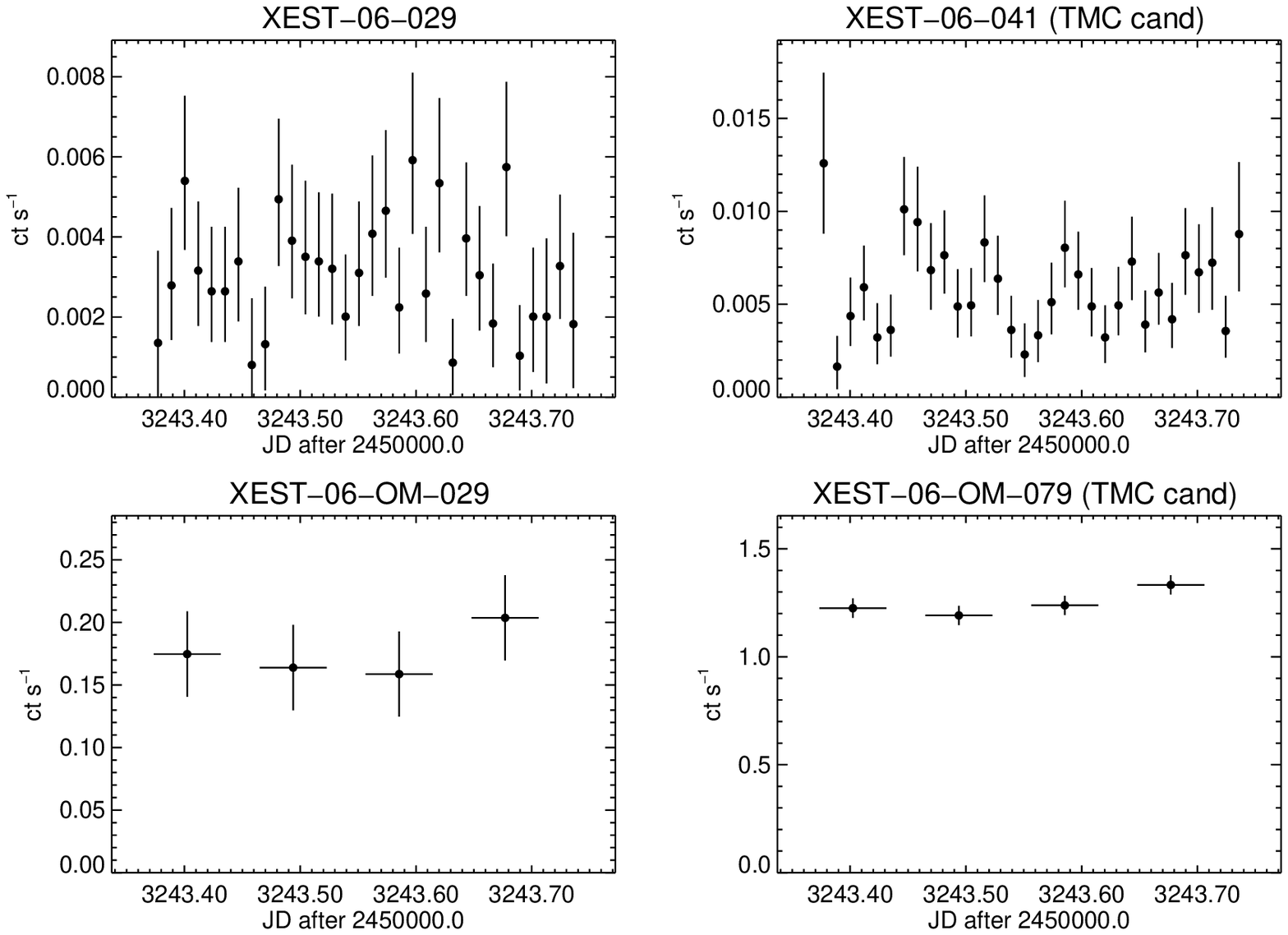}}
\caption{Light curves (continued).}
\end{figure*}

\clearpage\addtocounter{figure}{-1}

\begin{figure*}
\centering
\resizebox{.9\textwidth}{!}{\includegraphics{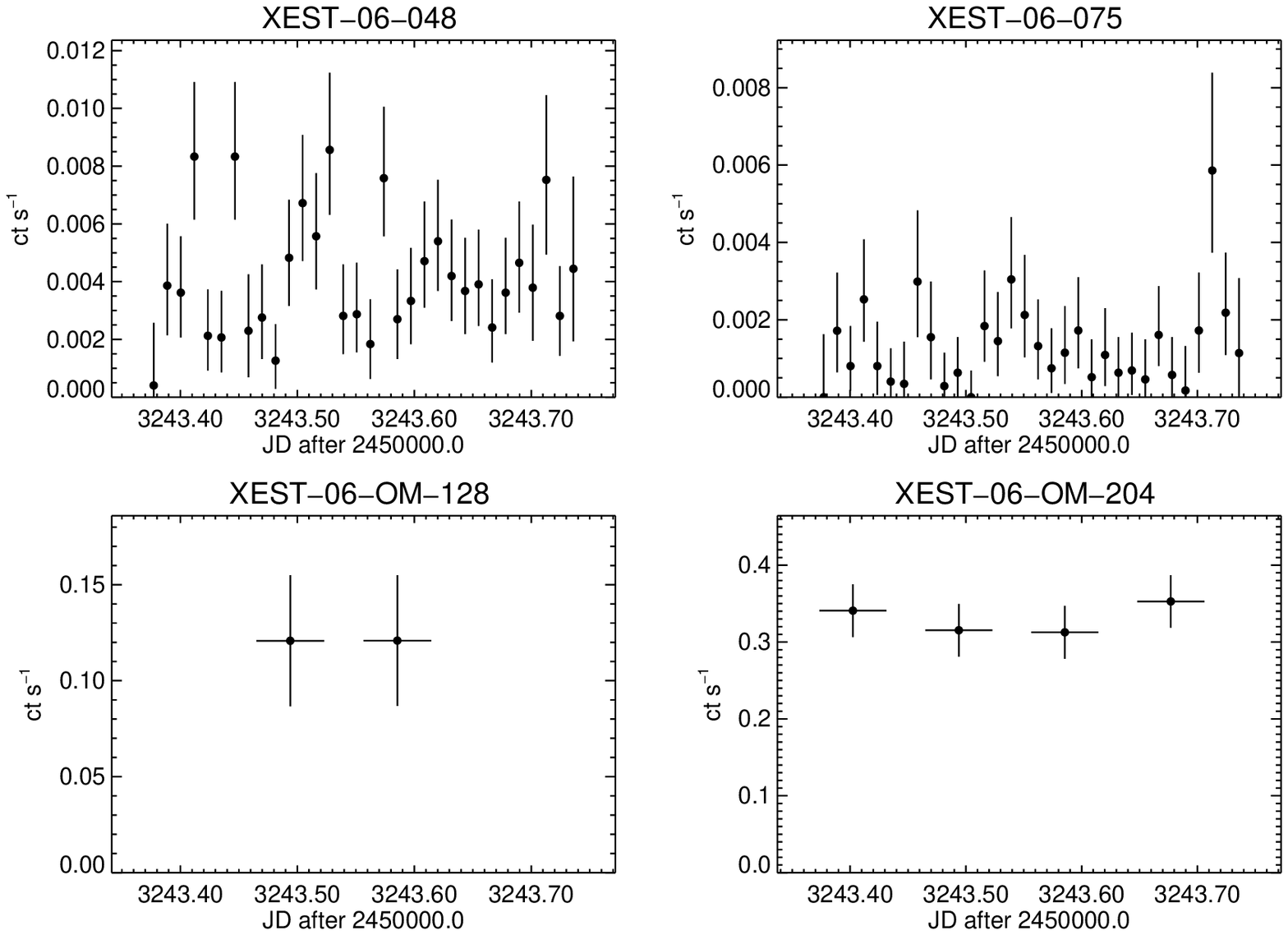}}
\resizebox{.9\textwidth}{!}{\includegraphics{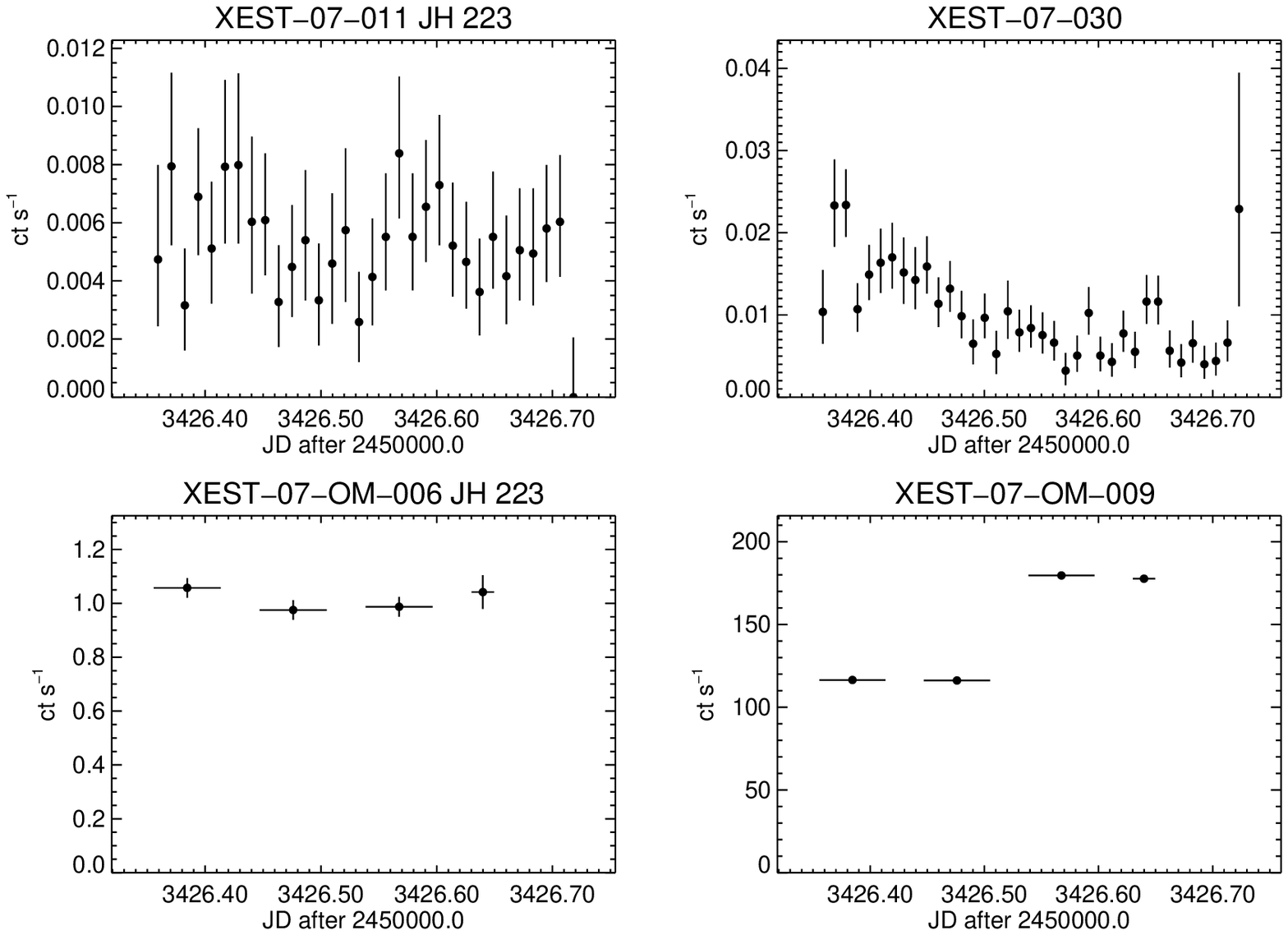}}
\caption{Light curves (continued).}
\end{figure*}

\clearpage\addtocounter{figure}{-1}

\begin{figure*}
\centering
\resizebox{.9\textwidth}{!}{\includegraphics{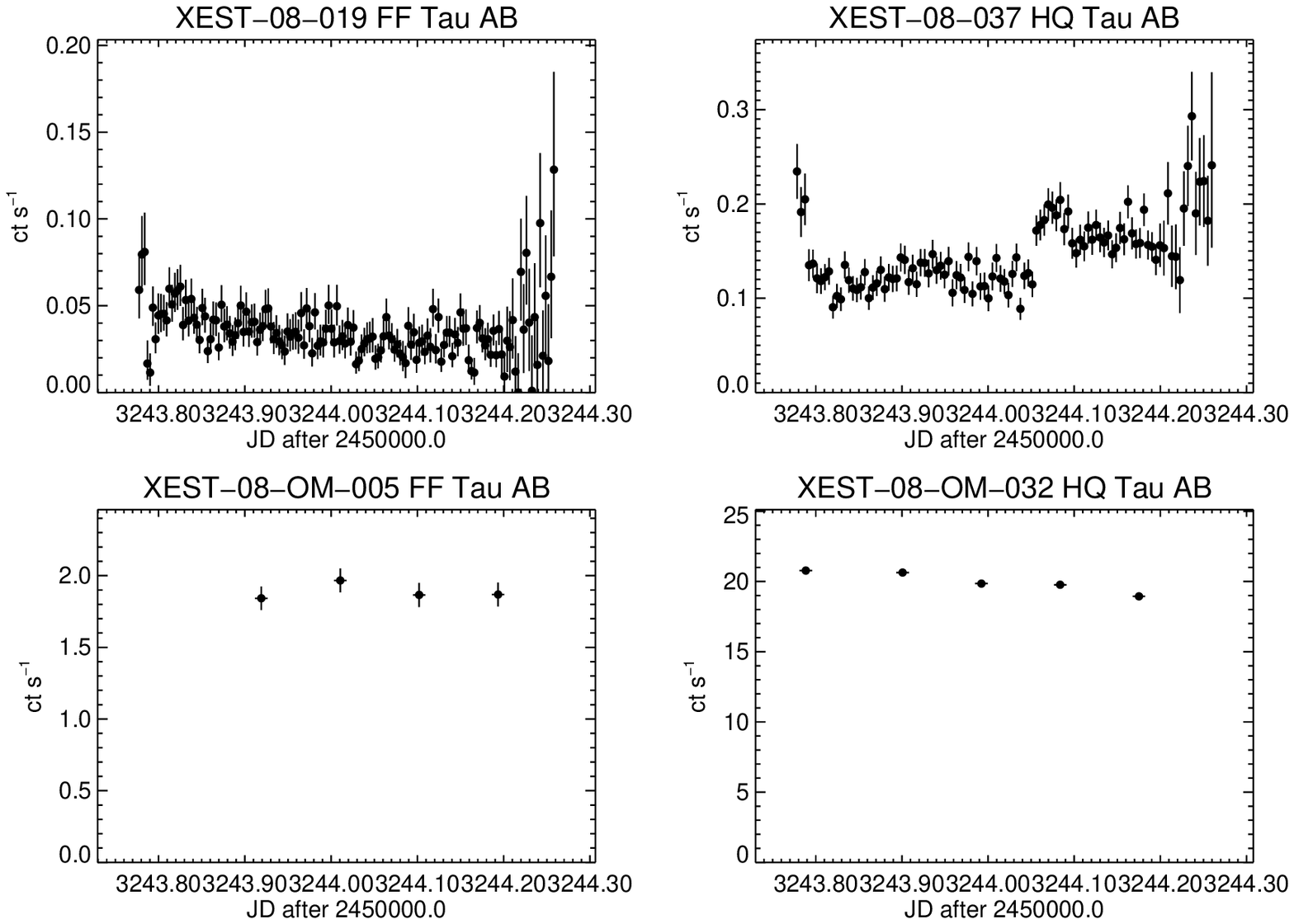}}
\resizebox{.9\textwidth}{!}{\includegraphics{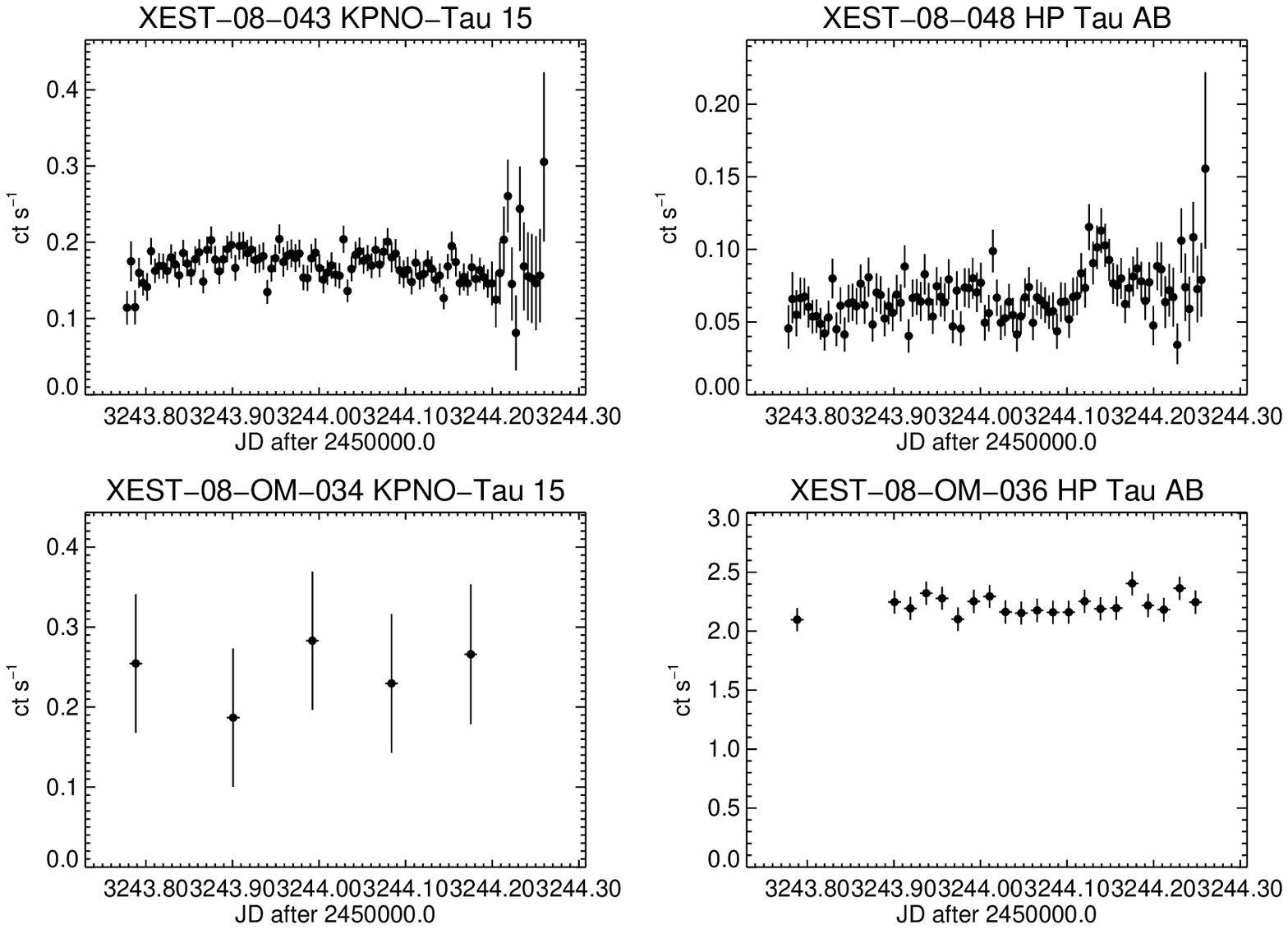}}
\caption{Light curves (continued).}
\end{figure*}

\clearpage\addtocounter{figure}{-1}

\begin{figure*}
\centering
\resizebox{.9\textwidth}{!}{\includegraphics{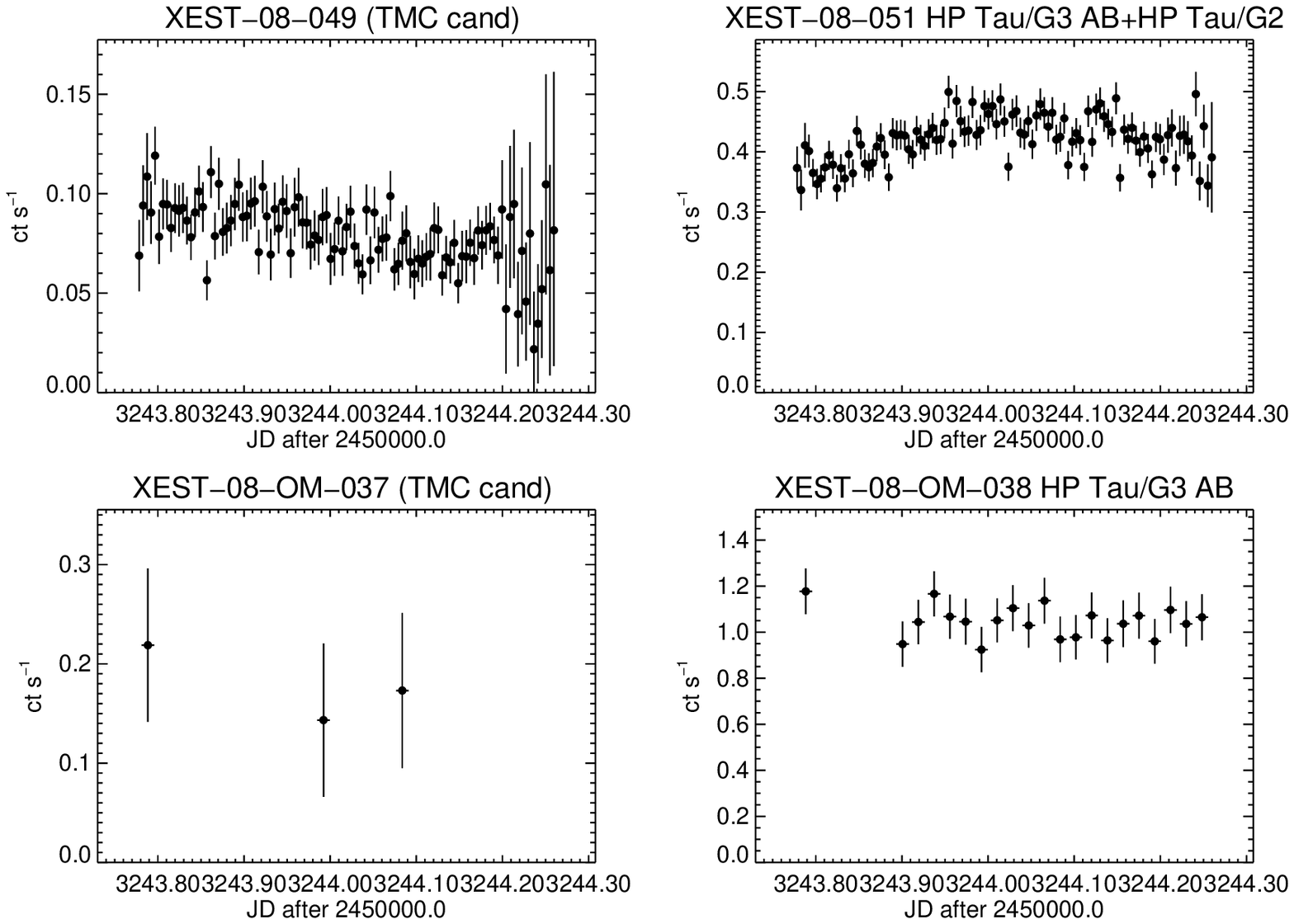}}
\resizebox{.9\textwidth}{!}{\includegraphics{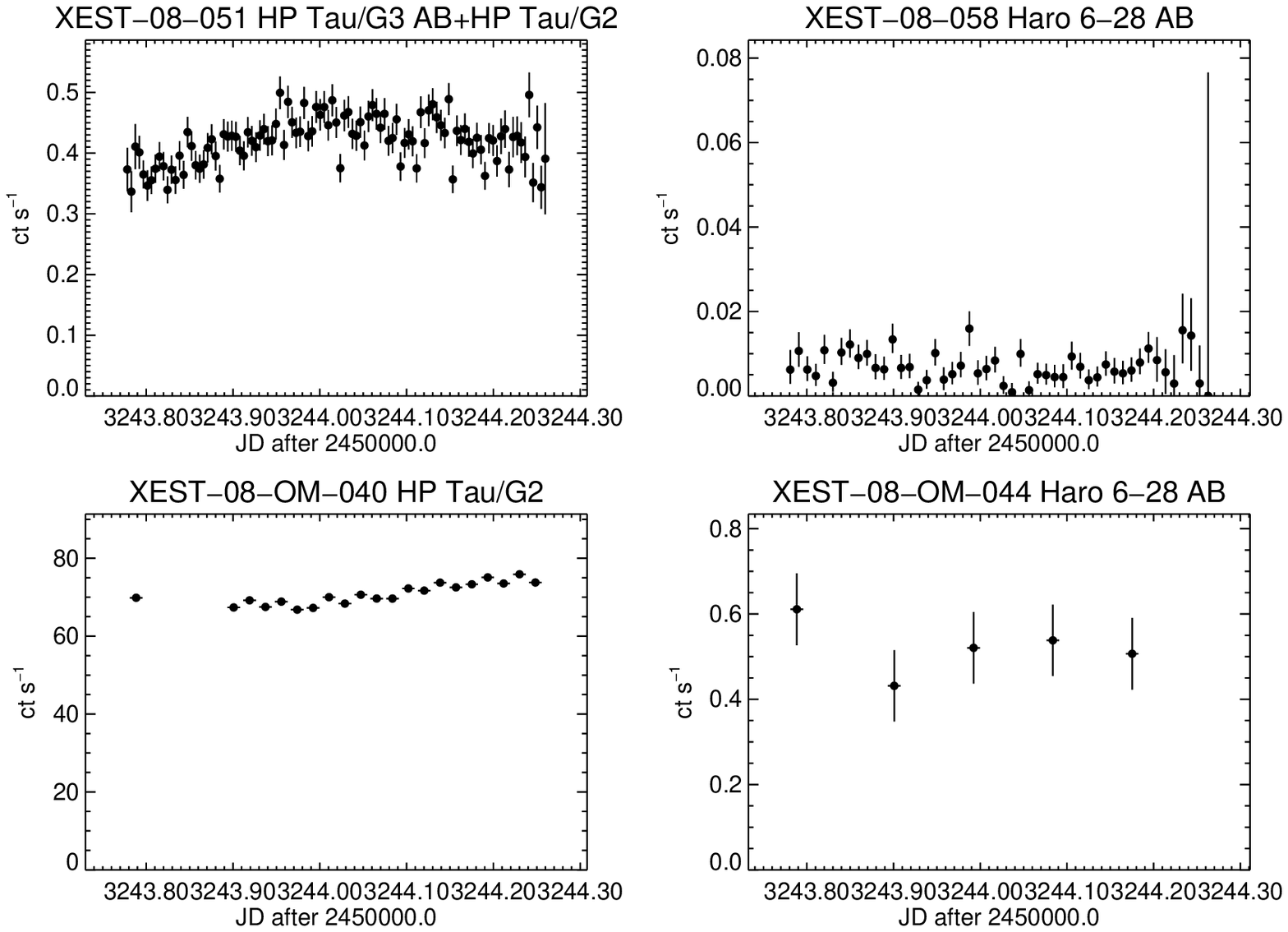}}
\caption{Light curves (continued).}
\end{figure*}

\clearpage\addtocounter{figure}{-1}

\begin{figure*}
\centering
\resizebox{.9\textwidth}{!}{\includegraphics{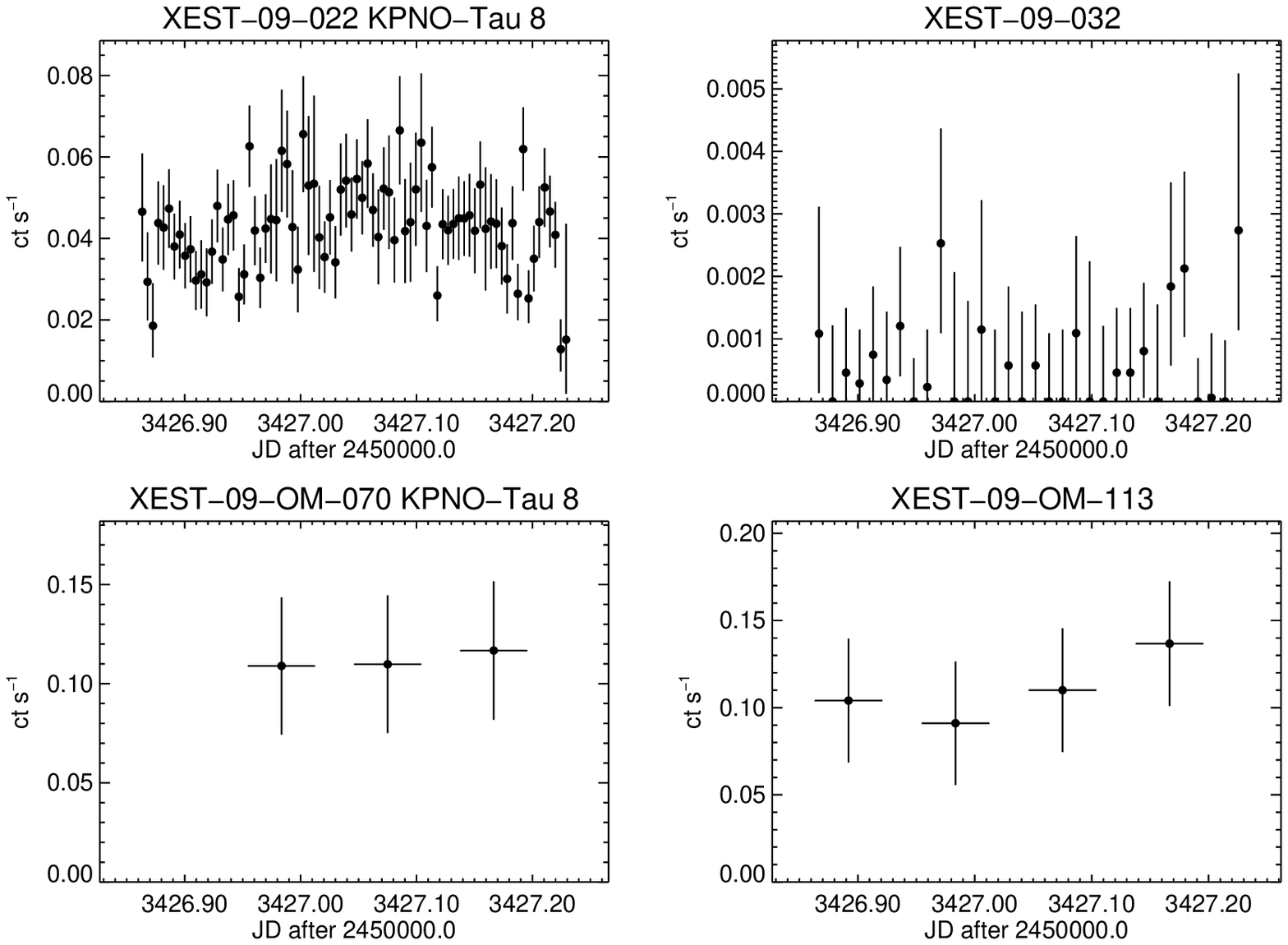}}
\resizebox{.9\textwidth}{!}{\includegraphics{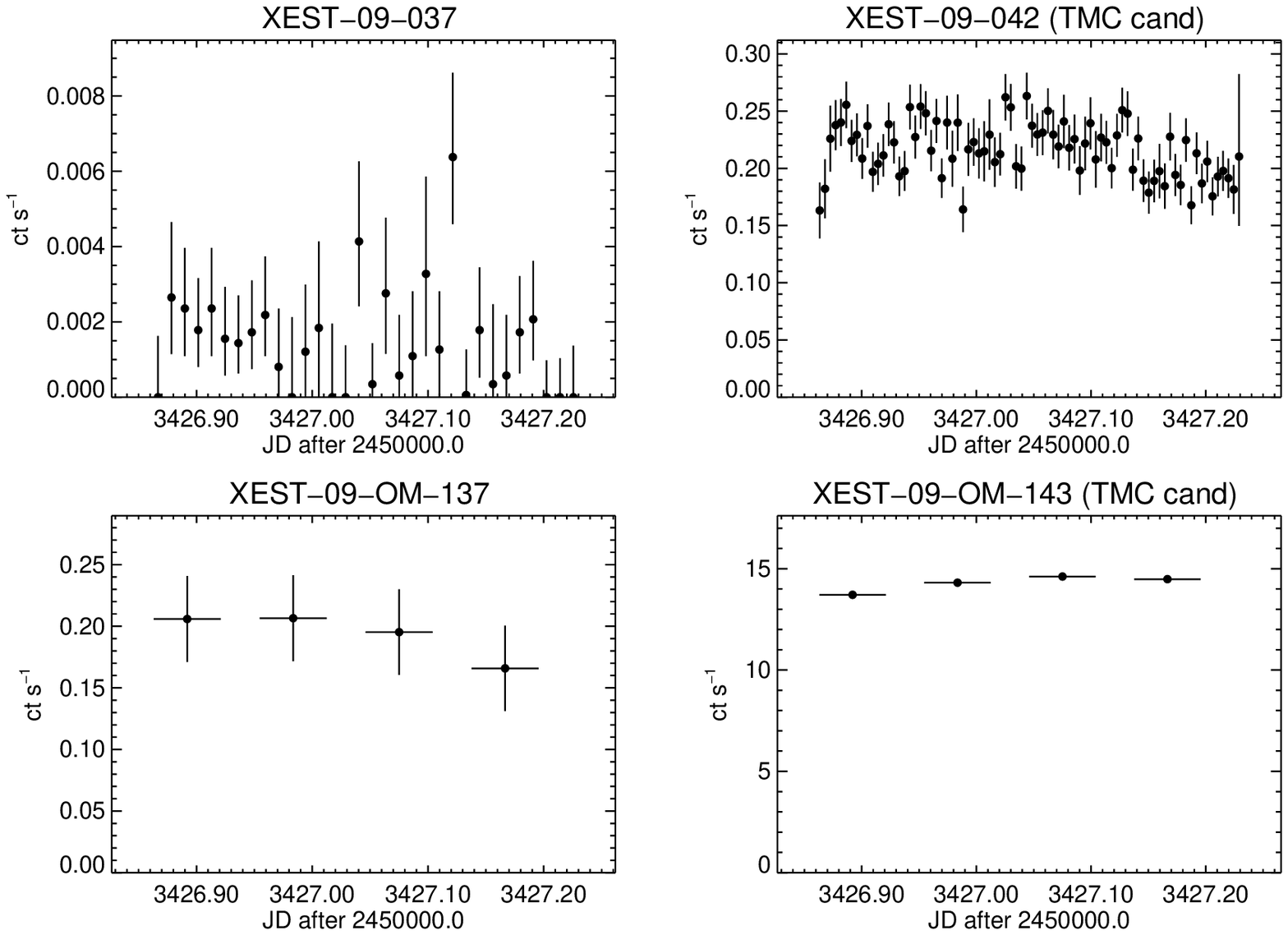}}
\caption{Light curves (continued).}
\end{figure*}

\clearpage\addtocounter{figure}{-1}

\begin{figure*}
\centering
\resizebox{.9\textwidth}{!}{\includegraphics{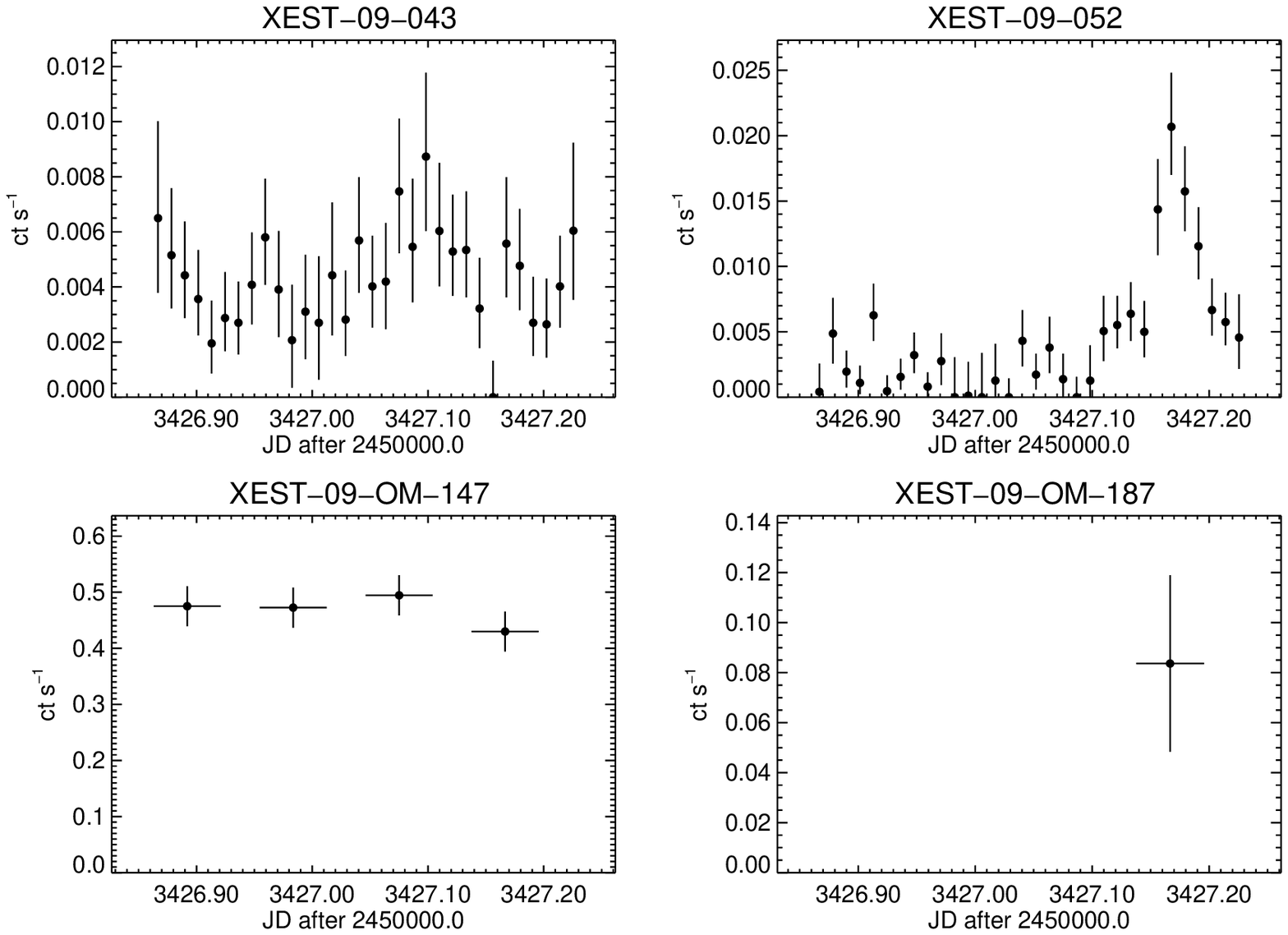}}
\resizebox{.9\textwidth}{!}{\includegraphics{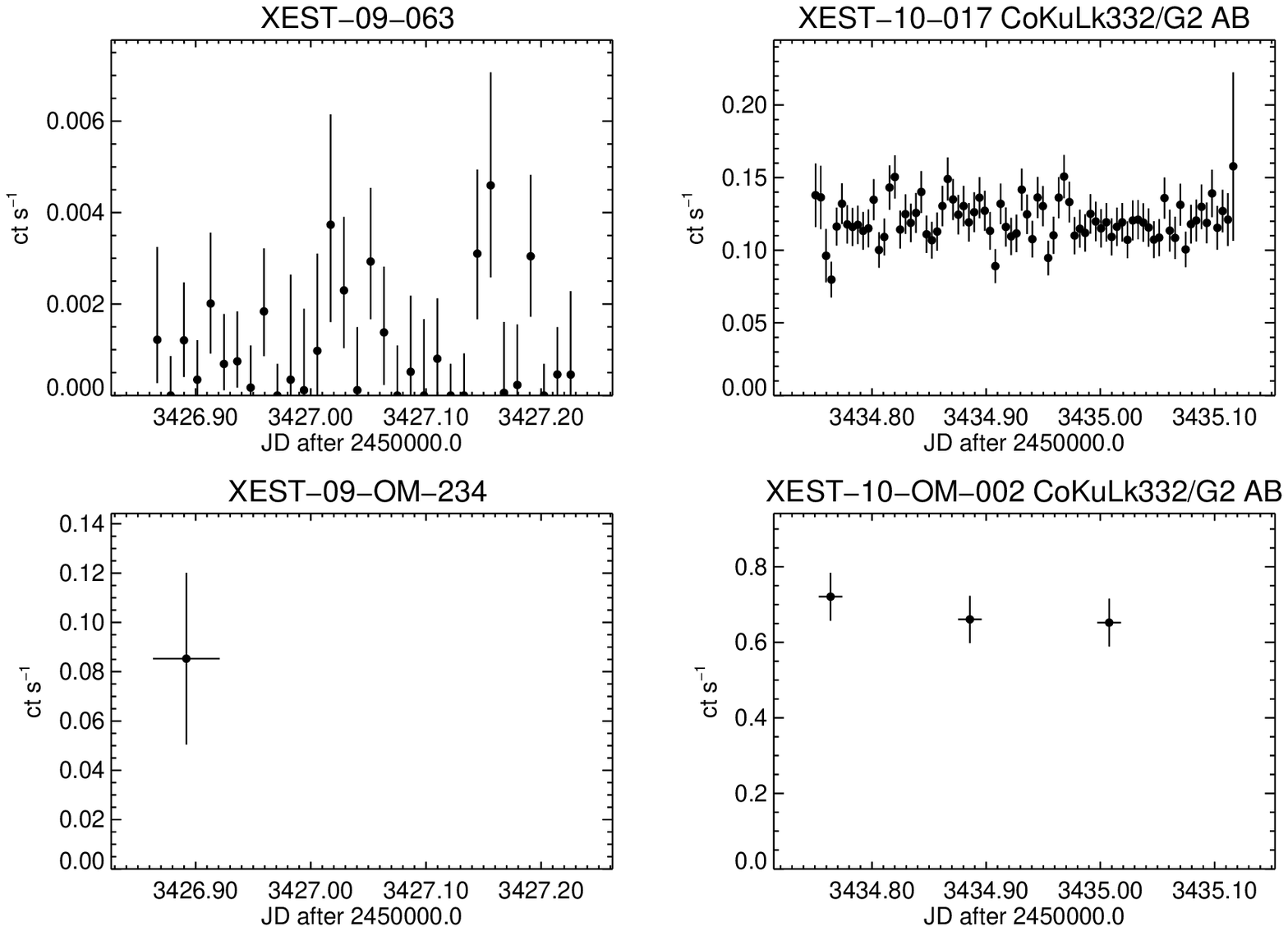}}
\caption{Light curves (continued).}
\end{figure*}

\clearpage\addtocounter{figure}{-1}

\begin{figure*}
\centering
\resizebox{.9\textwidth}{!}{\includegraphics{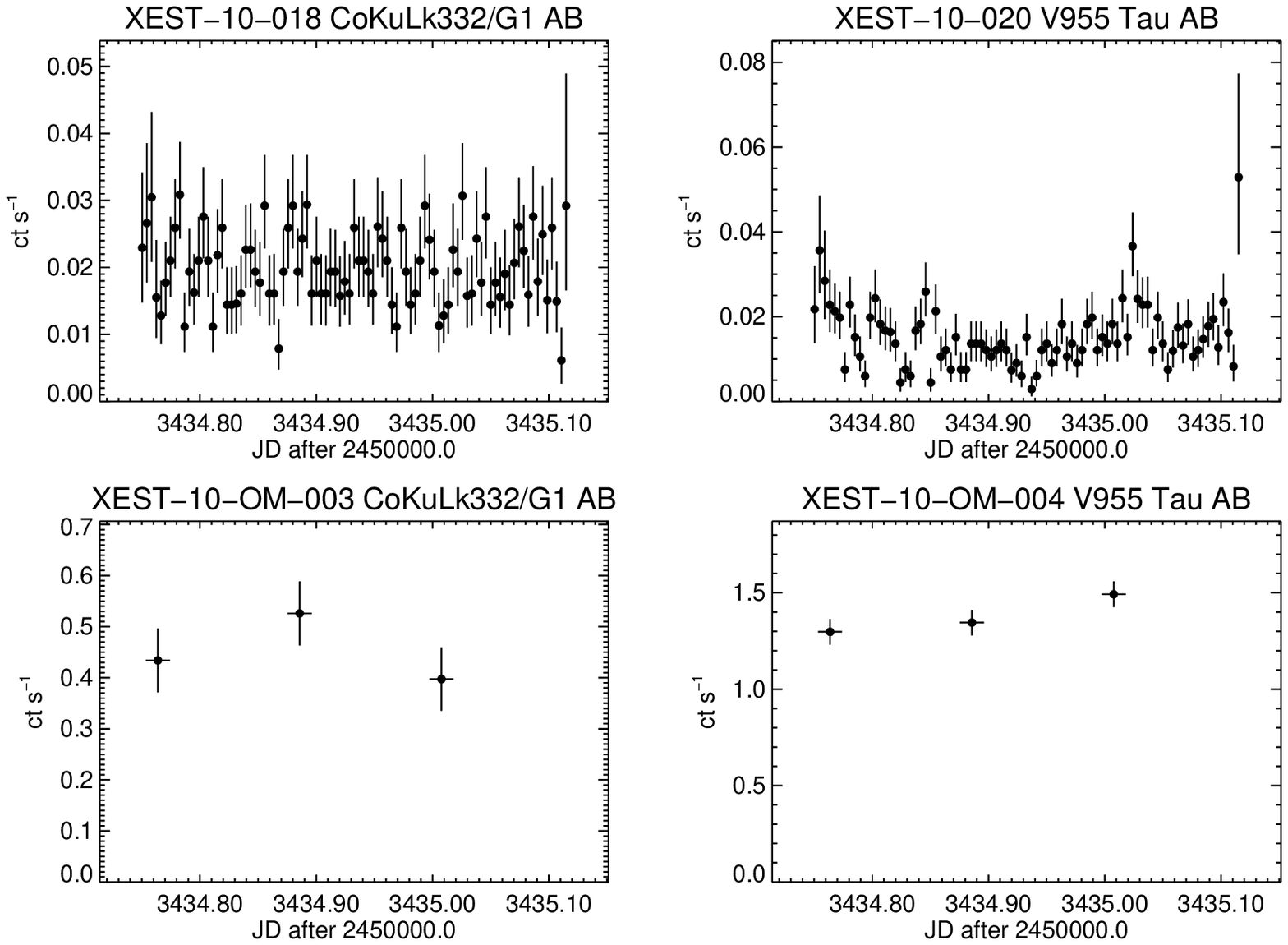}}
\resizebox{.9\textwidth}{!}{\includegraphics{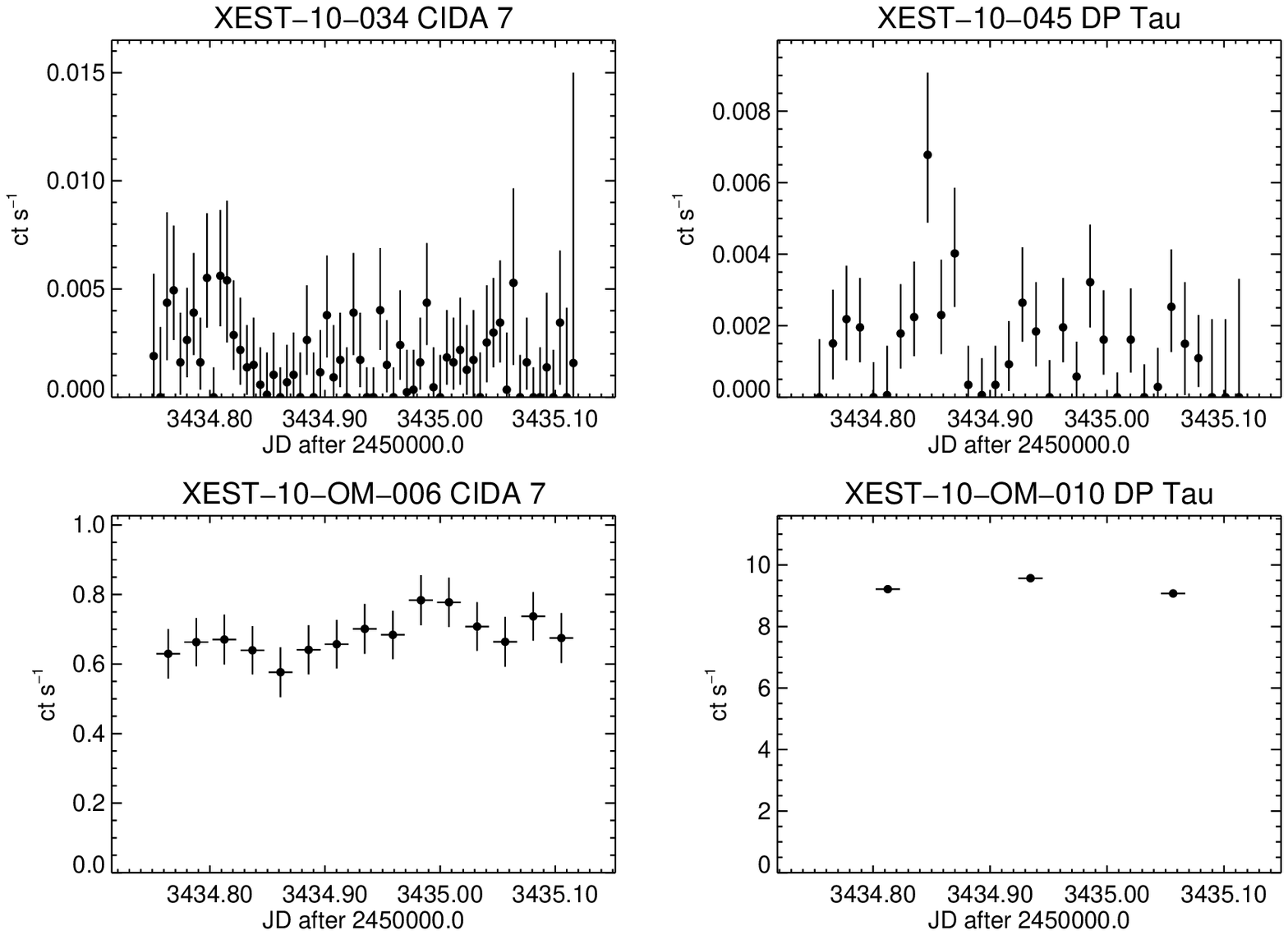}}
\caption{Light curves (continued).}
\end{figure*}

\clearpage\addtocounter{figure}{-1}

\begin{figure*}
\centering
\resizebox{.9\textwidth}{!}{\includegraphics{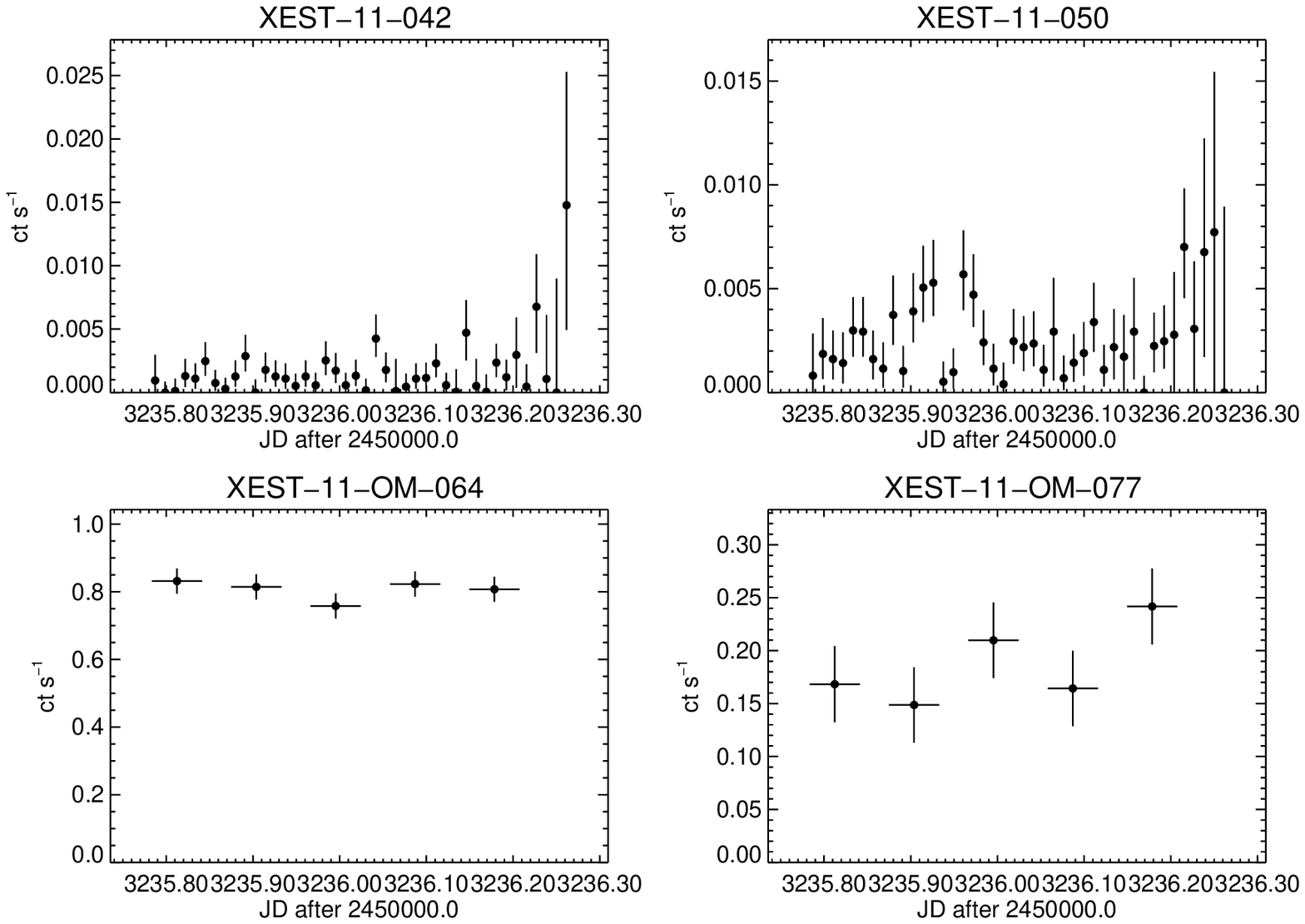}}
\resizebox{.9\textwidth}{!}{\includegraphics{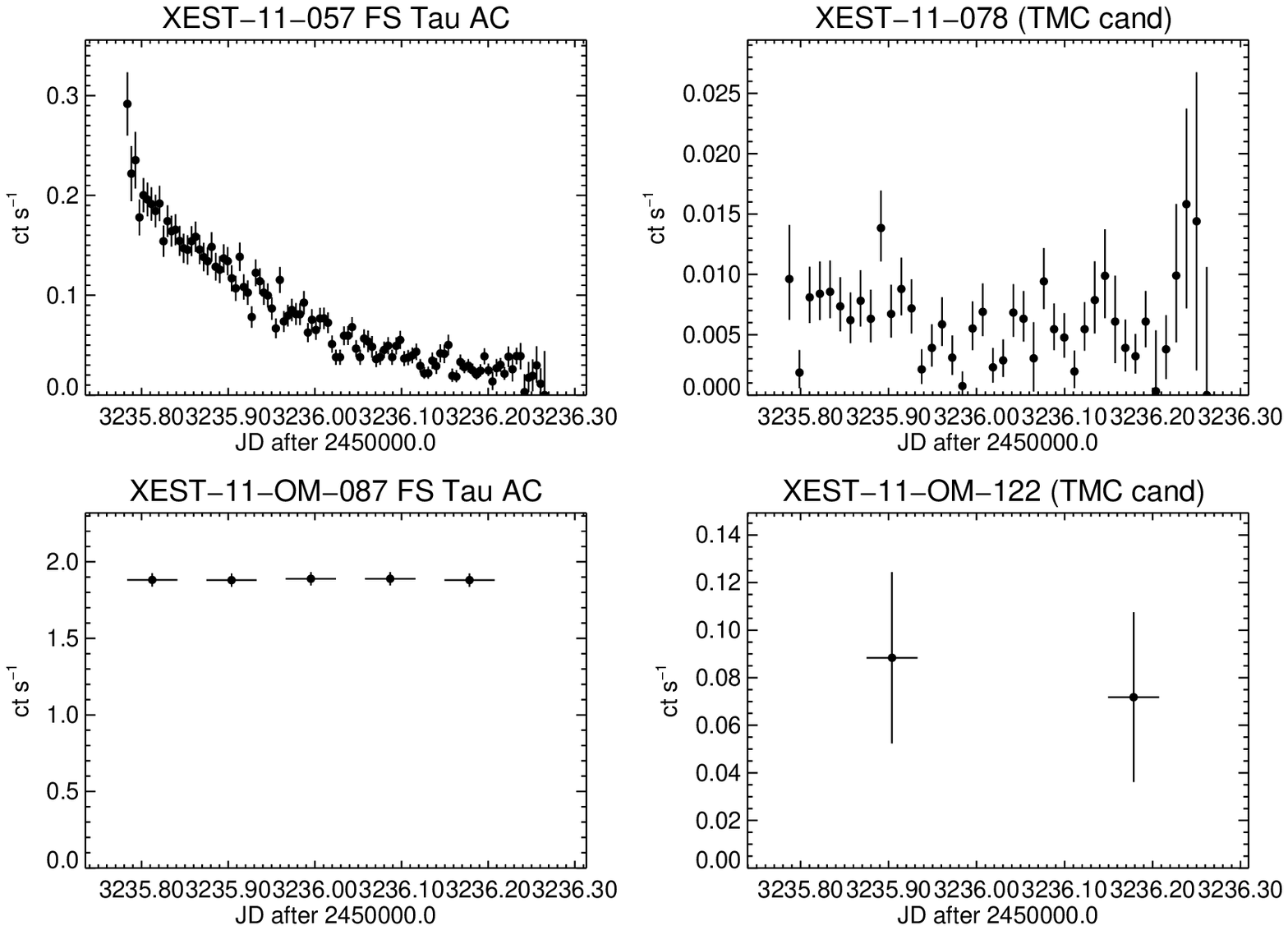}}
\caption{Light curves (continued).}
\end{figure*}

\clearpage\addtocounter{figure}{-1}

\begin{figure*}
\centering
\resizebox{.9\textwidth}{!}{\includegraphics{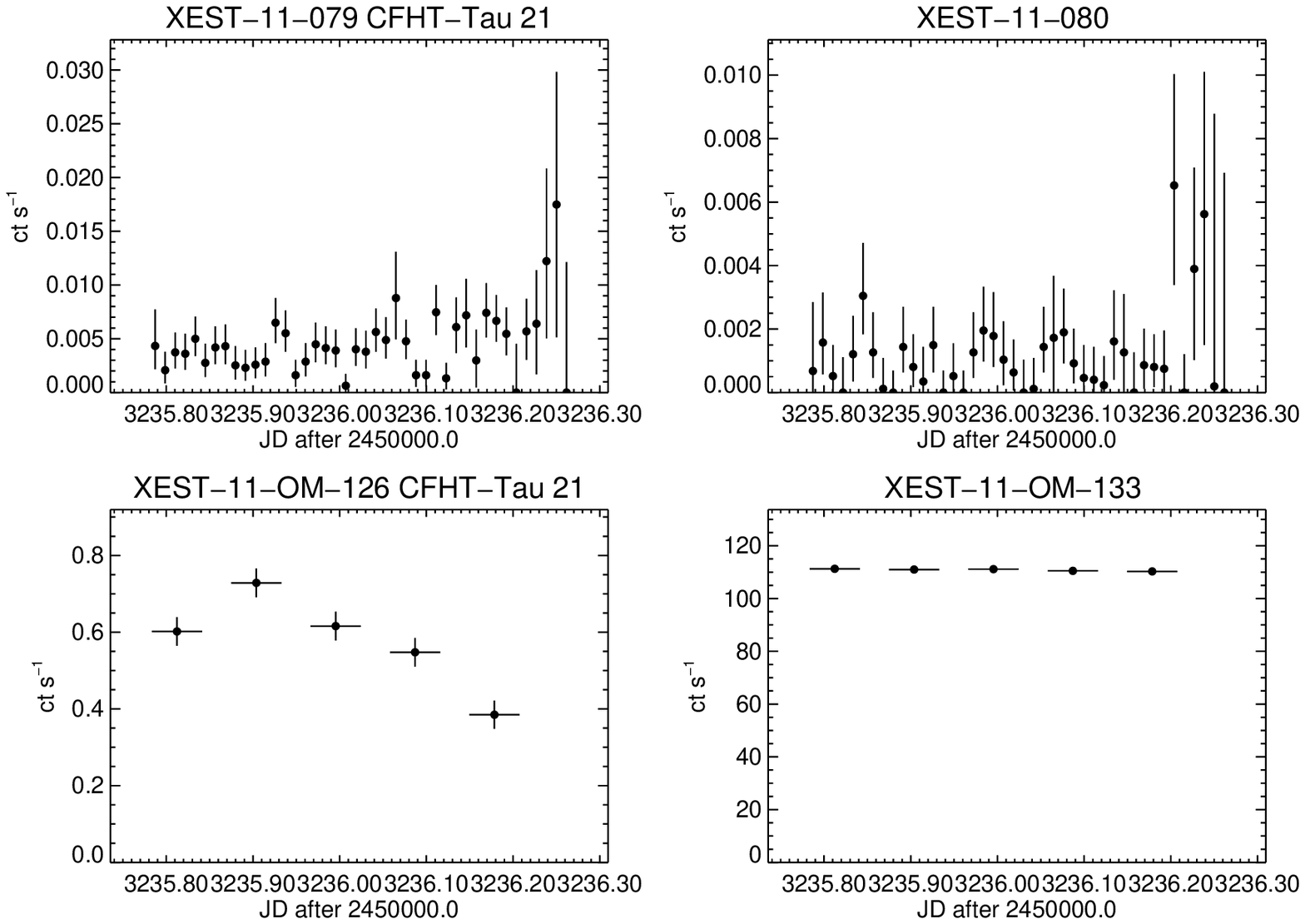}}
\resizebox{.9\textwidth}{!}{\includegraphics{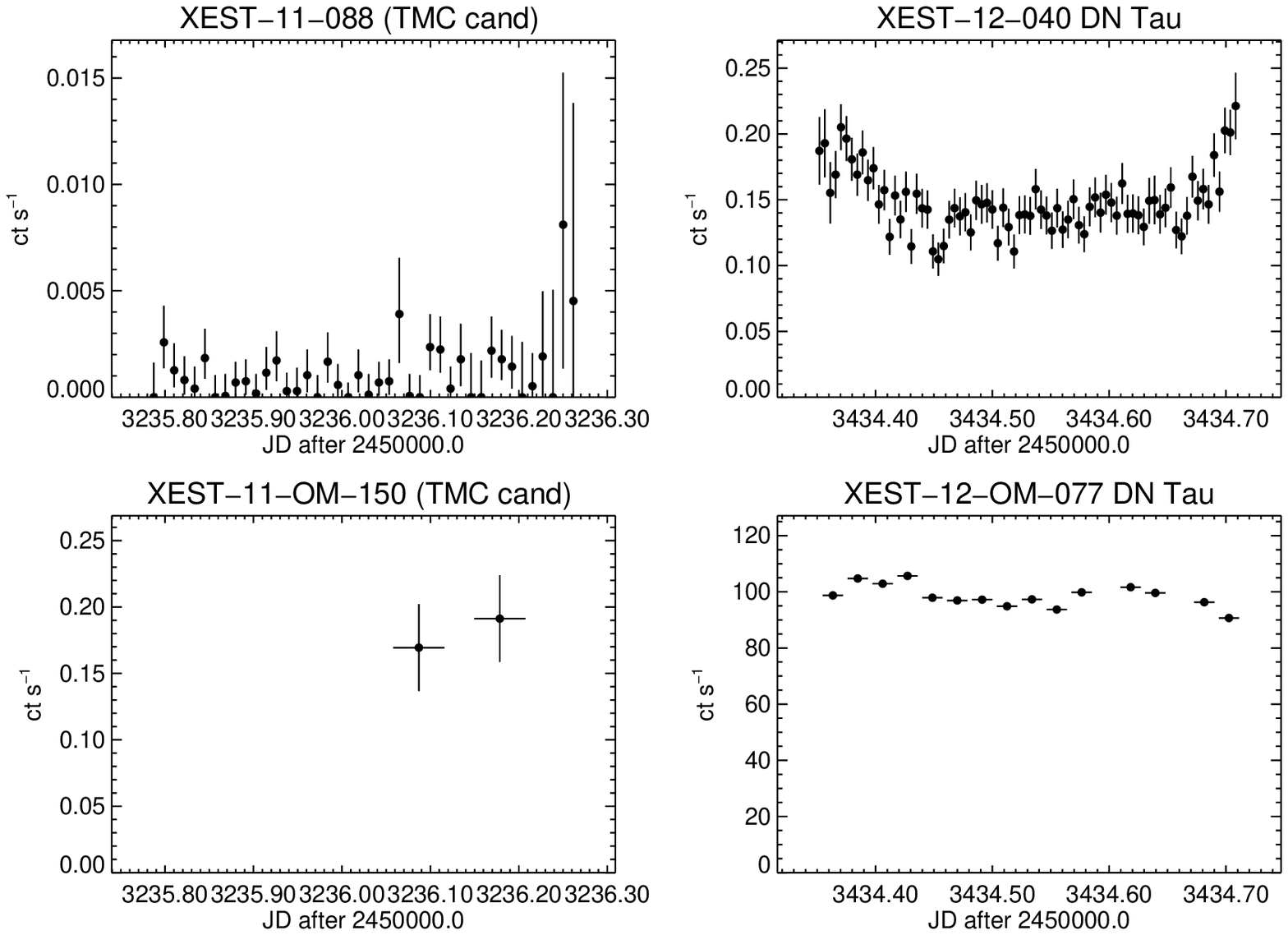}}
\caption{Light curves (continued).}
\end{figure*}

\clearpage\addtocounter{figure}{-1}

\begin{figure*}
\centering
\resizebox{.9\textwidth}{!}{\includegraphics{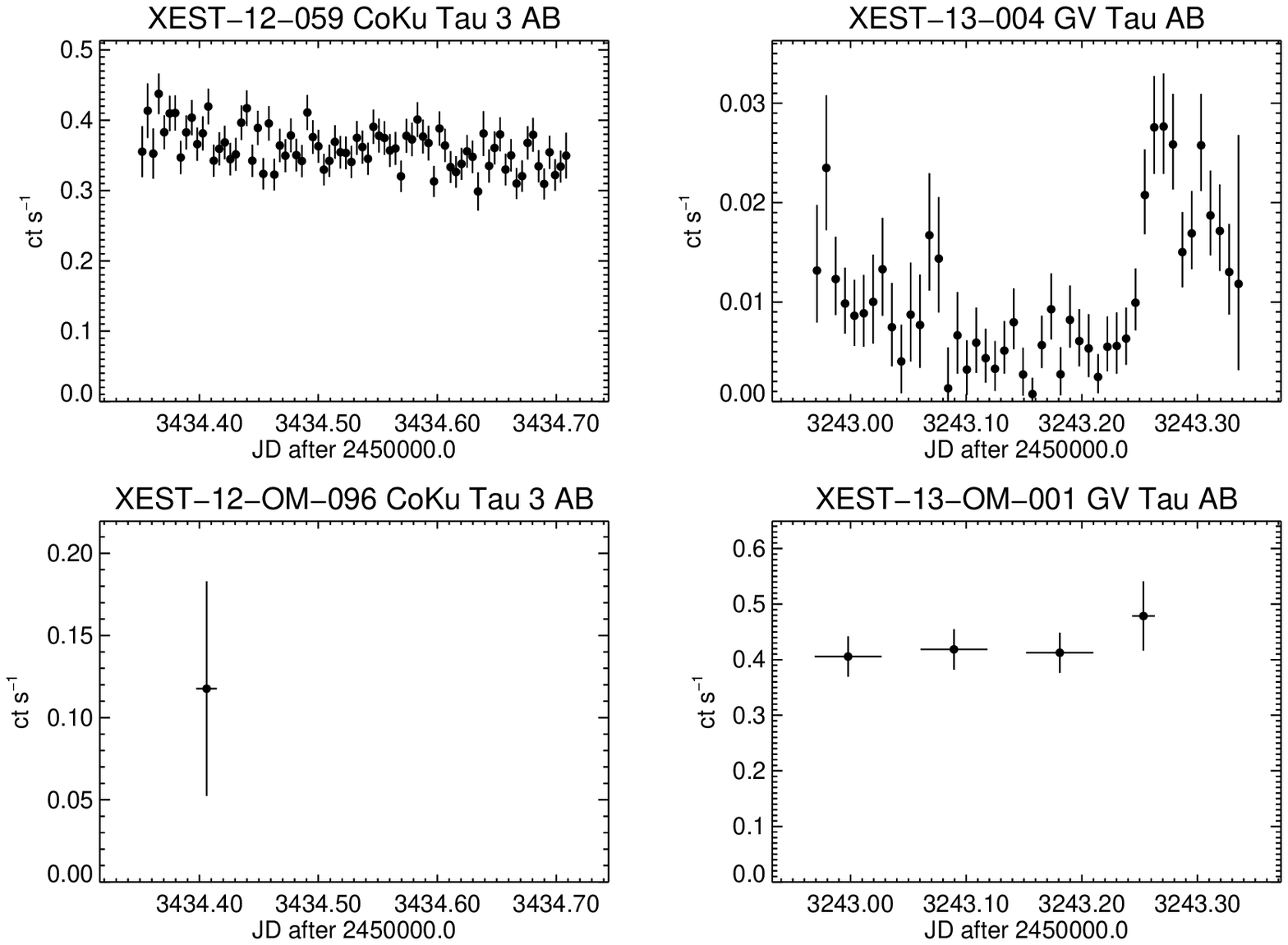}}
\resizebox{.9\textwidth}{!}{\includegraphics{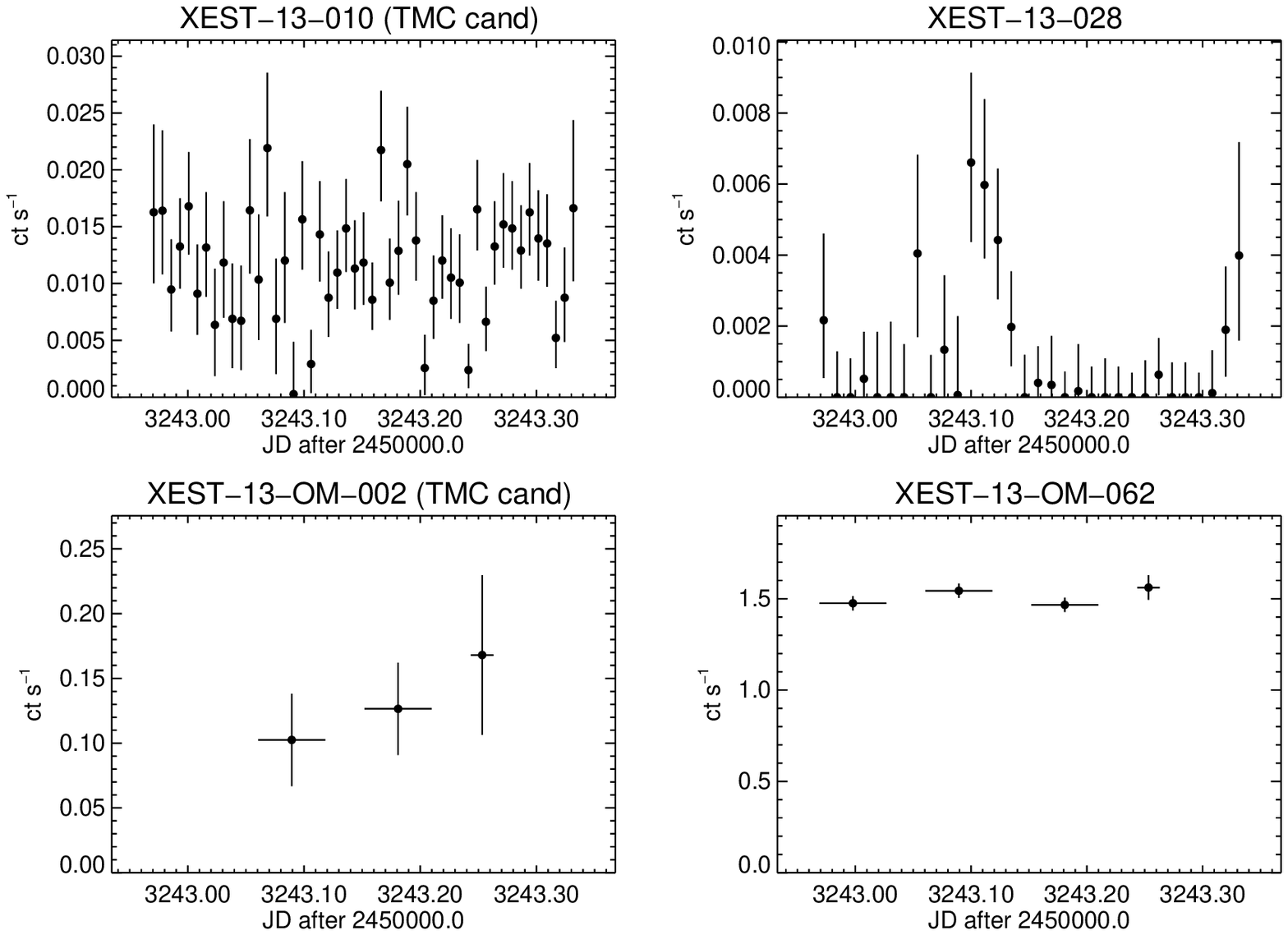}}
\caption{Light curves (continued).}
\end{figure*}

\clearpage\addtocounter{figure}{-1}

\begin{figure*}
\centering
\resizebox{.9\textwidth}{!}{\includegraphics{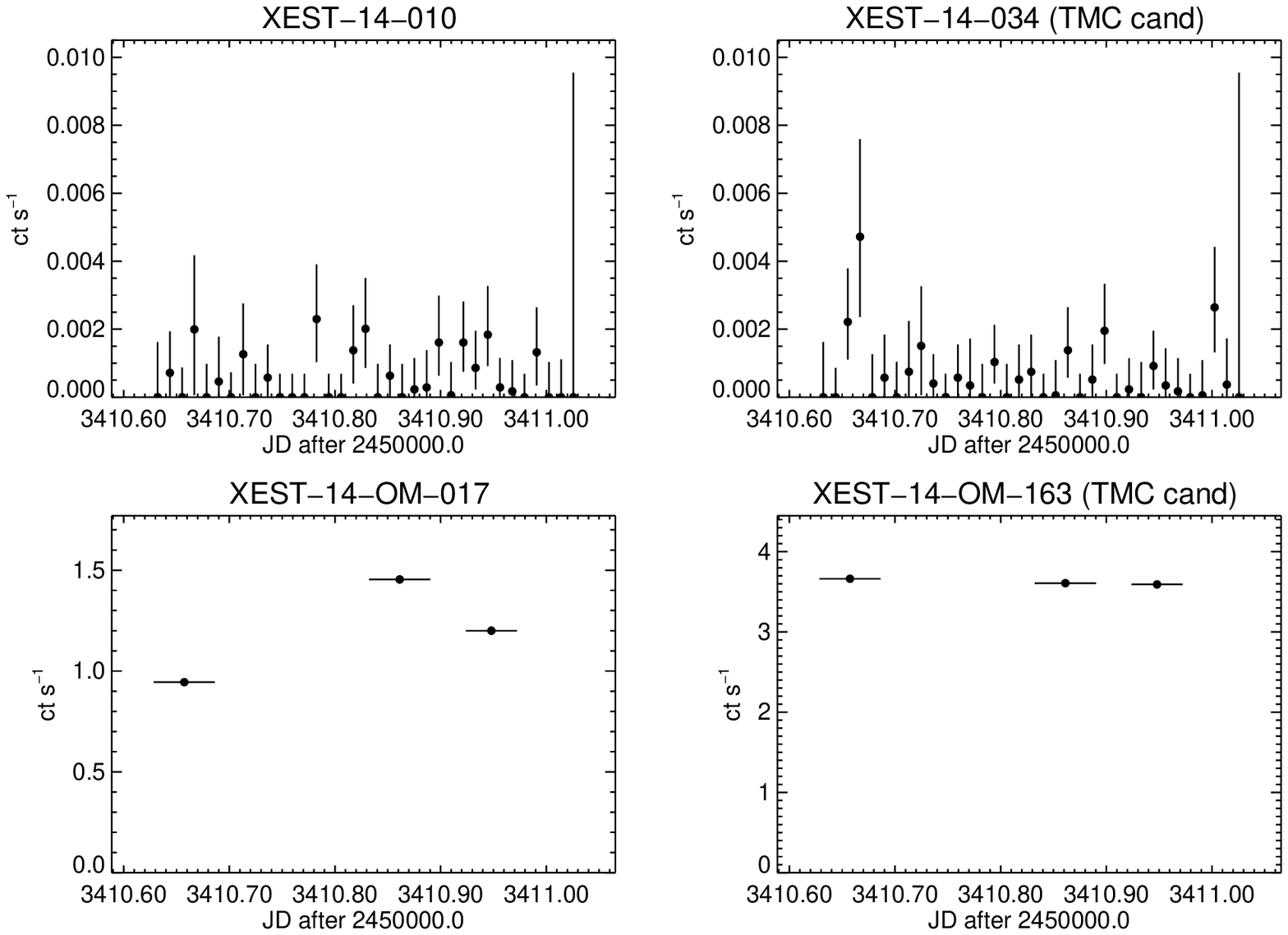}}
\resizebox{.9\textwidth}{!}{\includegraphics{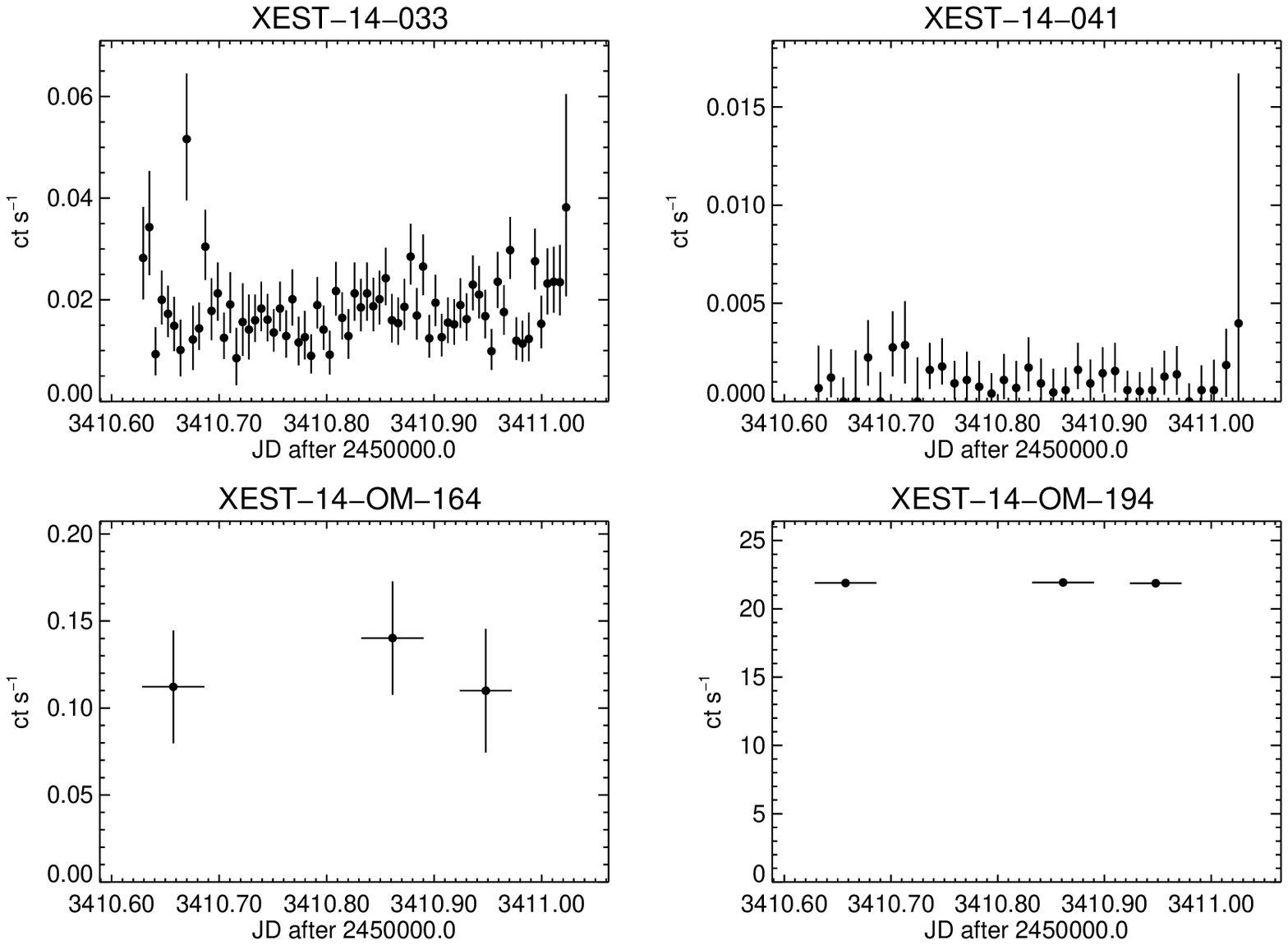}}
\caption{Light curves (continued).}
\end{figure*}

\clearpage\addtocounter{figure}{-1}

\begin{figure*}
\centering
\resizebox{.9\textwidth}{!}{\includegraphics{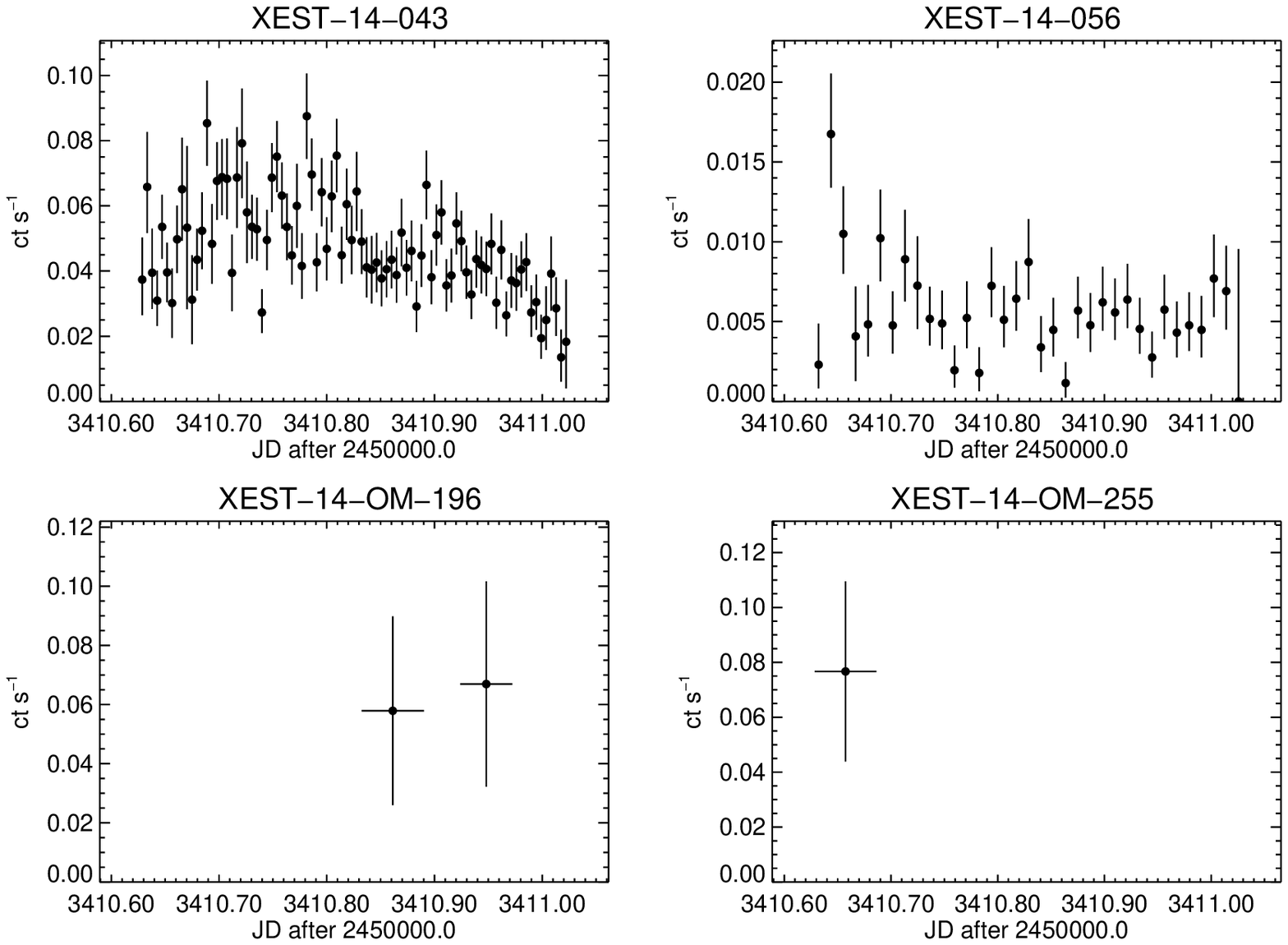}}
\resizebox{.9\textwidth}{!}{\includegraphics{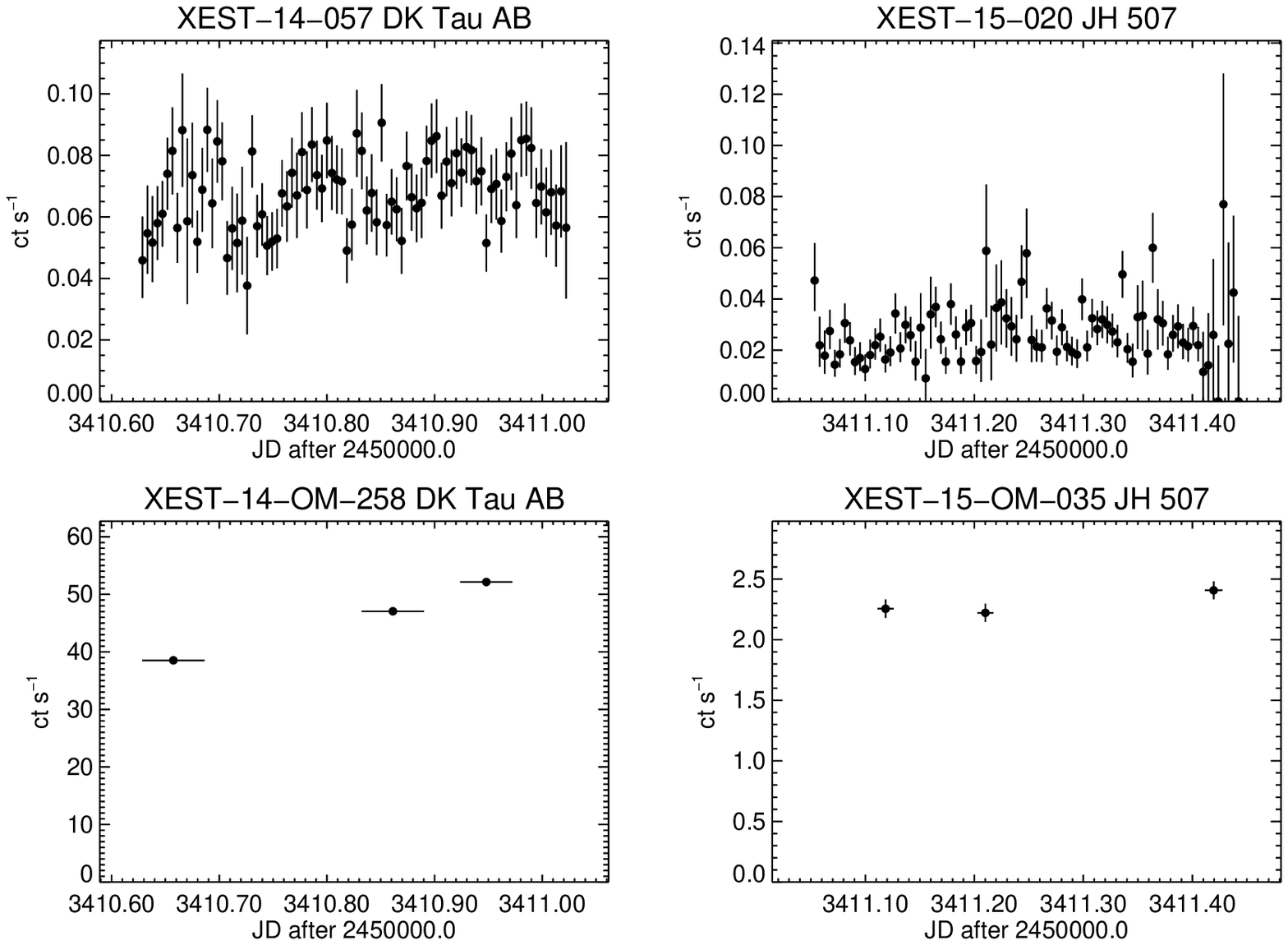}}
\caption{Light curves (continued).}
\end{figure*}

\clearpage\addtocounter{figure}{-1}

\begin{figure*}
\centering
\resizebox{.9\textwidth}{!}{\includegraphics{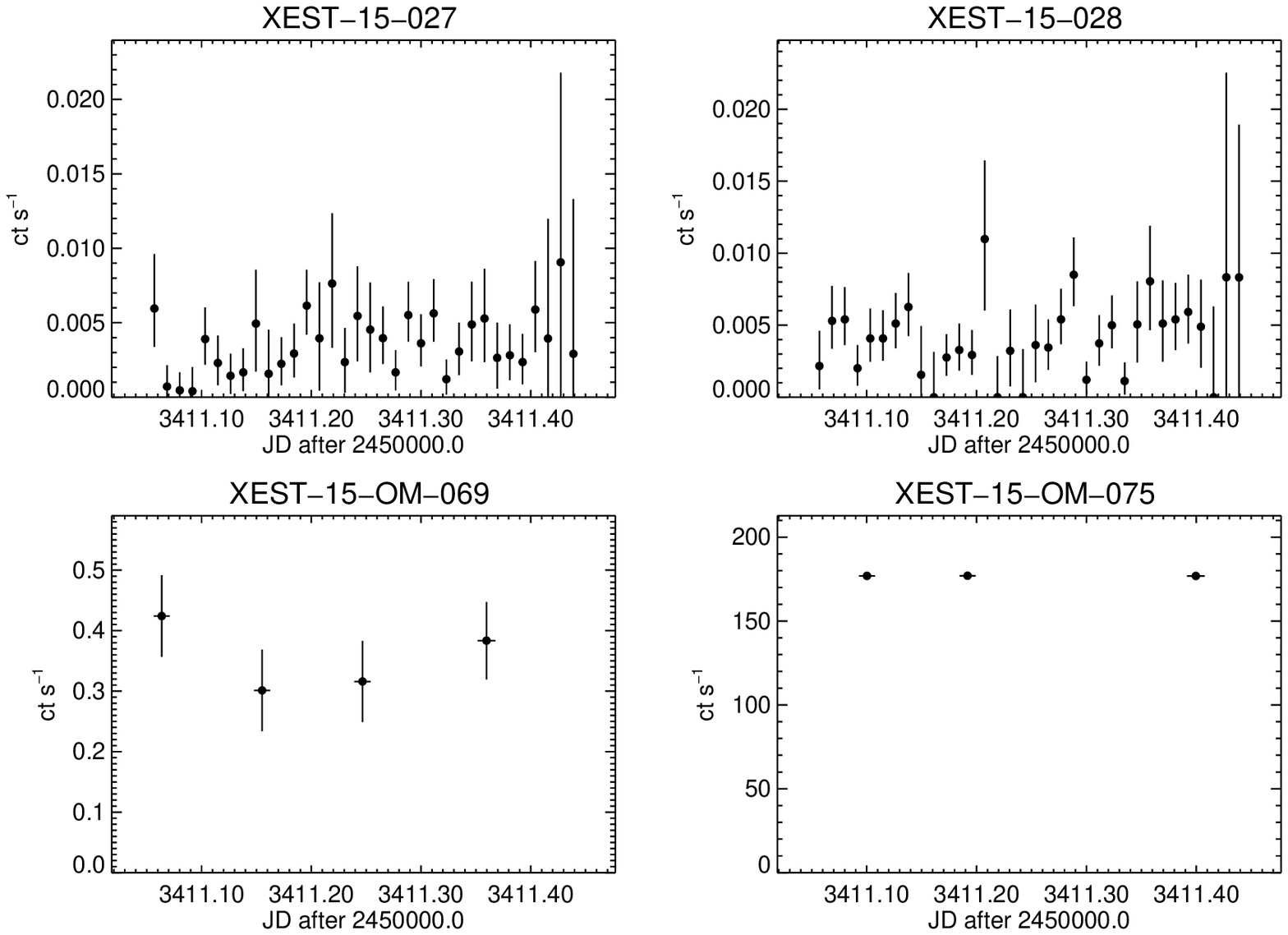}}
\resizebox{.9\textwidth}{!}{\includegraphics{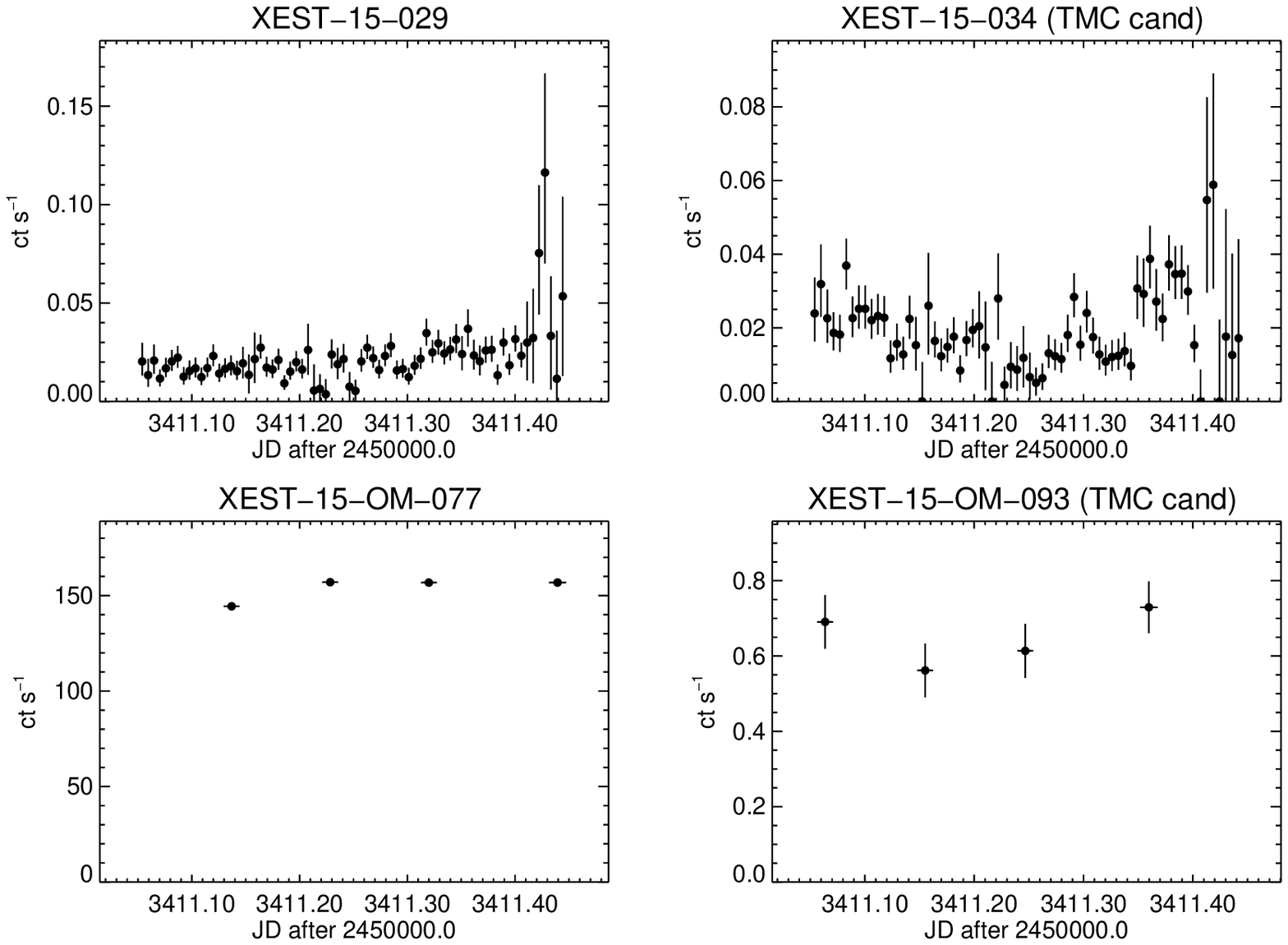}}
\caption{Light curves (continued).}
\end{figure*}

\clearpage\addtocounter{figure}{-1}

\begin{figure*}
\centering
\resizebox{.9\textwidth}{!}{\includegraphics{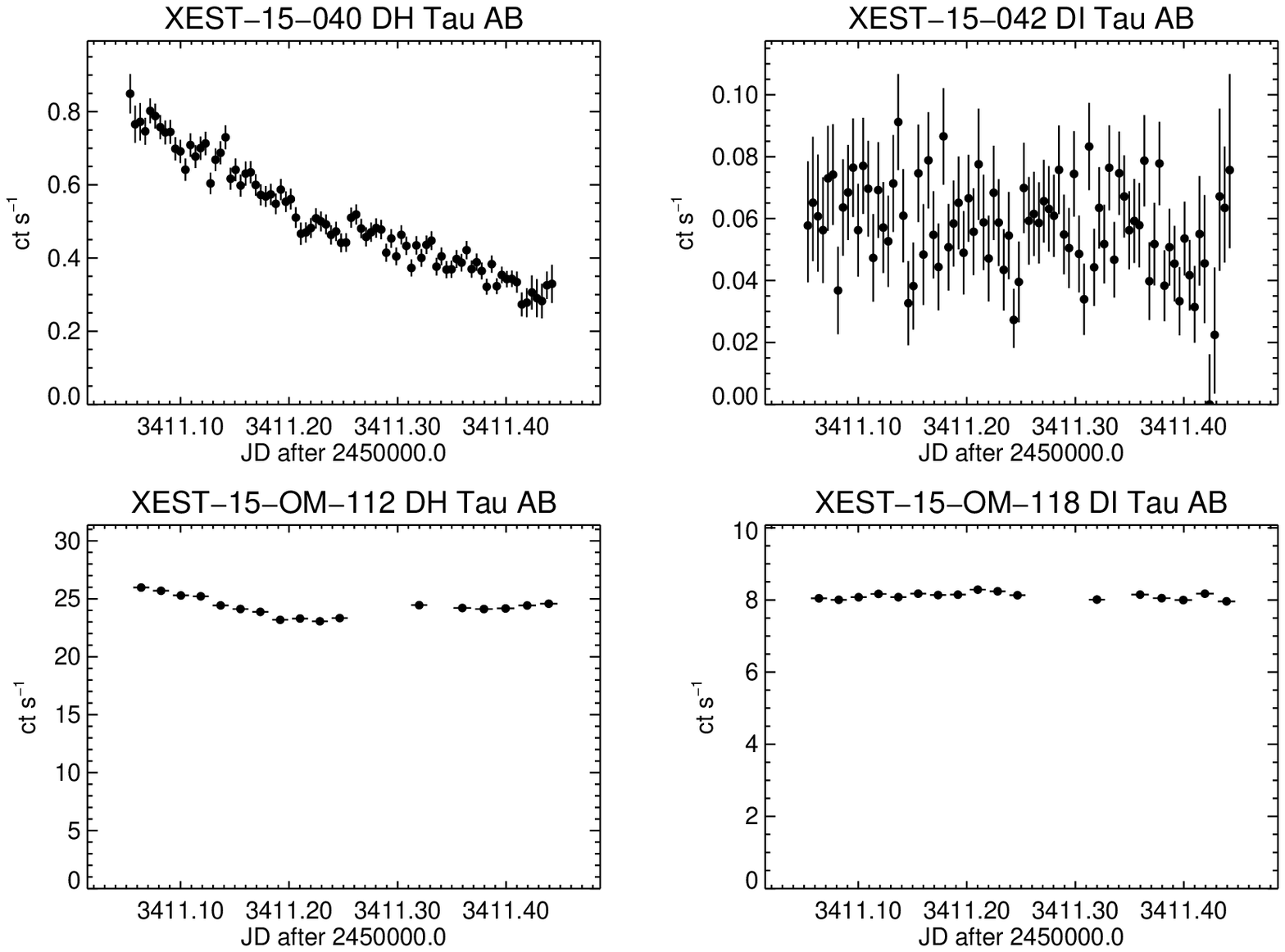}}
\resizebox{.9\textwidth}{!}{\includegraphics{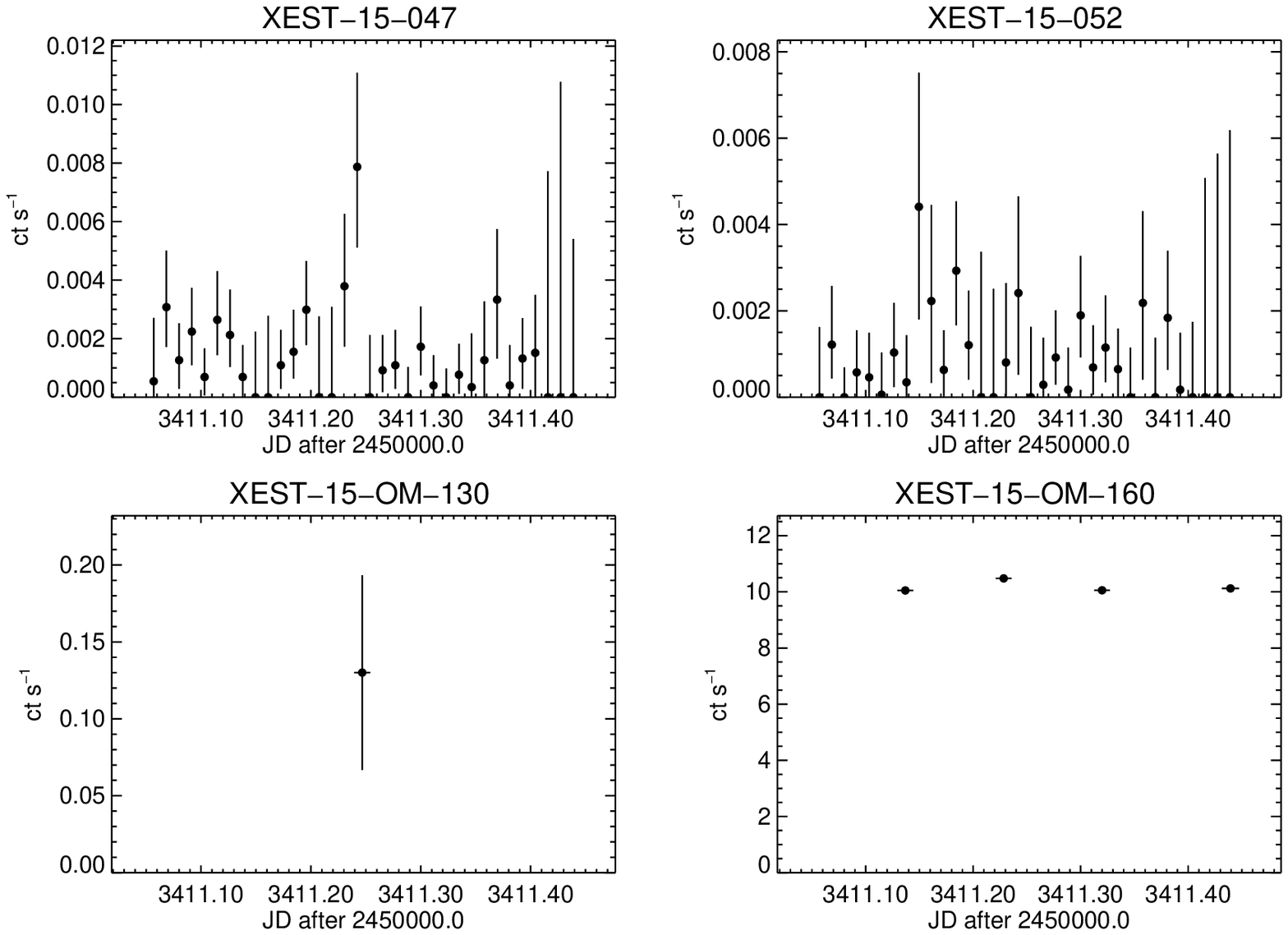}}
\caption{Light curves (continued).}
\end{figure*}

\clearpage\addtocounter{figure}{-1}

\begin{figure*}
\centering
\resizebox{.9\textwidth}{!}{\includegraphics{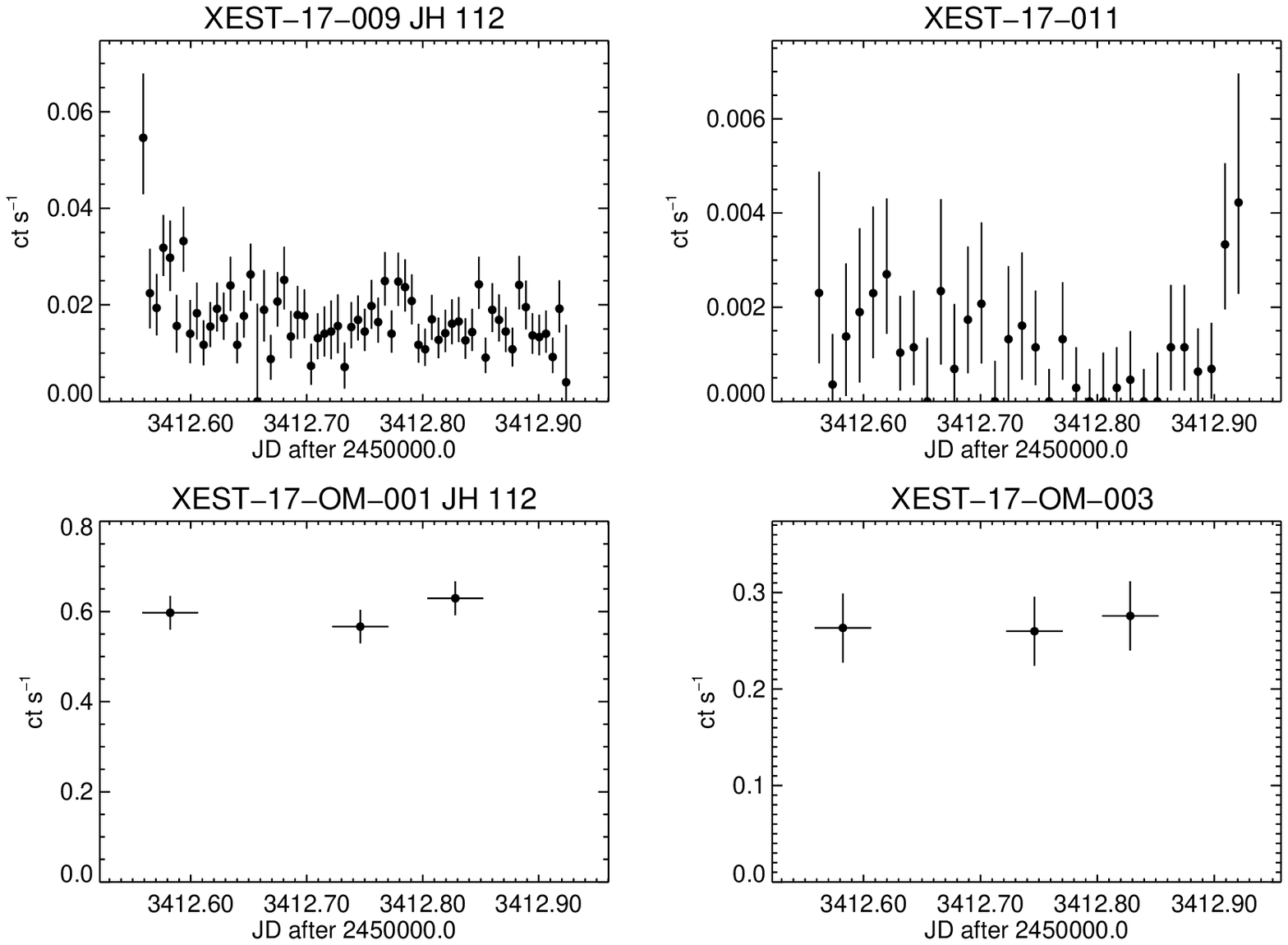}}
\resizebox{.9\textwidth}{!}{\includegraphics{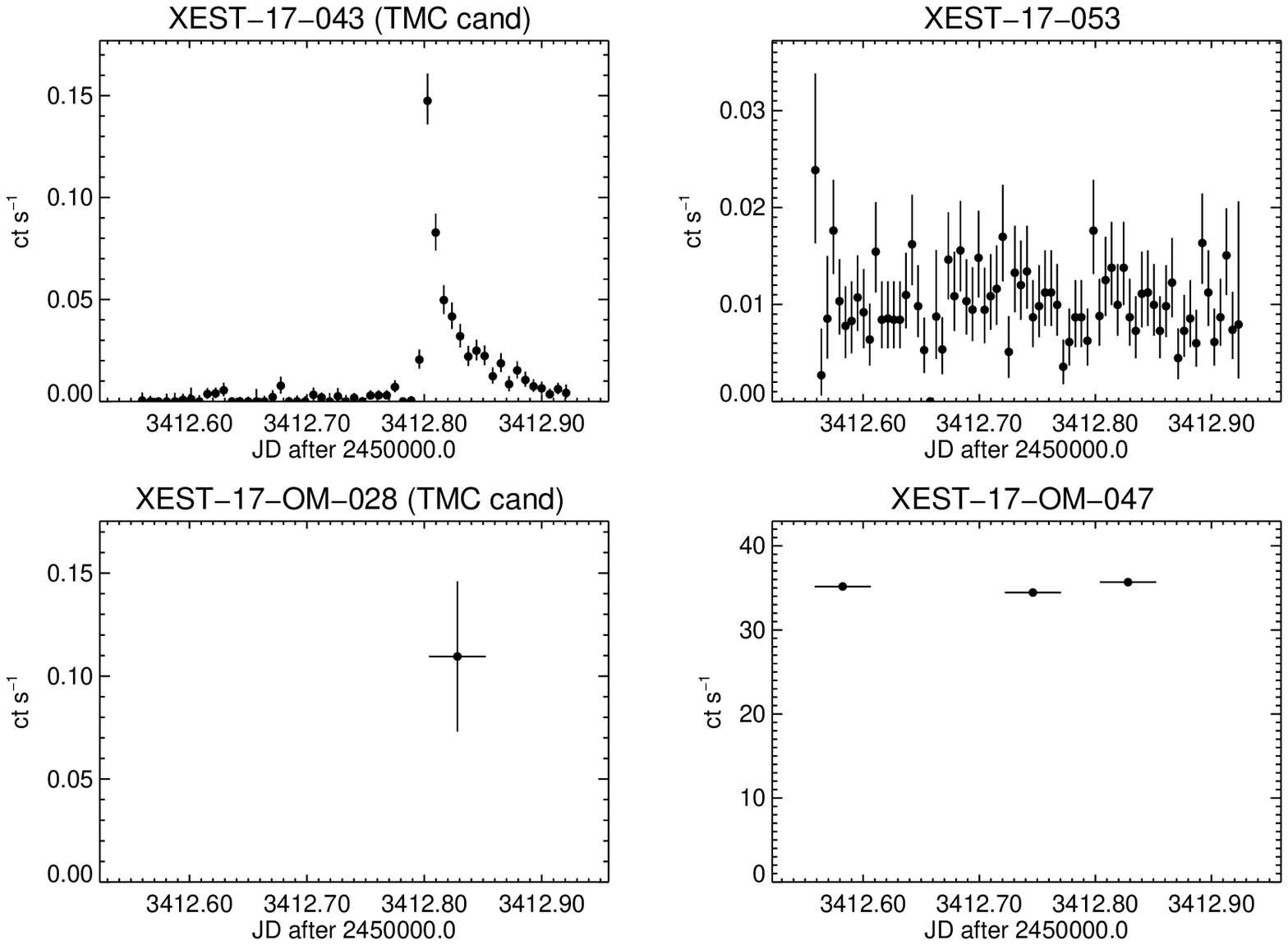}}
\caption{Light curves (continued).}
\end{figure*}

\clearpage\addtocounter{figure}{-1}

\begin{figure*}
\centering
\resizebox{.9\textwidth}{!}{\includegraphics{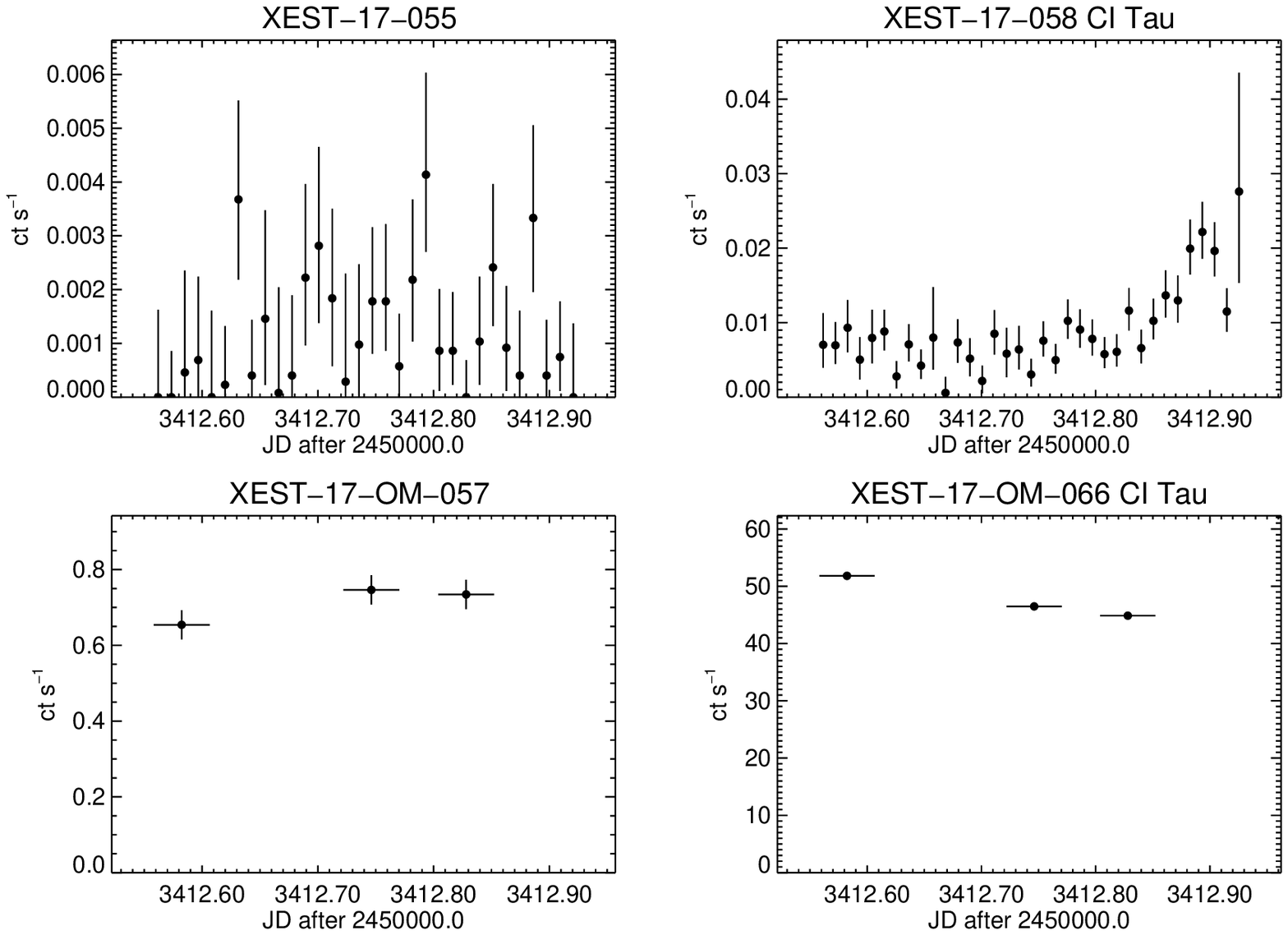}}
\resizebox{.9\textwidth}{!}{\includegraphics{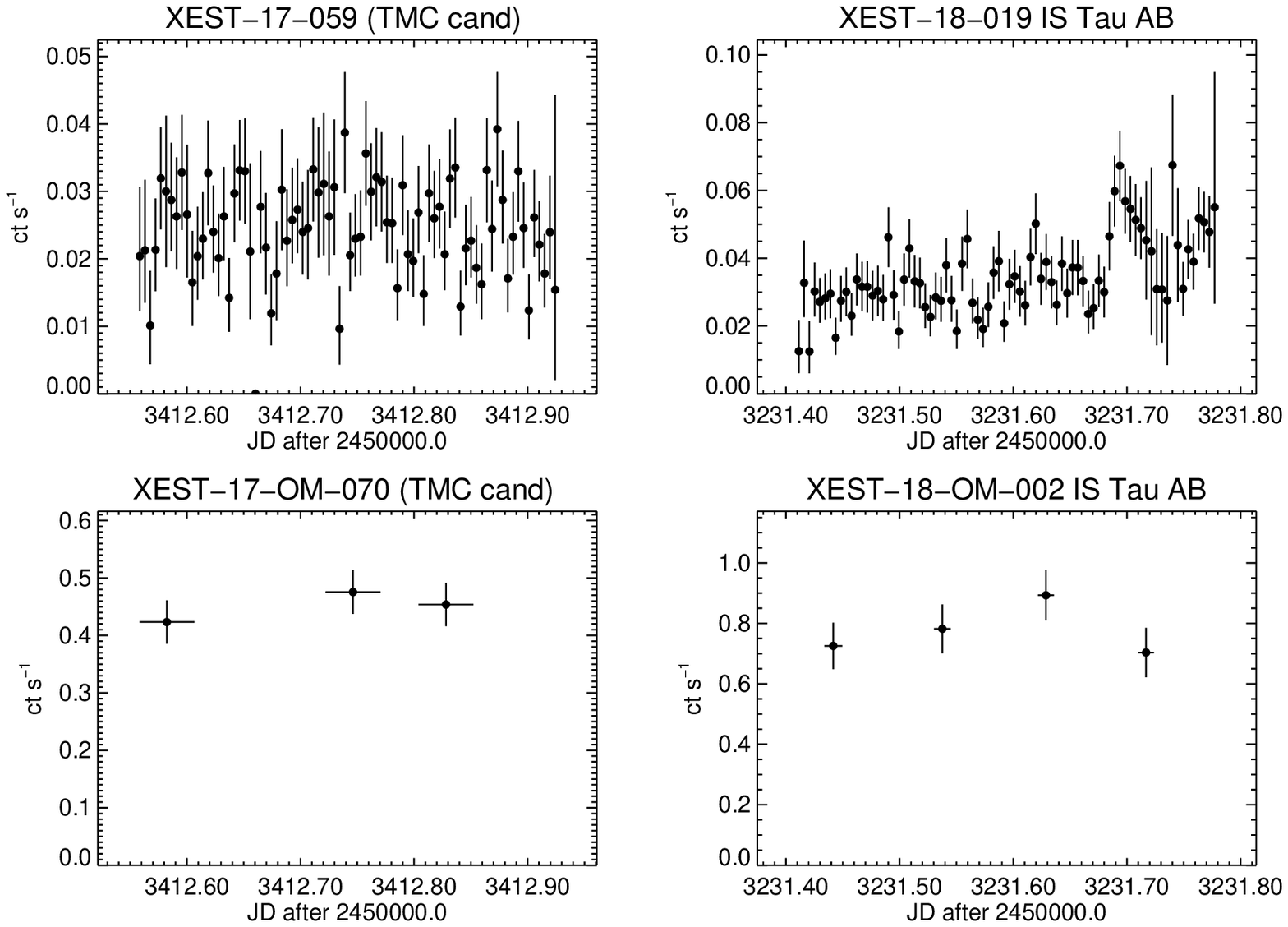}}
\caption{Light curves (continued).}
\end{figure*}

\clearpage\addtocounter{figure}{-1}

\begin{figure*}
\centering
\resizebox{.9\textwidth}{!}{\includegraphics{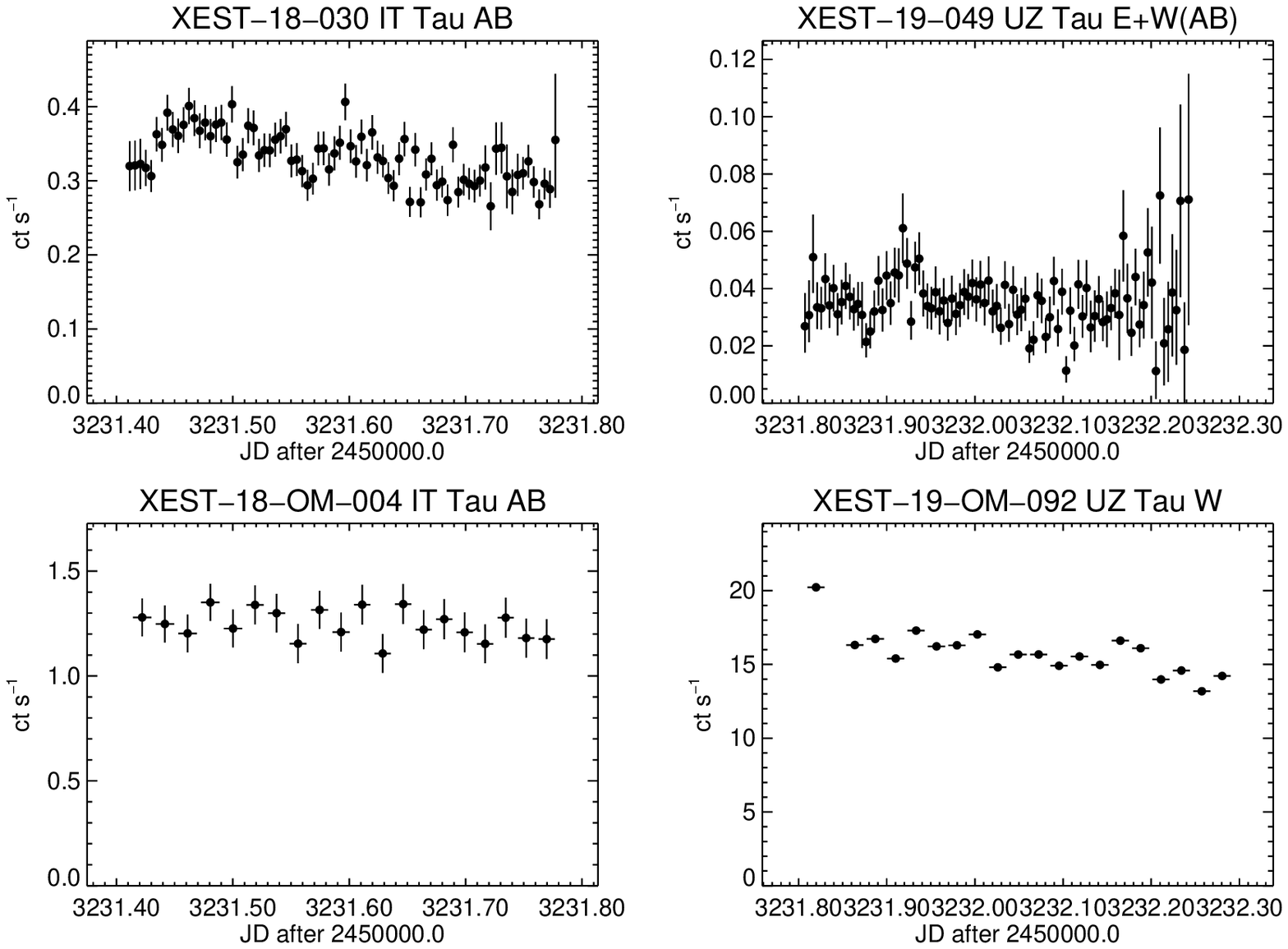}}
\resizebox{.9\textwidth}{!}{\includegraphics{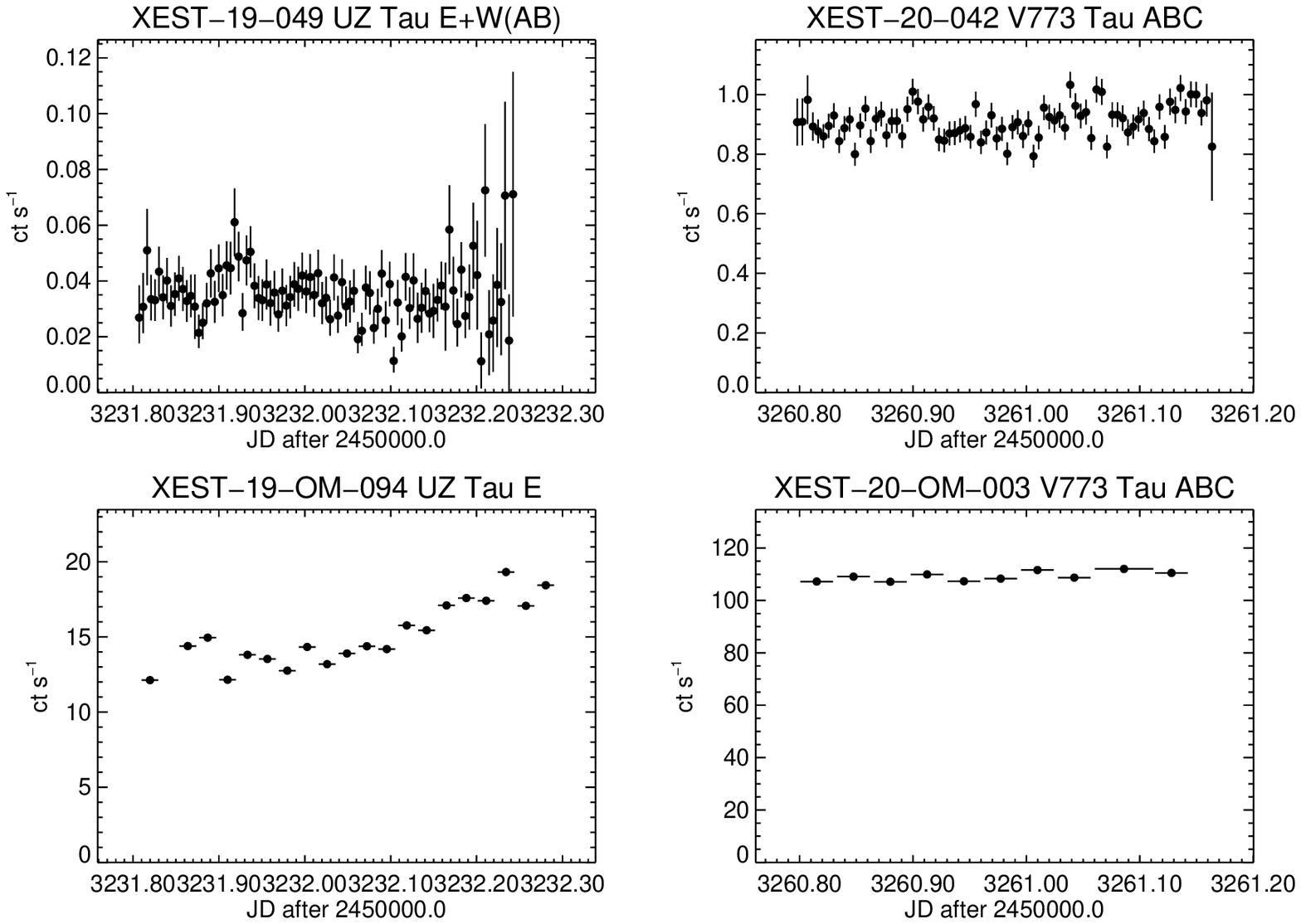}}
\caption{Light curves (continued).}
\end{figure*}

\clearpage\addtocounter{figure}{-1}

\begin{figure*}
\centering
\resizebox{.9\textwidth}{!}{\includegraphics{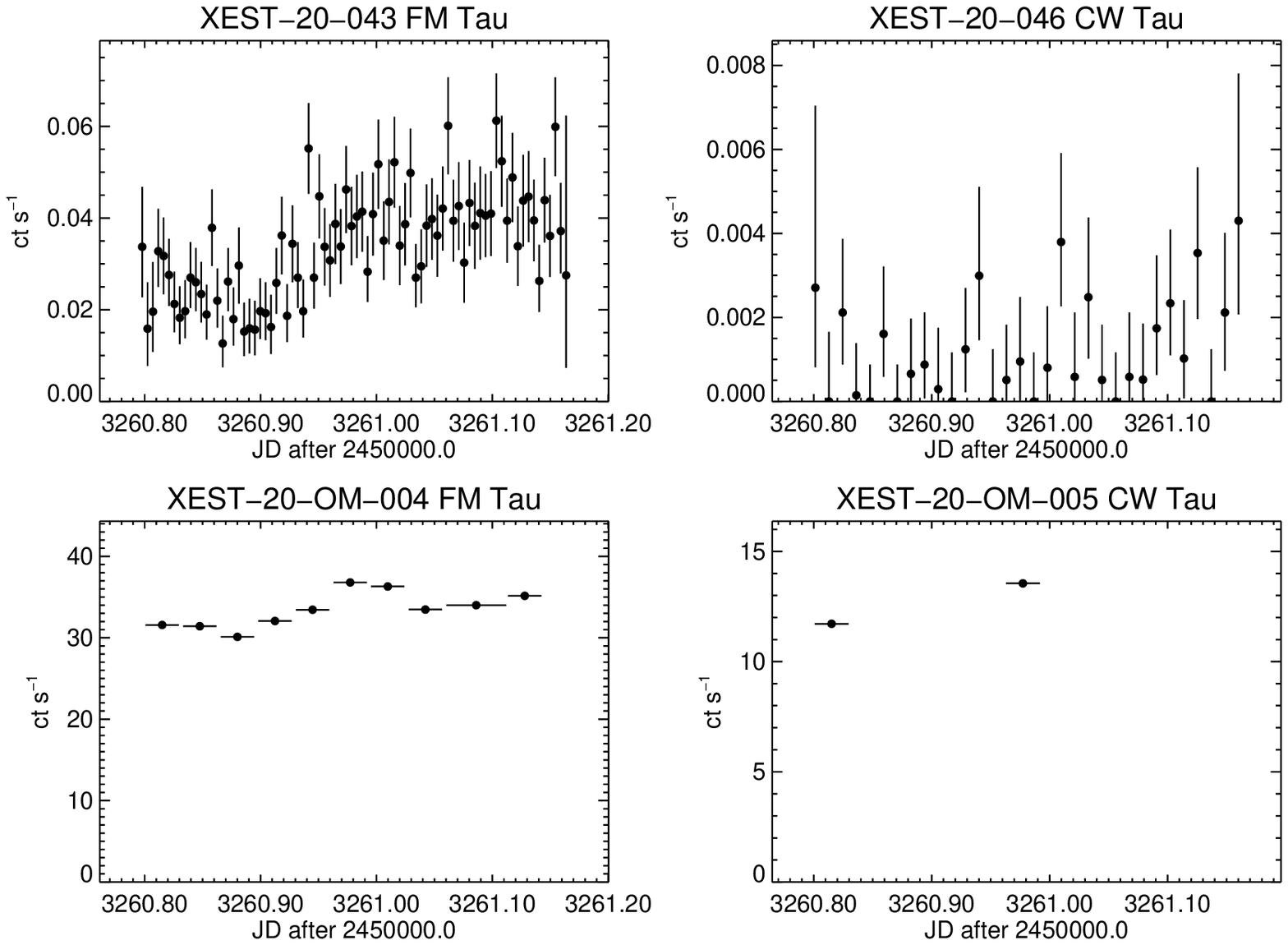}}
\resizebox{.9\textwidth}{!}{\includegraphics{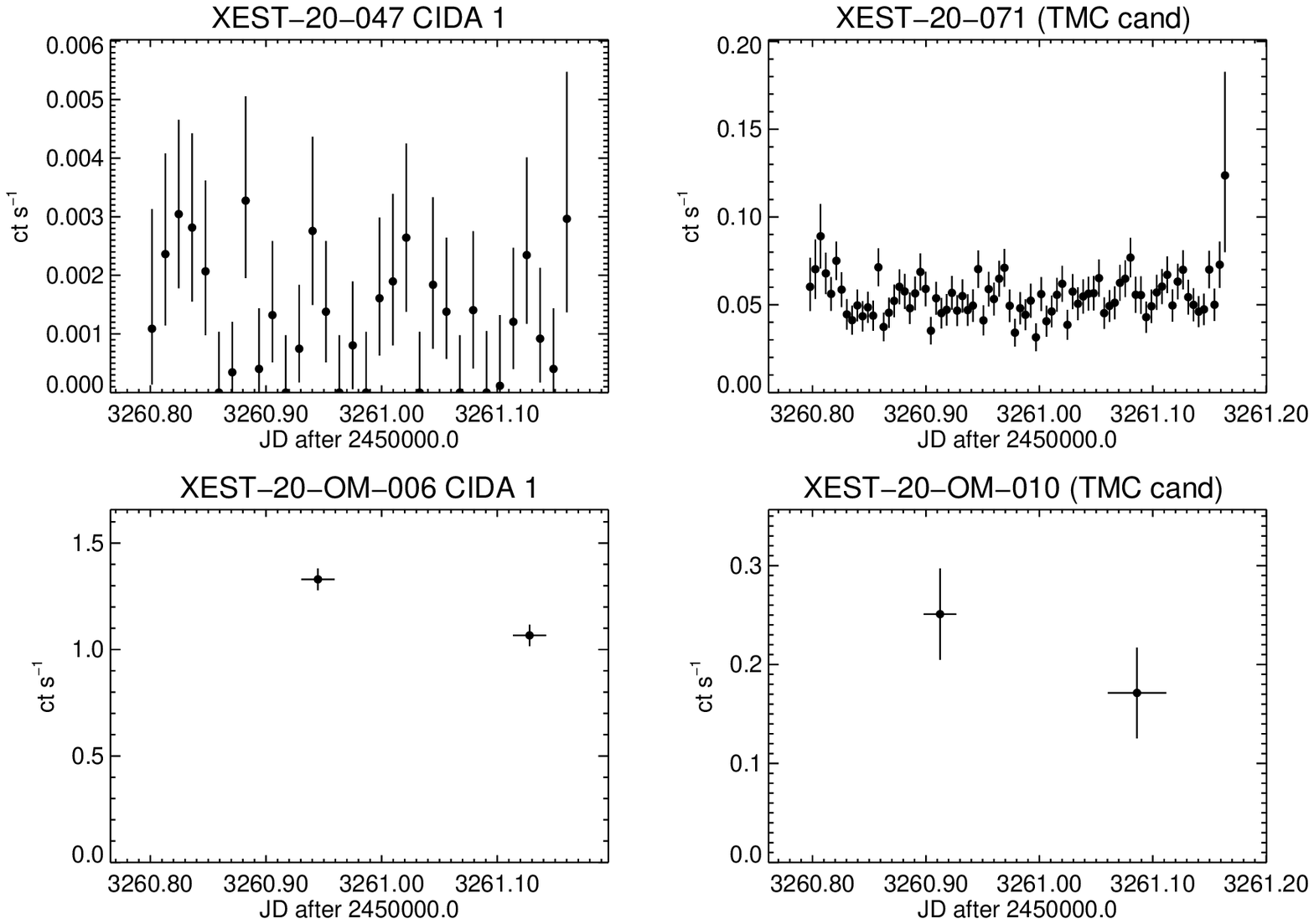}}
\caption{Light curves (continued).}
\end{figure*}

\clearpage\addtocounter{figure}{-1}

\begin{figure*}
\centering
\resizebox{.9\textwidth}{!}{\includegraphics{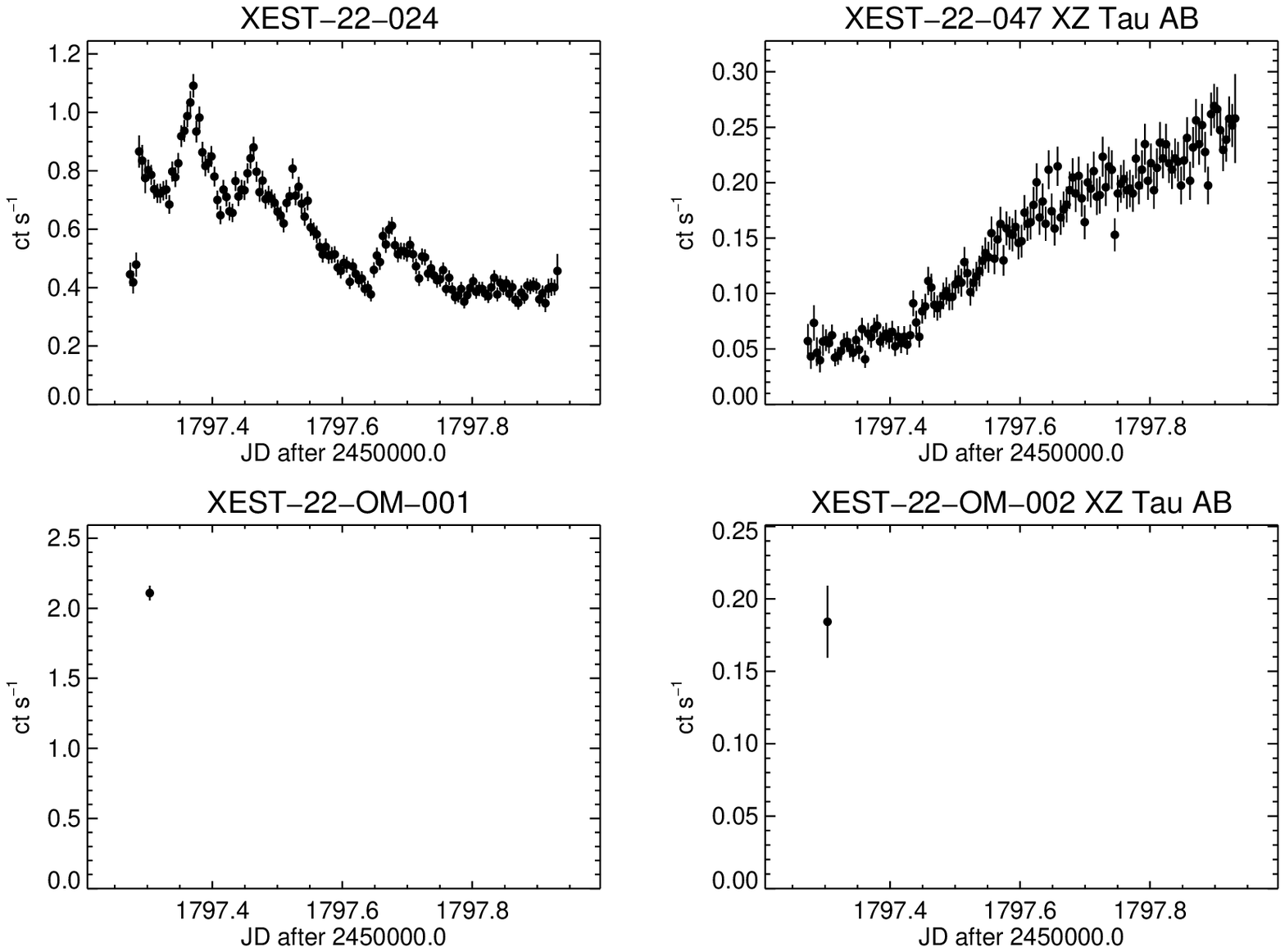}}
\resizebox{.9\textwidth}{!}{\includegraphics{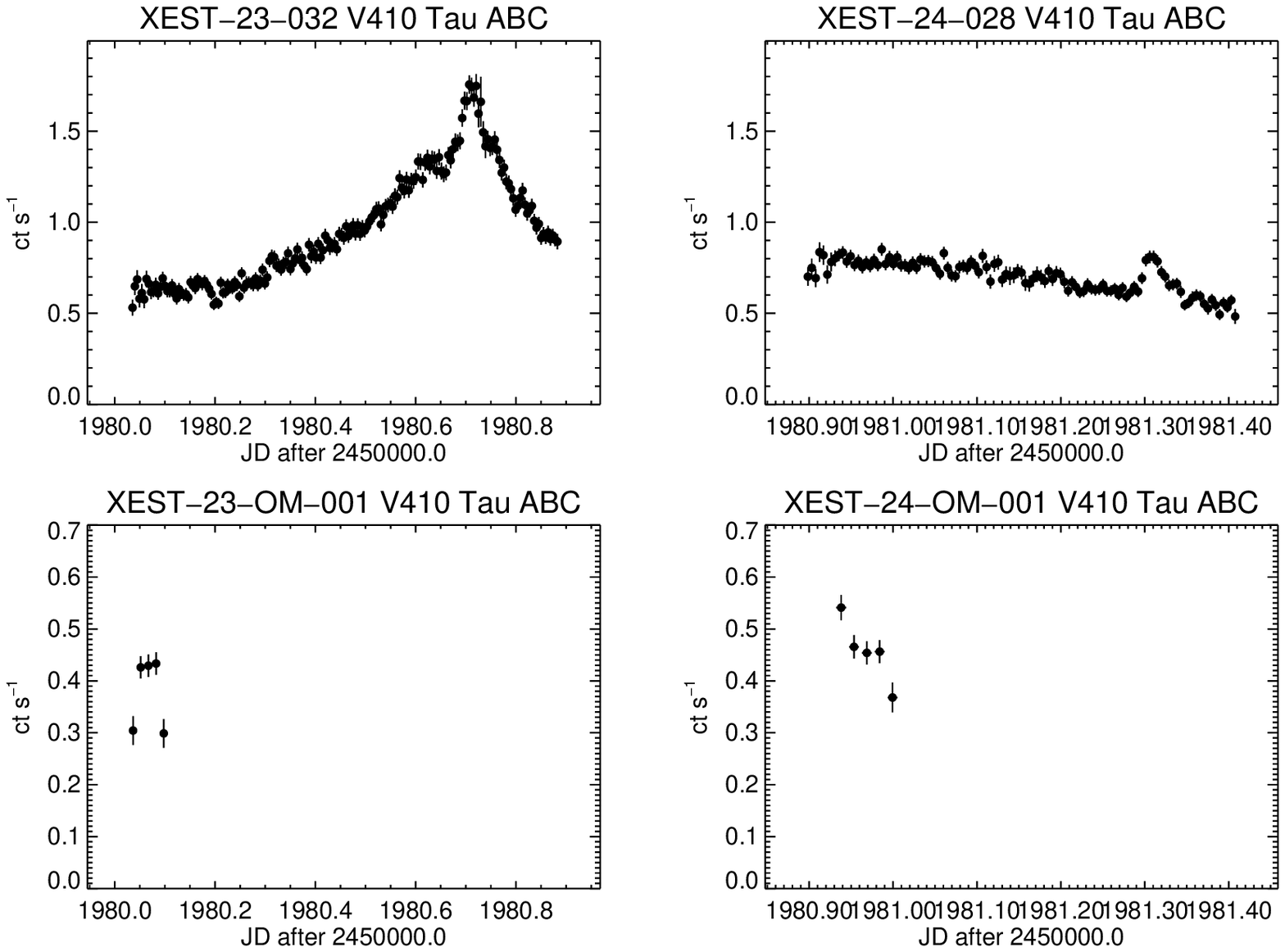}}
\caption{Light curves (continued).}
\end{figure*}

\clearpage\addtocounter{figure}{-1}

\begin{figure*}
\centering
\resizebox{.9\textwidth}{!}{\includegraphics{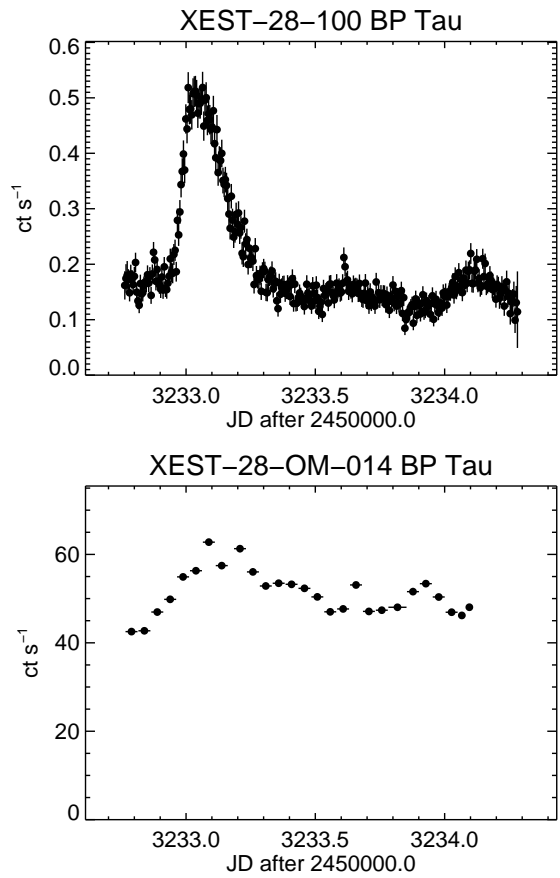}}
\caption{Light curves (continued).}
\end{figure*}

\clearpage\addtocounter{figure}{-1}

\end{document}

%% file: OMcat_paper.tex
\begin{table*}
\caption{XEST OM Catalogue (first thirty entries)}
\label{table:OMcat}
\centering
\begin{tabular}{r c c c c c c c c c r}     
\hline\hline
\#    & $\mathrm{RA_{corr}}$ & $\mathrm{\delta_{corr}}$ & $\Delta$ &  XEST       &       2MASS      & $\rho_1$ &   XEST   & $\rho_2$ &   Mag$^a$   & Signif.\\ 
      &          h m s       & $\degr$ $\arcmin$ $\arcsec$ & ($\arcsec$) &   OM          &           & ($\arcsec$) &       & ($\arcsec$) &    &         \\ 
\hline                    
   1  &      04 03 59.95     &      +26 23 31.8         &   0.60    &  06-OM-001  & 04035989+2623319 &  0.73  &          &        &$  20.62 \pm    0.09         $      &  15.2   \\
   2  &      04 04 00.62     &      +26 24 43.9         &   0.60    &  06-OM-002  & 04040056+2624441 &  0.81  &          &        &$  19.05 \pm    0.04         $      &  21.4   \\
   3  &      04 04 00.78     &      +26 23 29.8         &   0.57    &  06-OM-003  & 04040075+2623295 &  0.45  &          &        &$  16.79 \pm    0.01         $      & 114.3   \\
   4  &      04 04 02.24     &      +26 24 59.7         &   0.57    &  06-OM-004  & 04040217+2624595 &  0.84  &          &        &$  16.02 \pm    0.00         $      & 176.6   \\
   5  &      04 04 03.27     &      +26 22 24.6         &   0.63    &  06-OM-005  & 04040327+2622244 &  0.13  &          &        &$  19.95 \pm    0.09         $      &  10.4   \\
   6  &      04 04 03.59     &      +26 25 02.8         &   0.60    &  06-OM-006  &                  &        &          &        &$  18.33 \pm    0.03         $      &  38.5   \\
   7  &      04 04 04.69     &      +26 24 48.7         &   0.63    &  06-OM-007  & 04040467+2624478 &  0.84  &          &        &$  20.15 \pm    0.10         $      &   8.5   \\
   8  &      04 04 07.46     &      +26 19 10.4         &   0.61    &  06-OM-008  & 04040741+2619102 &  0.60  &          &        &$  19.19 \pm    0.04         $      &  19.6   \\
   9  &      04 04 08.17     &      +26 17 50.7         &   0.59    &  06-OM-009  & 04040814+2617508 &  0.33  &          &        &$  18.38 \pm    0.01         $      &  37.5   \\
  10  &      04 04 08.31     &      +26 19 18.6         &   0.63    &  06-OM-010  &                  &        &          &        &$  20.46 \pm    0.17         $      &   6.2   \\
  11  &      04 04 08.43     &      +26 19 49.0         &   0.63    &  06-OM-011  &                  &        &          &        &$  20.15 \pm    0.12         $      &   8.4   \\
  12  &      04 04 09.26     &      +26 19 10.1         &   0.61    &  06-OM-012  & 04040919+2619098 &  0.88  &          &        &$  19.26 \pm    0.12         $      &  18.1   \\
  13  &      04 04 09.35     &      +26 14 06.2         &   0.59    &  06-OM-013  & 04040936+2614059 &  0.32  &          &        &$  18.65 \pm    0.03         $      &  29.7   \\
  14  &      04 04 09.50     &      +26 15 16.7         &   0.64    &  06-OM-014  &                  &        &          &        &$  20.87 \pm    0.08         $      &   4.4   \\
  15  &      04 04 09.57     &      +26 22 03.8         &   0.58    &  06-OM-015  & 04040952+2622034 &  0.62  &          &        &$  17.38 \pm    0.02         $      &  78.6   \\
  16  &      04 04 09.87     &      +26 19 00.6         &   0.64    &  06-OM-016  &                  &        &          &        &$  20.51 \pm    0.16         $      &   6.0   \\
  17  &      04 04 10.24     &      +26 21 58.7         &   0.60    &  06-OM-017  & 04041018+2621581 &  0.98  &          &        &$  18.94 \pm    0.05         $      &  24.4   \\
  18  &      04 04 10.44     &      +26 22 57.3         &   0.64    &  06-OM-018  &                  &        &          &        &$  20.84 \pm    0.14         $      &   4.7   \\
  19  &      04 04 10.53     &      +26 19 01.3         &   0.58    &  06-OM-019  & 04041051+2619011 &  0.33  &          &        &$  17.67 \pm    0.03         $      &  63.8   \\
  20  &      04 04 11.02     &      +26 10 57.7         &   0.61    &  06-OM-020  & 04041104+2610573 &  0.51  &          &        &$  19.19 \pm    0.06         $      &  19.6   \\
  21  &      04 04 11.19     &      +26 20 04.0         &   0.61    &  06-OM-021  & 04041116+2620036 &  0.50  &          &        &$  19.57 \pm    0.05         $      &  14.3   \\
  22  &      04 04 11.22     &      +26 10 47.7         &   0.63    &  06-OM-022  &                  &        &          &        &$  20.21 \pm    0.19         $      &   7.7   \\
  23  &      04 04 11.34     &      +26 25 16.4         &   0.57    &  06-OM-023  & 04041129+2625162 &  0.65  &          &        &$  16.77 \pm    0.01         $      & 115.6   \\
  24  &      04 04 11.82     &      +26 11 41.1         &   0.64    &  06-OM-024  &                  &        &          &        &$  20.92 \pm    0.17         $      &   4.1   \\
  25  &      04 04 11.83     &      +26 25 03.8         &   0.62    &  06-OM-025  & 04041178+2625035 &  0.64  &          &        &$  19.62 \pm    0.13         $      &  13.3   \\
  26  &      04 04 11.85     &      +26 12 03.2         &   0.62    &  06-OM-026  &                  &        &          &        &$  19.63 \pm    0.09         $      &  13.5   \\
  27  &      04 04 11.89     &      +26 11 24.4         &   0.59    &  06-OM-027  & 04041191+2611239 &  0.58  &          &        &$  18.59 \pm    0.04         $      &  32.1   \\
  28  &      04 04 11.93     &      +26 19 12.0         &   0.60    &  06-OM-028  & 04041190+2619118 &  0.25  &          &        &$  19.00 \pm    0.05         $      &  22.8   \\
  29  &      04 04 12.66     &      +26 12 44.4         &   0.63    &  06-OM-029  &                  &        &  06-029  &  2.86  &$  20.13 \pm    0.12         $      &   8.7   \\
  30  &      04 04 13.03     &      +26 11 41.6         &   0.58    &  06-OM-030  & 04041305+2611414 &  0.27  &          &        &$  18.00 \pm    0.02         $      &  50.3   \\
\hline
\multicolumn{11}{l}{$^a$ $U$-band except if identified with $^\dagger$ (UVW1) or $^\ddagger$ (UVW2).}\\
\end{tabular}
\end{table*}

%% file: OM_TMConly_paper.tex
\begin{table*}
\caption{XEST OM catalogue of TMC members detected in X-rays and in the OM}
\label{table:OMTMConlycat}
\centering
\begin{tabular}{c c c c l c c r r r}     
\hline\hline
$\mathrm{RA_{corr}}$ & $\mathrm{\delta_{corr}}$ &   XEST OM      & XEST   & Name  &   Type & Mag$^a$  & \multicolumn{1}{c}{Rate$_\mathrm{OM}$} & \multicolumn{1}{c}{$L_\mathrm{X}^b$} & \multicolumn{1}{c}{Rate$_\mathrm{X}$}\\ 
\hline                    
04 14 12.92     &      +28 12 12.6         &      20-OM-003  &      20-042  &          V773 Tau ABC  &    3  &$  13.16 \pm  0.02         $      &  109.13  &        9.49        &    1.2113   \\
04 14 13.58     &      +28 12 49.4         &      20-OM-004  &      20-043  &                FM Tau  &    2  &$  14.44 \pm  0.07         $      &   33.76  &        0.53        &    0.0611   \\
04 14 17.02     &      +28 10 57.9         &      20-OM-005  &      20-046  &                CW Tau  &    2  &$  15.50 \pm  0.10         $      &   12.73  &        2.84        &    0.0015   \\
04 14 17.66     &      +28 06 09.8         &      20-OM-006  &      20-047  &                CIDA 1  &    2  &$  18.04 \pm  0.15         $      &    1.23  &        0.03        &    0.0031   \\
04 18 31.10     &      +28 27 16.2         &      23-OM-001  &      23-032  &          V410 Tau ABC  &    3  &$  15.85 \pm  0.15^\ddagger$      &    0.38  &        3.76        &    1.2910   \\
04 18 31.10     &      +28 27 16.2         &      24-OM-001  &      24-028  &          V410 Tau ABC  &    3  &$  15.69 \pm  0.11^\ddagger$      &    0.44  &        4.66        &    0.8534   \\
04 19 15.84     &      +29 06 26.7         &      28-OM-014  &      28-100  &                BP Tau  &    2  &$  13.01 \pm  0.10^\dagger $      &   46.80  &        1.37        &    0.2343   \\
04 21 43.27     &      +19 34 12.9         &      01-OM-022  &      01-028  &       IRAS 04187+1927  &    2  &$  19.91 \pm  0.31^\dagger $      &    0.08  &        0.91        &    0.0481   \\
04 21 59.45     &      +19 32 06.4         &      01-OM-039  &      01-045  &         T Tau N(+Sab)  &    2  &$  12.00 \pm  0.04^\dagger $      &  119.13  &        8.05        &    0.8189   \\
04 22 02.17     &      +26 57 31.5         &      11-OM-087  &      11-057  &             FS Tau AC  &    2  &$  17.57 \pm  0.00         $      &    1.88  &        3.22        &    0.1449   \\
04 22 04.86     &      +19 34 48.3         &      01-OM-042  &      01-054  &       RX J0422.1+1934  &    3  &$  19.58 \pm  0.08^\dagger $      &    0.11  &        3.11        &    0.3390   \\
04 22 16.76     &      +26 54 56.6         &      11-OM-126  &      11-079  &           CFHT-Tau 21  &    2  &$  18.78 \pm  0.18         $      &    0.62  &        0.15        &    0.0070   \\
04 26 53.56     &      +26 06 54.6         &      02-OM-017  &      02-013  &             FV Tau AB  &    2  &$  19.58 \pm  0.09         $      &    0.30  &        0.53        &    0.0212   \\
04 27 04.69     &      +26 06 15.9         &      02-OM-028  &      02-022  &              DG Tau A  &    2  &$  13.57 \pm  0.05         $      &   75.08  &        0.47$^{c}$  &    0.0290   \\
04 29 20.73     &      +26 33 40.3         &      15-OM-035  &      15-020  &                JH 507  &    3  &$  17.35 \pm  0.04         $      &    2.30  &        0.46        &    0.0382   \\
04 29 23.71     &      +24 33 00.2         &      13-OM-001  &      13-004  &             GV Tau AB  &    1  &$  19.20 \pm  0.06         $      &    0.42  &        0.95$^{c}$  &    0.0293   \\
04 29 41.59     &      +26 32 58.0         &      15-OM-112  &      15-040  &             DH Tau AB  &    2  &$  14.79 \pm  0.04         $      &   24.38  &        8.46        &    1.0729   \\
04 29 42.52     &      +26 32 49.1         &      15-OM-118  &      15-042  &             DI Tau AB  &    3  &$  15.99 \pm  0.01         $      &    8.11  &        1.57        &    0.0979   \\
04 30 44.25     &      +26 01 24.4         &      14-OM-258  &      14-057  &             DK Tau AB  &    2  &$  14.09 \pm  0.14         $      &   46.54  &        0.92        &    0.0989   \\
04 31 40.07     &      +18 13 57.2         &      22-OM-002  &      22-047  &             XZ Tau AB  &    2  &$  16.70 \pm  0.15^\ddagger$      &    0.17  &        0.96        &    0.2440   \\
04 31 50.61     &      +24 24 17.7         &      03-OM-001  &      03-005  &             HK Tau AB  &    2  &$  16.81 \pm  0.09         $      &    3.81  &        0.08        &    0.0045   \\
04 32 15.39     &      +24 28 59.3         &      03-OM-007  &      03-016  &             Haro 6-13  &    2  &$  19.22 \pm  0.19         $      &    0.41  &        0.80        &    0.0141   \\
04 32 18.84     &      +24 22 27.5         &      03-OM-008  &      03-019  &           V928 Tau AB  &    3  &$  17.65 \pm  0.04         $      &    1.76  &        1.05        &    0.0732   \\
04 32 30.58     &      +24 19 57.4         &      03-OM-015  &      03-022  &                FY Tau  &    2  &$  17.83 \pm  0.17         $      &    1.48  &        0.81        &    0.0822   \\
04 32 31.77     &      +24 20 02.9         &      03-OM-016  &      03-023  &                FZ Tau  &    2  &$  16.15 \pm  0.06         $      &    6.99  &        0.64        &    0.0232   \\
04 32 42.84     &      +25 52 31.3         &      19-OM-092  &      19-049  &              UZ Tau W  &    2  &$  14.23 \pm  0.04         $      &   40.86  &        0.89        &    0.0465   \\
04 32 43.05     &      +25 52 31.1         &      19-OM-094  &      19-049  &              UZ Tau E  &    2  &$  14.33 \pm  0.02         $      &   37.37  &        0.89        &    0.0465   \\
04 32 49.10     &      +22 53 03.0         &      17-OM-001  &      17-009  &                JH 112  &    2  &$  18.81 \pm  0.05         $      &    0.60  &        0.82        &    0.0357   \\
04 33 34.05     &      +24 21 16.9         &      04-OM-018  &      04-034  &                GI Tau  &    2  &$  14.34 \pm  0.07         $      &   36.97  &        0.83        &    0.0364   \\
04 33 34.54     &      +24 21 05.6         &      04-OM-020  &      04-035  &             GK Tau AB  &    2  &$  15.50 \pm  0.09         $      &   12.73  &        1.47        &    0.1404   \\
04 33 36.80     &      +26 09 49.6         &      18-OM-002  &      18-019  &             IS Tau AB  &    2  &$  18.52 \pm  0.11         $      &    0.79  &        0.66        &    0.0586   \\
04 33 51.98     &      +22 50 30.3         &      17-OM-066  &      17-058  &                CI Tau  &    2  &$  14.06 \pm  0.08         $      &   48.00  &        0.19        &    0.0204   \\
04 33 54.61     &      +26 13 26.8         &      18-OM-004  &      18-030  &             IT Tau AB  &    2  &$  18.01 \pm  0.05         $      &    1.25  &        6.49        &    0.4041   \\
04 34 55.42     &      +24 28 52.7         &      25-OM-003  &      25-026  &                AA Tau  &    2  &$  15.67 \pm  0.18^\ddagger$      &    0.45  &        1.24        &    0.0597   \\
04 35 20.93     &      +22 54 23.4         &      08-OM-005  &      08-019  &             FF Tau AB  &    3  &$  17.57 \pm  0.03         $      &    1.89  &        0.80        &    0.0681   \\
04 35 27.39     &      +24 14 58.5         &      12-OM-077  &      12-040  &                DN Tau  &    2  &$  13.27 \pm  0.04         $      &   98.91  &        1.15        &    0.1955   \\
04 35 40.94     &      +24 11 09.2         &      12-OM-096  &      12-059  &         CoKu Tau 3 AB  &    3  &$  20.58 \pm  0.60         $      &    0.12  &        5.85        &    0.5224   \\
04 35 41.80     &      +22 34 11.8         &      09-OM-070  &      09-022  &            KPNO-Tau 8  &    3  &$  20.64 \pm  0.04         $      &    0.11  &        0.50        &    0.0668   \\
04 35 47.36     &      +22 50 20.6         &      08-OM-032  &      08-037  &             HQ Tau AB  &    3  &$  15.01 \pm  0.04         $      &   20.03  &        2.48        &    0.3050   \\
04 35 51.10     &      +22 52 39.7         &      08-OM-034  &      08-043  &           KPNO-Tau 15  &    3  &$  19.75 \pm  0.13         $      &    0.25  &        2.62        &    0.2232   \\
04 35 52.77     &      +22 54 22.7         &      08-OM-036  &      08-048  &             HP Tau AB  &    2  &$  17.39 \pm  0.04         $      &    2.22  &        2.55        &    0.1156   \\
04 35 53.52     &      +22 54 08.6         &      08-OM-038  &      08-051  &          HP Tau/G3 AB  &    3  &$  18.21 \pm  0.07         $      &    1.05  &        1.29        &    0.0648   \\
04 35 54.16     &      +22 54 12.9         &      08-OM-040  &      08-051  &             HP Tau/G2  &    3  &$  13.63 \pm  0.04         $      &   70.80  &        9.65        &    0.8136   \\
04 35 56.84     &      +22 54 34.9         &      08-OM-044  &      08-058  &          Haro 6-28 AB  &    2  &$  18.94 \pm  0.12         $      &    0.53  &        0.25        &    0.0124   \\
04 39 20.87     &      +25 45 02.8         &      05-OM-002  &      05-013  &             GN Tau AB  &    2  &$  17.22 \pm  0.11         $      &    2.61  &        0.79        &    0.0108   \\
04 40 49.61     &      +25 51 19.2         &      07-OM-006  &      07-011  &                JH 223  &    3  &$  18.25 \pm  0.04         $      &    1.01  &        0.06        &    0.0131   \\
04 42 05.46     &      +25 22 56.0         &      10-OM-002  &      10-017  &       CoKuLk332/G2 AB  &    3  &$  18.68 \pm  0.06         $      &    0.68  &        3.26        &    0.1826   \\
04 42 07.33     &      +25 23 03.8         &      10-OM-003  &      10-018  &       CoKuLk332/G1 AB  &    3  &$  19.09 \pm  0.15         $      &    0.47  &        0.49        &    0.0470   \\
04 42 07.77     &      +25 23 12.3         &      10-OM-004  &      10-020  &           V955 Tau AB  &    2  &$  17.90 \pm  0.08         $      &    1.39  &        1.62        &    0.0576   \\
04 42 21.06     &      +25 20 34.1         &      10-OM-006  &      10-034  &                CIDA 7  &    2  &$  18.67 \pm  0.09         $      &    0.68  &        0.04        &    0.0028   \\
04 42 37.66     &      +25 15 37.2         &      10-OM-010  &      10-045  &                DP Tau  &    2  &$  15.84 \pm  0.03         $      &    9.29  &        0.10$^{c}$  &    0.0031   \\
\hline
\multicolumn{10}{l}{$^a$ $U$-band except if identified with $^\dagger$ (UVW1) or $^\ddagger$ (UVW2).}\\
\multicolumn{10}{l}{$^b$ $L_\mathrm{X}$ ($0.3-10$~keV) in 10$^{30}$ erg~s$^{-1}$ from the DEM method.}\\
\multicolumn{10}{l}{$^c$ $L_\mathrm{X}$ ($0.3-10$~keV) of the hard component from \citet{guedel06b}.}\\
\end{tabular}
\end{table*}

%% file: OM_TMCcand_paper.tex
\begin{table*}
\caption{XEST OM catalogue of TMC candidates detected in X-rays and in the OM}
\label{table:OMTMCcandcat}
\centering
\begin{tabular}{c c c c l  c r r r}     
\hline\hline
$\mathrm{RA_{corr}}$ & $\mathrm{\delta_{corr}}$ &   XEST OM      & XEST   & \multicolumn{1}{c}{Name}  & Mag$^a$  & \multicolumn{1}{c}{Rate$_\mathrm{OM}$} & \multicolumn{1}{c}{$L_\mathrm{X}^b$} & \multicolumn{1}{c}{Rate$_\mathrm{X}$}\\ 
\hline                    
04 04 24.45     &      +26 11 12.0         &      06-OM-079  &      06-041  &            (TMC cand)  &$  18.02 \pm  0.05         $      &    1.25  &         7.3        &      17.1   \\
04 14 52.31     &      +28 05 59.7         &      20-OM-010  &      20-071  &            (TMC cand)  &$  19.87 \pm  0.25         $      &    0.23  &         171        &       159   \\
04 22 15.72     &      +26 57 05.3         &      11-OM-122  &      11-078  &            (TMC cand)  &$  20.98 \pm  0.14         $      &    0.08  &          23        &       9.4   \\
04 22 27.29     &      +26 59 49.4         &      11-OM-150  &      11-088  &            (TMC cand)  &$  20.11 \pm  0.08         $      &    0.18  &         1.6        &       2.1   \\
04 29 35.98     &      +24 35 57.0         &      13-OM-002  &      13-010  &            (TMC cand)  &$  20.47 \pm  0.23         $      &    0.13  &          68        &      22.1   \\
04 29 36.29     &      +26 34 23.2         &      15-OM-093  &      15-034  &            (TMC cand)  &$  18.70 \pm  0.11         $      &    0.66  &          10        &      24.4   \\
04 30 25.21     &      +26 02 56.9         &      14-OM-163  &      14-034  &            (TMC cand)  &$  16.86 \pm  0.01         $      &    3.62  &        0.35        &       0.8   \\
04 33 33.07     &      +22 52 50.7         &      17-OM-028  &      17-043  &            (TMC cand)  &$  20.66 \pm  0.36         $      &    0.11  &         1.2        &      20.4   \\
04 33 52.54     &      +22 56 27.0         &      17-OM-070  &      17-059  &            (TMC cand)  &$  19.12 \pm  0.06         $      &    0.45  &          40        &      57.4   \\
04 33 55.64     &      +24 25 01.7         &      04-OM-044  &      04-060  &            (TMC cand)  &$  17.30 \pm  0.01         $      &    2.42  &        0.74        &       1.3   \\
04 35 52.88     &      +22 50 57.8         &      08-OM-037  &      08-049  &            (TMC cand)  &$  20.06 \pm  0.21         $      &    0.19  &         115        &       118   \\
04 35 58.93     &      +22 38 35.2         &      09-OM-143  &      09-042  &            (TMC cand)  &$  15.37 \pm  0.03         $      &   14.29  &         172        &       269   \\
\hline
\multicolumn{9}{l}{$^a$ $U$-band.}\\
\multicolumn{9}{l}{$^b$ $L_\mathrm{X}$ ($0.3-10$~keV) in 10$^{30}$ erg~s$^{-1}$ from the DEM method.}\\
\end{tabular}
\end{table*}